\documentclass[12pt]{iopart}

\usepackage{iopams}

\expandafter\let\csname equation*\endcsname=\relax
\expandafter\let\csname endequation*\endcsname=\relax

\usepackage[utf8]{inputenc}
\usepackage[T1]{fontenc}
\usepackage{amsmath,amssymb}
\usepackage{amsmath,amssymb}
\usepackage{mathrsfs}
\usepackage{physics}
\usepackage{graphicx}
\usepackage{bm}
\usepackage{hyperref}
\usepackage{xcolor}
\usepackage{wrapfig}

\newcommand{\de}{\mathrm{d}}

\newcommand{\avg}[1]{\left\langle #1 \right\rangle}

\newcommand{\ie}{{\it i.e.}}

\begin{document}

\title{Anderson localization on the Bethe lattice}

\author{Dafne Prado Bandeira$^{1}$, Angelica Vecchiarelli$^{2}$, Marco Tarzia$^{1,3}$}
\address{$^1$ Sorbonne Université, Laboratoire de Physique Théorique de la Matière Condensée, CNRS-UMR 7600, 4 Place Jussieu, 75252 Paris Cedex 05, France}
\address{$^2$ Dipartimento di Fisica, Sapienza Università di Roma, P.le A. Moro 5, 00185 Rome, Italy}
\address{$^3$ Institut Universitaire de France, 1 rue Descartes, 75005 Paris, France}

\begin{abstract}
Anderson localization---the suppression of wave propagation by interference in a sufficiently disordered medium---is one of the most remarkable phenomena in condensed matter physics. These lecture notes provide a pedagogical introduction to Anderson localization on the Bethe lattice, an infinite tree with fixed connectivity and no loops, and one of the few settings in which the transition can be studied exactly. The problem is important for two main reasons. First, the Anderson model on the Bethe lattice provides a mean-field, infinite-dimensional limit in which the transition point and critical behavior can be determined exactly. Second, it provides a simplified description of the high-dimensional configuration spaces of interacting quantum systems, and thus a useful toy model for many-body localization.

We focus on the Green's function formalism and the cavity method. After recalling standard diagnostics of localization based on transport, spectral statistics, and eigenfunction moments, we show that they are encoded in the statistics of the Green's functions, and that the full distribution of the local density of states is the natural order parameter of the transition. We derive the exact self-consistent cavity equations on the Bethe lattice, use them to characterize the two phases and critical behavior---including an analytic solution at large connectivity---and describe population-dynamics algorithms that solve them with high precision. We explain how an exponentially large correlation volume and the localized nature of the critical point account for the strong finite-size effects on random regular graphs, which led to a long debate over a possible intermediate non-ergodic phase. Finally, we discuss the mapping onto directed polymers in random media, interpreting localization as a freezing transition dominated by a few rare resonant paths.

\end{abstract}

\tableofcontents

%=== INTRO ===
\section{Introduction}\label{sec:intro}

Can disorder alone bring a quantum particle to a halt? In a perfectly periodic crystal the answer is clearly no: Bloch's theorem dictates that single-particle eigenstates are extended plane waves modulated by the lattice periodicity~\cite{bloch1928}, and an electron propagates ballistically through the whole sample. Real materials, however, inevitably contain disorder---impurities, defects, or structural imperfections---that scatter electrons and modify their propagation. The classical intuition, embodied in the Drude--Sommerfeld model~\cite{Drude1900,sommerfeld1927elektronentheorie}, is that scattering only slows the particle down. Electrons are treated as independent particles undergoing random collisions, characterized phenomenologically by a relaxation time; their motion becomes diffusive, and the conductivity decreases as the impurity concentration increases, but it never vanishes. Disorder can weaken conduction, but never halt it entirely.

In 1958 Anderson showed that this conclusion fails when quantum coherence is fully accounted for~\cite{anderson1958absence}. A quantum particle does not follow a single trajectory: the amplitudes associated with the many multiple-scattering paths interfere, and this interference can be destructive, enhancing the probability that the particle returns to its starting point. Beyond a critical disorder strength, interference becomes strong enough to completely trap the wavefunction, and the eigenstates become exponentially localized around a center,
\begin{equation}\label{eq:exp_loc}
    |\psi(\mathbf{r})|^2 \sim e^{-|\mathbf{r}-\mathbf{r}_0|/\xi},
\end{equation}
where $\xi$ is the localization length. Anderson localization is therefore not a quantitative crossover, as in the Drude picture, but a genuine quantum phase transition between a metal and an insulator, driven purely by disorder and interference, without any change in the band structure.

Subsequent work clarified the nature of this transition (see Refs.~\cite{lee1985disordered,evers2008anderson,lagendijk2009fifty,scardicchio2017perturbation,pascazio2023anderson} for reviews). Mott introduced the concept of the mobility edge, an energy separating localized from extended states~\cite{Mott1967}. The scaling theory of localization~\cite{abrahams1979scaling} revealed the crucial role of dimensionality: in one and two dimensions all states are localized for arbitrarily weak disorder (in the presence of time-reversal symmetry and in the absence of spin--orbit coupling), while in three dimensions~\cite{slevin1999corrections,slevin2014critical} and above~\cite{tarquini2017critical} a true metal--insulator transition occurs at a finite disorder strength. These ideas were confirmed by renormalization group calculations in dimension $2+\epsilon$, based on the nonlinear $\sigma$ model~\cite{hikami1992localization,wegner1979mobility,schafer1980disordered}. The interference mechanism is universal and applies to classical waves such as light and sound as well~\cite{John1984,Anderson1985}. Anderson localization has indeed been observed in a remarkably diverse range of platforms, from ultracold atomic gases~\cite{billy2008direct,roati2008anderson,kondov2011three,jendrzejewski2012three,semeghini2015measurement} to kicked rotors~\cite{chabe2008experimental} and classical elastic waves~\cite{hu2008localization}.

Despite this progress, an exact analytical description of the transition in finite dimensions is still lacking. The aim of these lecture notes is to present, in a pedagogical way, the one setting in which the problem can be solved exactly: the Anderson model on the Bethe lattice, an infinite tree in which every site has the same number $k+1$ of neighbors and there are no loops. There are three main reasons why this problem deserves a detailed study:
\begin{enumerate}
    \item[(i)] \emph{It is exactly solvable.} Because the Bethe lattice has no loops, the diagonal elements of the resolvent obey closed and exact self-consistent equations~\cite{abou1973selfconsistent}. These equations give access to the critical disorder, the critical exponents, and the full statistics of the local density of states, and can be both analyzed analytically and solved numerically with very high precision. This analytic solution has served for five decades as the mathematical sandbox that has shaped our physical intuition about the localization transition.
    \item[(ii)] \emph{It provides a mean-field description of the problem.} On a hypercubic lattice, the number of sites at distance $r$ from a given site grows only as a power of $r$, whereas the number of self-avoiding paths of length $r$ grows exponentially, as on a tree. In high dimension, these paths also rarely form short loops. Anderson's original analysis of the transition neglected these loops~\cite{anderson1958absence}, and Abou-Chacra, Anderson, and Thouless later showed that this approximation becomes exact on the Bethe lattice~\cite{abou1973selfconsistent}. The Anderson model on the Bethe lattice thus plays the role of a mean-field theory of the localization problem. As we will see, however, its critical behavior is very different from the one found in finite dimensions, which is consistent with the conjecture that the upper critical dimension of Anderson localization is infinite~\cite{tarquini2017critical,Castellani1986,mirlin1994distribution}.
    \item[(iii)] \emph{It is a toy model of many-body configuration spaces.} The quantum dynamics of an interacting many-body system can be recast as the hopping of a single fictitious particle on the graph of its many-body configurations~\cite{altshuler1997quasiparticle}. Such graphs are effectively high-dimensional---the number of configurations reachable with a given number of spin flips grows exponentially---and Anderson localization on the Bethe lattice provides a simplified mean-field description of many-body localization~\cite{basko2006metal,gornyi2005interacting,tikhonov2021anderson}.
\end{enumerate}

The history of the problem reflects these three aspects. The self-consistent equations were derived in 1973 by Abou-Chacra, Anderson, and Thouless~\cite{abou1973selfconsistent}, who used them to locate the transition. In the 1980s and early 1990s, the supersymmetric approach developed by Efetov, Zirnbauer, Verbaarschot, Mirlin, Fyodorov, and others~\cite{efetov1985anderson,efetov1987density,efetov1987anderson,zirnbauer1986localization,zirnbauer1986anderson,verbaarschot1988graded,mirlin1991localization,mirlin1991universality,fyodorov1991localization,fyodorov1992novel,mirlin1994statistical,mirlin1994distribution} unveiled the highly unconventional critical behavior of the model, and identified the full probability distribution of the local density of states as the natural order parameter of the transition. In parallel, the Bethe lattice became the setting in which the existence of extended states in the presence of disorder was first rigorously proven~\cite{klein1994absolutely,aizenman2006absolutely,froese2006transfer,aizenman2011extended}, a program that has recently led to rigorous proofs of the existence of a mobility edge~\cite{aggarwal2025mobility,liu2026mobility}. Finally, over the last fifteen years interest in the problem has surged again, largely due to its connection with many-body localization (MBL)~\cite{gornyi2005interacting,basko2006metal,Nandkishore2015,abanin2019MBLcolloquium,alet2018many,sierant2025many,Huse2014}---the phenomenon by which interacting disordered quantum systems can fail to thermalize---which can be viewed as Anderson localization in the many-body configuration space~\cite{altshuler1997quasiparticle,tikhonov2021anderson,tikhonov2021eigenstate,tarzia2020many,de2013ergodicity,biroli2017delocalized,biroli2020anomalous,logan2019many,garcia2022critical,herre2023ergodicity}.

This renewed interest came with a surprise. Numerical studies of Anderson localization have been extensively performed on random regular graphs (RRGs)~\cite{wormald1999models}, finite random graphs in which every node has exactly $k+1$ neighbors, and which converge locally to the Bethe lattice as their size $N\to\infty$ (see Sec.~\ref{sec:graph} for a precise definition). For a broad range of disorder strengths below the predicted transition, exact diagonalizations on RRGs found eigenstates that appeared to be neither localized nor ergodic: they seemed to occupy a vanishing fraction of the system, with multifractal statistics~\cite{biroli2012difference,de2013ergodicity}. These observations were interpreted as evidence for an intermediate delocalized but non-ergodic phase~\cite{de2013ergodicity,Kravtsov2018nonergodic,pino2020scaling,bera2018return,de2020subdiffusion}, which would have been at odds with the analytical solution, and which would have had important consequences for MBL.

The resolution of this puzzle is one of the main threads of these notes. It came from the realization that two concomitant and subtle features of the Bethe lattice solution produce dramatic finite-size effects~\cite{tikhonov2016anderson,garcia2017scaling,tarquini2016level,biroli2018delocalization}. First, the delocalized phase is characterized by a correlation volume $\Lambda_c$, which diverges exponentially fast upon approaching the transition and is already huge far from it: eigenstates look ergodic only on graphs much larger than $\Lambda_c$. Second, in the limit of infinite dimension the critical point has a \emph{localized} character~\cite{tarquini2017critical,zirnbauer1986localization,zirnbauer1986anderson,mirlin1991localization,mirlin1991universality,fyodorov1991localization,fyodorov1992novel,mirlin1994statistical}, so that systems smaller than $\Lambda_c$ inherit localized-like properties from it. The combination of these two elements produces a strongly non-ergodic behavior in a crossover regime which is often larger than the accessible system sizes, but which eventually disappears in the thermodynamic limit~\cite{tikhonov2016anderson,biroli2018delocalization,garcia2017scaling,tikhonov2019critical,biroli2022critical}. Several questions concerning the precise finite-size scaling on RRGs, however, are still debated~\cite{garcia2017scaling,garcia2022critical,sierant2023universality,PhysRevResearch.2.012020,vanoni2024renormalization}.

The second surprise is the critical behavior itself, which at first sight seems to contradict the scaling theory of localization. In finite dimensions, the diffusion coefficient vanishes at the transition as a power law, and the inverse participation ratio (IPR), which measures the inverse of the volume occupied by an eigenstate, vanishes continuously at the critical point~\cite{lee1985disordered,evers2008anderson,mirlin1994distribution}. On the Bethe lattice, instead, the diffusion coefficient vanishes exponentially when the transition is approached from the metallic side, while the IPR jumps discontinuously from $O(1/N)$ to a finite value at the transition~\cite{efetov1985anderson,efetov1987density,efetov1987anderson,zirnbauer1986localization,zirnbauer1986anderson,verbaarschot1988graded,mirlin1991localization,mirlin1991universality,fyodorov1991localization,fyodorov1992novel,mirlin1994statistical,mirlin1994distribution,tikhonov2019statistics,rizzo2024localized}.
Both intuitive arguments and quantitative calculations~\cite{mirlin1994distribution,baroni2024corrections}
have shown that these apparent discrepancies arise because the Bethe lattice corresponds to a
\emph{singular} infinite-dimensional limit of Anderson localization. This interpretation is
consistent with the conjecture that the upper critical dimension of Anderson localization is
infinite, $d_u=\infty$~\cite{tarquini2017critical,Castellani1986,mirlin1994distribution}.

\paragraph{The physical picture in a nutshell.} Before entering the technical details, let us anticipate in words the physical picture that emerges from the exact solution, which the reader may keep in mind throughout these notes:
\begin{itemize}
    \item What distinguishes a metal from an Anderson insulator is not the average density of states, which is smooth in both phases, but the way the spectral weight at a given energy is shared among the sites, as measured by the \emph{local} density of states. Deep in the metal, eigenstates are spread over the whole system, with squared amplitudes of order $1/N$, and every site receives a comparable share of the spectral weight. In the insulator, a typical site receives an exponentially small weight, and the spectral weight is concentrated on the rare sites close to which an eigenstate with that energy is localized. The order parameter of the transition is therefore a whole probability distribution rather than a single number.
    \item Localization is controlled by resonances, i.e., rare groups of sites whose energies happen to be very close to each other, between which the particle can tunnel efficiently. In the insulator, resonances are isolated and do not proliferate. The transition takes place when chains of resonances can extend arbitrarily far along the branches of the tree.
    \item On the metallic side, close to the transition, extended eigenstates are built by hybridizing many resonant clusters, each of them occupying a large region of the lattice, whose size is the correlation volume mentioned above. This volume, rather than a correlation length, is the relevant scale of the metallic phase. Since it diverges exponentially at the transition, the metal looks localized on all scales smaller than this huge volume.
    \item The whole critical behavior can be obtained from a single linear problem, which describes how resonances propagate along the branches of the tree, and which leads to the unconventional critical properties anticipated above.
    \item Finally, localization can be understood as the freezing transition of a directed polymer: in the metal, a particle created on a given site escapes through an exponentially large number of paths, whereas in the insulator only a few rare resonant paths survive, and even along these the particle cannot reach infinity.
\end{itemize}
These statements will be made precise in the following sections, and are summarized quantitatively in Sec.~\ref{sec:conclusions} (see Table~\ref{tab:main_results}).

\emph{How to read these notes.} These notes are aimed at graduate students and researchers who are familiar with basic quantum mechanics and with the notion of Green's function, but not necessarily with disordered systems or with the cavity method, and they are largely self-contained. Readers already familiar with random graphs can skip Sec.~\ref{sec:graph}. Sections~\ref{sec:popdyn}, devoted to numerical algorithms, and~\ref{sec:connection_DP}, devoted to the connection with directed polymers, are more specialized and can be read independently of each other. We do not attempt an exhaustive account of all approaches to the problem. In particular, we do not discuss the supersymmetric approach in detail, and only quote some of its main results (see Refs.~\cite{efetov1983supersymmetry,Mirlin_2000,evers2008anderson} for reviews). We focus on the analytical solution in the infinite-volume limit, and discuss numerical results on finite RRGs only when necessary (see, e.g., Refs.~\cite{tikhonov2016anderson,de2013ergodicity,Kravtsov2018nonergodic,pino2020scaling,bera2018return,de2020subdiffusion,vanoni2024renormalization,garcia2017scaling,garcia2022critical,sierant2023universality,PhysRevResearch.2.012020} for this extensive literature). Finally, we do not discuss trees with a finite number of generations (Cayley trees), for which boundary effects are dramatic and persist in the thermodynamic limit~\cite{biroli2020anomalous,sonner2017multifractality,tikhonov2016fractality}.

These notes are organized as follows. In Sec.~\ref{sec:model}, we introduce the Anderson model, discuss the standard diagnostics of the localization transition, summarize the known results in finite dimensions, and explain how MBL can be mapped onto a hopping problem in configuration space. In Sec.~\ref{sec:resolvent}, we introduce the resolvent and show that all the diagnostics of Sec.~\ref{sec:model} can be expressed in terms of it. Section~\ref{sec:graph} is a brief introduction to random graphs, which explains why the RRG---and not the Cayley tree---is the appropriate finite-size counterpart of the Bethe lattice. The cavity equations are derived in Sec.~\ref{sec:cavity}, and used in Sec.~\ref{sec:critical} to characterize the two phases and the critical behavior, including an analytic solution in the limit of large connectivity. Population dynamics algorithms to solve the cavity equations numerically are described in Sec.~\ref{sec:popdyn}. In Sec.~\ref{sec:connection_DP}, we discuss the analogy between Anderson localization and directed polymers in random media~\cite{derrida1988polymers}. Finally, Sec.~\ref{sec:conclusions} contains a summary and a discussion of open directions.

%=== MODEL ===
\section{Anderson Localization}\label{sec:model}

In this section, we introduce the Anderson model and establish the basic physical
picture of its localization transition that will be used throughout these notes.
We begin from the model's two trivial limits, the clean limit, in which the
eigenstates are extended and transport is possible, and the atomic limit, in which
the eigenstates are perfectly localized, and use the return probability as a first
dynamical diagnostic of these two regimes. Understanding the transition between
them is more subtle than either limit alone suggests: perturbing around the clean
limit breaks down at strong disorder, while perturbing around the atomic limit
requires care to avoid spurious divergences from resonances. We indicate how
Anderson resolved this difficulty through the locator expansion, without entering
into its derivation. We then turn to a more precise characterization of the
localized and delocalized phases in terms of transport, eigenfunction statistics,
spectral properties, and level statistics. Finally, we summarize some basic
results concerning the dependence of Anderson localization on spatial dimension
and its connection with many-body localization, one of the main motivations for
studying the problem on the Bethe lattice.

The minimal model exhibiting Anderson localization is the tight-binding Hamiltonian~\cite{ashcroft1976solid} with a random potential, known as the Anderson model. This describes non-interacting electrons hopping between lattice sites with random site energies. In second quantization for spinless fermions, the Hamiltonian takes the form
\begin{equation}
    H = \sum_i \varepsilon_i c_i^\dagger c_i - t \sum_{\langle i, j \rangle}(c_i^\dagger c_j + \text{h.c.}) ,\label{eq:H_AL_2nd}
\end{equation}
where $\langle i,j \rangle$ denotes the sum over nearest-neighbor pairs on the lattice. The on-site energies $\varepsilon_i$ are independent identically distributed random variables, typically drawn from a uniform distribution
\begin{equation} \label{eq:gamma_epsilon}
    \gamma(\varepsilon) = 
    \begin{cases}
        \frac{1}{W}, \quad \text{for} \quad \varepsilon \in \left[-\frac{W}{2},\frac{W}{2}\right]\\
        \,\,\,0\,, \quad \text{otherwise}
    \end{cases}
\end{equation}
with $W$ controlling the disorder strength. The hopping amplitude $t$ is taken to be constant between nearest neighbors, and we work in units where $\hbar = 1$. 

If we restrict to the one electron sector, an equivalent representation for the Hamiltonian is:
\begin{equation}
    H = \sum_i \varepsilon_i |i\rangle \langle i| - t \sum_{\langle i, j \rangle} \big( |i\rangle \langle j| + |j\rangle \langle i| \big) \label{eq:H_AL_1st}\,.
\end{equation}
This representation shows that $H$ is an $N \times N$ real symmetric random matrix written as the sum of ($-t$ times) the adjacency matrix of the underlying lattice plus a random diagonal matrix. It admits a complete set of normalized eigenstates with corresponding eigenvalues $\{\lambda_n, \psi_n\}$ satisfying $H|\psi_n\rangle = \lambda_n |\psi_n\rangle$.

The states $|i\rangle$ form an orthonormal basis corresponding to an electron localized on site $i$ of the lattice. The hopping term $t$ allows the electron to tunnel between neighboring sites, while the random on-site energies $\varepsilon_i$ break translational symmetry and introduce disorder into the system. The precise shape of the distribution $\gamma(\varepsilon)$ does not affect the universal physical properties of the localization transition; the uniform distribution is chosen for convenience.

Before analyzing the Anderson model in full generality, it is instructive to examine its two opposite limits on $d$-dimensional
hypercubic Euclidean lattices of linear size $L$ (so $N=L^d$ sites) with periodic
boundary conditions: the clean system ($W=0$) and the infinite disorder limit ($t=0$). These extremes capture the essential dichotomy between extended and localized behavior and provide intuition for the competition between hopping and disorder.

In the absence of disorder, the Hamiltonian in Eq.~\eqref{eq:H_AL_1st} becomes
translationally invariant and the hopping amplitude $t$ is the only energy scale.
The eigenstates are plane waves, which form an orthonormal basis of extended states:
\begin{equation}
    \psi_{\mathbf{k}}(\mathbf{r}) = \frac{1}{\sqrt{L^d}} e^{i\mathbf{k}\cdot\mathbf{r}}, \qquad
    \mathbf{k} = \frac{2\pi}{L}(n_1,\ldots,n_d), \quad n_{\mu}\in\{0,1,\ldots,L-1\}.
\end{equation}
The allowed momenta are those compatible with periodic boundary conditions,
$e^{ik_\mu L}=1$, which gives $L$ distinct values of $k_\mu$ per direction. Since
$e^{i\mathbf{k}\cdot\mathbf{r}}$ is $2\pi$-periodic in each $k_\mu$ for integer
lattice coordinates $\mathbf{r}$, these can equivalently be taken in the first
Brillouin zone $k_\mu\in[-\pi,\pi)$. The corresponding energy eigenvalues for each vector $\textbf{k}$ are then,
\begin{equation}
    E_{\mathbf{k}} = -2t\sum_{\mu=1}^{d} \cos(k_{\mu})
\end{equation}
(where the lattice spacing has been set to one), which fill a band of width $4dt$ centered at zero.
These eigenstates are extended over the entire system, with $|\psi_{\mathbf{k}}(\mathbf{r})|^2 = 1/L^d$.

A useful dynamical probe of localization is the return probability. Consider a
particle initialized at site $i=0$ at time $\tau=0$, i.e., $\psi(\tau=0,i)=\delta_{i,0}$,
and let $\psi(\tau,i)=\langle i|e^{-iH\tau}|0\rangle$ be its time-evolved wavefunction.
The return probability,
\begin{equation}\label{eq:retunr_prob}
    R(\tau) = |\psi(\tau,0)|^2,
\end{equation}
is the probability of finding the particle back at its starting point after a time $\tau$.
If the particle spreads throughout the system, $R(\tau)$ decays to zero at long times,
whereas if it remains localized around its initial position, $R(\tau)$ stays finite.
We compute it in the two limits introduced above, starting with the clean system.

In the absence of disorder, expanding $|0\rangle$ in plane waves gives the return
amplitude $\psi(\tau,0)=L^{-d}\sum_{\mathbf{k}}e^{-iE_{\mathbf{k}}\tau}$, which in the
thermodynamic limit $L\to\infty$ becomes an integral over the Brillouin zone. For large
times ($t\tau\gg1$), the integral is dominated by stationary points. Expanding around
$\mathbf{k}=0$, where $\cos(k_\mu)\approx1-k_\mu^2/2$ and
$E_{\mathbf{k}}\approx-2dt+t\,\mathbf{k}^2$, we obtain:
\begin{equation}
    \psi(\tau,0) = \int_{[-\pi,\pi]^d} \frac{d^d k}{(2\pi)^d} e^{-iE_{\mathbf{k}} \tau}
               \propto \frac{e^{2idt\tau}}{(t\tau)^{d/2}}.
\end{equation}
Hence, the return probability decays as
\begin{equation}
    R(\tau)  \sim \frac{1}{\tau^{d}},
\end{equation}
vanishing at long times. This power-law decay signals ballistic transport (for more details see \cite{scardicchio2017perturbation}), characteristic of a translationally invariant system where eigenstates are extended and the particle spreads indefinitely. Indeed, in the clean system a wavepacket spreads ballistically, $|\mathbf{r}| \sim t\tau$, so that at time $\tau$ the particle is spread over a volume $\sim (t\tau)^d$ and $R(\tau) \sim (t\tau)^{-d}$. As discussed below, a weak disorder changes this behavior at long times into diffusive spreading, $|\mathbf{r}|^2 \sim D \tau$, for which $R(\tau) \sim (D\tau)^{-d/2}$. Throughout these notes, when we refer to transport in the extended phase of the disordered system, we will always mean diffusive transport; ballistic motion only occurs in the clean limit, or at times shorter than the scattering time.

In the opposite limit ($t=0$), the hopping amplitude vanishes and the Hamiltonian becomes purely diagonal. Therefore, the eigenstates are perfectly localized on single sites:
\begin{equation}
    \psi_n(j) = \delta_{nj}, \qquad H|n\rangle = \varepsilon_n|n\rangle.
\end{equation}
The density of states inherits the distribution of the random energies, $\rho(E) = \gamma (E)$.

Dynamics in this limit are trivial: a particle initially at site $|0\rangle$ remains there indefinitely,
\begin{equation}
    |\psi(\tau)\rangle = e^{-iH\tau}|0\rangle = e^{-i\varepsilon_0 \tau}|0\rangle,
\end{equation}
so the return probability is constant,
\begin{equation}
    R(\tau) = 1.
\end{equation}

To understand the transition between these two opposite limits, the clean system ($W=0$) and the atomic limit ($t=0$), we must explore the intermediate regime where both hopping and disorder are present. The natural approach is perturbation theory, but the choice of the unperturbed Hamiltonian is crucial.

One might attempt to treat the disorder as a perturbation to the clean system~\cite{scardicchio2017perturbation}. Starting from the Bloch states of the periodic lattice, Fermi's golden rule gives a scattering rate $1/\tau_{\rm s} \sim W^2 \rho(E)$, where $\rho(E)$ is the clean density of states,
leading to a diffusion coefficient $D \sim v^2 \tau_{\rm s} \sim t^3/W^2$ that decreases with disorder but remains finite for any $W$. This approach corresponds physically to a semiclassical description of motion:
ballistic propagation between elastic scattering events off impurities. At times much longer than the scattering time $\tau_{\rm s}$, the resulting random walk is diffusive, with the
diffusion coefficient $D$ given above. Such a picture neglects the quantum interference between multiple scattering paths that is essential for localization, and implicitly assumes continuity of the spectrum---an assumption that fails in the localized phase, as we will see in future sections. 
As a consequence, this expansion breaks down at strong disorder and is therefore inadequate for describing the localized phase or locating the transition.

Rather than perturbing in the disorder, one may instead treat the hopping as a perturbation about the atomic limit of infinite disorder. In this picture, the unperturbed Hamiltonian $H_0 = \sum_i \varepsilon_i |i\rangle\langle i|$ consists of isolated sites with random on-site energies $\varepsilon_i$; its eigenstates are strictly localized, $|\psi_i\rangle = |i\rangle$, with eigenvalues $\varepsilon_i$. Writing the full Hamiltonian as $H = H_0 - tT$, where $T$ is the adjacency operator, standard perturbation theory yields the following expansion for the amplitude of the $i$-th eigenstate on site $j$:
\begin{equation}
    \psi_i(j) = \delta_{j,i}
    - t\,\frac{T_{ij}}{\varepsilon_{i} - \varepsilon_{j}}
    + t^2 \sum_{k} \frac{T_{ik}\,T_{kj}}
      {(\varepsilon_{i} - \varepsilon_{j})(\varepsilon_{i} - \varepsilon_{k})}
    - \cdots
    \label{eq:eigenfunction_expansion}
\end{equation}
Since the matrix elements $T_{ij} = \langle j|T|i\rangle$ equal one if $i$ and $j$ are nearest neighbors and zero otherwise, the first non-vanishing term in the expansion of $\psi_i(j)$ appears at order equal to the length of the shortest path between $i$ and $j$. The perturbative series can therefore be recast as a sum over all paths on the lattice connecting $i$ to $j$:
\begin{equation}
    \psi_i(j) = \sum_{\mathcal{P}:\, i \to j}
    \prod_{m \in \mathcal{P}\setminus \{i\}} \frac{-t}{\varepsilon_i - \varepsilon_m} \, .
    \label{eq:eigenfunction_expansion1}
\end{equation}
At high orders, this expansion includes paths that revisit the same site arbitrarily many times. Whenever a site $m$ hosts a \emph{resonance}, $|\varepsilon_i - \varepsilon_m| < t$, paths crossing $m$ repeatedly contribute factors $-t/(\varepsilon_i - \varepsilon_m)$ at each visit, which can become arbitrarily large. Since the number of resonant sites scales with the system volume in the large-$N$ limit, one might naively conclude that the perturbative expansion diverges even at arbitrarily small $t$, and that the localized phase is always unstable.

Anderson recognized, however, that these divergent contributions from repeated visits to resonant pairs of sites can be systematically resummed into a renormalized self-energy, which effectively removes the anomalously small energy denominators~\cite{anderson1958absence,Pietracaprina_2016,scardicchio2017perturbation}. This path expansion in Eq.~\eqref{eq:eigenfunction_expansion1} is commonly referred to as locator expansion. The resulting series runs only over \emph{self-avoiding paths}---walks that never revisit a site---and is therefore free of such divergences~\cite{anderson1958absence}. As the ratio $t/W$ increases, the probability of encountering resonances along a self-avoiding path grows, and the series eventually diverges at a critical value $(t/W)_c$. This divergence marks the Anderson transition from the localized to
the extended phase.

The full derivation of these results lies beyond the scope of this review. Nevertheless, the locator expansion can be resummed exactly on tree-like structures such as the Bethe lattice~\cite{abou1973selfconsistent}, where it becomes equivalent to the cavity method presented in Sec.~\ref{sec:cavity}.

\subsection{Diagnostics of Anderson localization}\label{sec:diagnostics}

The extreme limits examined previously suggest the two phases through their
transport properties. At weak disorder, the system is a conductor: the diffusion
coefficient $D$ is finite, and the conductivity $\sigma$ obeys the Einstein relation
$\sigma = e^2 D \rho(E_F)$, signaling diffusive transport. In the infinite disorder
limit, the system is an insulator: $D = 0$, $\sigma = 0$, and the return probability
remains constant, $R(\tau) = 1$, indicating complete absence of motion. Between these
extremes lies the Anderson transition, where the system crosses from delocalized to
localized behavior at a critical disorder strength $W_c$.

These dynamical signatures are intimately linked to the spectral properties of the Hamiltonian. A more fundamental characterization of the phases, therefore, comes from the classification of the spectrum into pure point, absolutely continuous, and singular continuous components. The nature of the spectrum determines the long‑time behavior of wavepackets \cite{morters2008growth}: the absolutely continuous spectrum is associated with extended states and transport, while the pure point spectrum corresponds to localized states and absence of transport. The singular continuous case occurs only at the critical point itself, where in finite dimensions eigenfunctions exhibit multifractal statistics \cite{evers2008anderson, Mirlin_2000}. As we will see in Sec.~\ref{sec:critical}, the Bethe lattice is peculiar in this respect, since there the critical point has a localized character.

A complementary characterization focuses on the spatial structure of individual eigenstates. For a normalized eigenstate $|\psi_{n}\rangle$, define the $q$-th moment of its amplitude
\begin{equation}
    I_q^{(n)} = \sum_i |\langle i | \psi_{n} \rangle|^{2q}, \qquad q \in \mathbb{R}_+.
\end{equation}
The case $q=2$ is known as the inverse participation ratio (IPR) and measures the inverse of the effective number of sites over which the eigenstate is spread. More generally, the scaling of $I_q$ (averaged over all eigenstates at fixed energy and over the disorder realizations) with system size $N$ defines the exponents $\zeta(q)$ via
\begin{equation}
    I_q \sim N^{-\zeta(q)},
\end{equation}
or equivalently the fractal dimensions $D_q = \zeta(q)/(q-1)$. 

These exponents have a simple geometric interpretation. Suppose that the sites where
the amplitude scales as $|\langle i|\psi_{n}\rangle|^2 \sim N^{-\alpha}$ form a set of
$\sim N^{f(\alpha)}$ sites, so that $f(\alpha)$, the multifractal spectrum, characterizes the scaling of the number of sites with exponent $\alpha$. Then
$I_q \sim \int d\alpha\, N^{f(\alpha)-q\alpha}$, which for large $N$ is dominated by
its saddle point:
\begin{equation}
    \zeta(q) = \min_{\alpha}\big\{q\alpha - f(\alpha)\big\}, \qquad
    D_q = \frac{1}{q-1}\min_{\alpha}\big\{q\alpha - f(\alpha)\big\},
\end{equation}
i.e., within the standard multifractal formalism, $\zeta(q)$ is the Legendre transform of $f(\alpha)$, and the two encode the same information. In terms of these quantities, the spectral regimes are characterized as follows:
\begin{itemize}
    \item In the localized phase (pure point spectrum), eigenfunctions are exponentially confined to a finite region, so $I_q^{(n)} \sim \text{const.}$ and $\zeta(q)=0$ for all $q>0$. Equivalently, $D_q = 0$.
    \item In the extended phase (absolutely continuous spectrum), eigenfunctions are uniformly delocalized with $|\langle i|\psi_{n}\rangle|^2 \sim 1/N$, giving $I_q \sim N^{1-q}$ and thus $\zeta(q) = q-1$, $D_q = 1$.
    \item At the critical point (singular continuous spectrum), eigenfunctions exhibit multifractality: they are neither localized nor uniformly spread over the system, with scale-invariant amplitude fluctuations. In this regime, $0 < \zeta(q) < q-1$ (for $q>1$) and $0 < D_q < 1$, with $\zeta(q)$ a nontrivial, $q$-dependent function encoding the multifractal properties of the states~\cite{rodriguez2011multifractal,evers2008anderson}. This is the generic situation in finite dimensions; on the Bethe lattice, instead, eigenstates at the critical point are localized-like, with $D_q=0$ and a finite IPR (see Sec.~\ref{sec:critical}).
\end{itemize}

In the Legendre transform picture, a uniformly extended state is characterized by the single exponent
$\alpha=1$, with $f(1)=1$, while localized states have their finite probability weight concentrated on $O(1)$ sites, corresponding to $\alpha=0$ and $f(0)=0$. At criticality,
by contrast, a wide range of exponents contributes and $f(\alpha)$ becomes a
nontrivial function.

Anderson's key insight was that in a disordered system, the pure point spectrum can persist even in the thermodynamic limit---the spectrum remains discrete, i.e., a sum of $\delta$-peaks, eigenfunctions remain localized, and transport is absent. This is the Anderson localized phase. The transition between spectral types at a critical disorder $W_c$ constitutes a genuine phase transition, driven purely by disorder without any change in band structure.

Another probe of the localization transition comes from the statistical properties of the energy spectrum itself. Consider the eigenvalues $\lambda_i$ of the Hamiltonian $H$, Eq.~\eqref{eq:H_AL_1st}, sorted in increasing order. Define the level spacings $s_i = (\lambda_{i+1} - \lambda_i)/\Delta$, where $\Delta$ is the mean level spacing in the energy window of interest. The distribution $P(s)$ of these spacings provides a sharp diagnostic that distinguishes the phases.

In the localized phase ($W > W_c$), eigenfunctions with nearby energies are typically localized around distant spatial regions and have negligible overlap, which vanishes exponentially with the system size. Consequently, their energy levels are effectively uncorrelated. 
Deep in the localized phase, where $t/W \ll 1$, the hopping is a small perturbation. To leading order, the eigenvalues are simply the on-site energies $\varepsilon_i$, which are independent random variables. Hopping induces corrections of order $t^2/W$ that do not introduce significant correlations between levels. The resulting level statistics are therefore Poissonian,
\begin{equation}
    P_{\mathrm{P}}(s) = e^{-s},
\end{equation}
characterized by the absence of level repulsion: $P(s)$ remains finite as $s\to 0$, reflecting the independence of nearby levels.

In the delocalized phase ($W < W_c$), eigenfunctions with nearby energies overlap significantly, inducing strong correlations between eigenvalues. The level statistics are then described by the universal ensembles of random matrix theory (see~ \cite{potters2020first, Livan_2018} for a comprehensive review). For systems with time-reversal symmetry (the case relevant to the $d$-dimensional Anderson model with real hoppings), the Hamiltonian belongs to the Gaussian Orthogonal Ensemble (GOE). The level spacing distribution is well approximated by the Wigner surmise
\begin{equation}
    P_{\mathrm{GOE}}(s) = \frac{\pi s}{2} e^{-\pi s^2/4},
\end{equation}
which exhibits level repulsion: $P(s) \sim s$ as $s\to 0$, indicating that eigenvalues cannot be arbitrarily close \cite{wigner1958distribution}. For systems without time-reversal symmetry, the Gaussian Unitary Ensemble (GUE) gives $P(s) \sim s^2$ at small $s$.

The transition from Poisson to GOE statistics at the Anderson critical point provides a precise numerical method to locate the mobility edge and has been extensively studied in the literature \cite{tarquini2017critical}.

\subsection{Known results for Anderson localization and motivation for the Bethe lattice limit}

Having established the diagnostics of localization, we now summarize the key known results for the Anderson model in different spatial dimensions. These results reveal a rich dependence on dimensionality that will motivate our focus on the Bethe lattice.

In one dimension, all single-particle states are localized for any nonzero disorder strength. This fundamental result was first established by Mott and Twose \cite{mott_twose_1961} and subsequently confirmed by rigorous mathematical proofs \cite{borland1963nature, ishii1973localization}. The absence of a delocalized phase in one dimension has also been demonstrated experimentally through the localization of ultracold atomic gases in disordered optical potentials \cite{Aspect2009anderson}.

The localization length in one dimension can be computed exactly using the transfer matrix method \cite{Thouless1972ARB}. The Schrödinger equation is recast as a product of $2\times 2$ matrices, and the growth rate of the wavefunction is determined by the Lyapunov exponent $\gamma$, defined as the limiting growth rate of the norm of this product \cite{comtet_lyapunov_2013}. For one-dimensional disordered systems, it can be proven that $\gamma > 0$ for any energy $E$ and any disorder strength $W>0$, implying exponential localization of all eigenstates \cite{ishii1973localization, Texier2013}. Consequently, the critical disorder strength is $W_c = 0$, so there is no metallic phase for any finite value of the disorder. 

The two-dimensional case is more subtle but shares the same ultimate fate: all states are localized for any finite disorder. According to the scaling theory of localization \cite{abrahams1979scaling}, $d=2$ is identified as the lower critical dimension for the Anderson transition in systems with time-reversal symmetry. This result follows from a one-parameter scaling hypothesis for the dimensionless conductance $g(L)$. The associated beta function,
\begin{equation}
    \beta(g) = \frac{d \ln g(L)}{d \ln L},
\end{equation}
describes how the conductance changes with linear system size $L$ (the total number of sites is $N = L^d$). A fixed point $\beta(g)=0$ would signal a metal-insulator transition, separating regimes where $g(L)$ grows with $L$ (metallic) from those where it decays (insulating). The scaling analysis shows that for $d \leq 2$, $\beta(g) < 0$ for all $g$, implying that all states are localized for any finite disorder. For $d > 2$, a fixed point appears at a finite $g_c$, corresponding to a true metal-insulator transition at a non-zero critical disorder $W_c$.

Despite the absence of a true metallic phase, in two dimensions the localization
length $\xi$ diverges exponentially in the weak-disorder limit, $\ln \xi \sim (t/W)^2$,
unlike the power-law divergence found in $d=1$.
Therefore, finite systems of experimentally or numerically accessible sizes may appear
delocalized, exhibiting diffusive transport on length scales shorter than $\xi$
\cite{tarquini2017critical}. This has historically led to debate, with some numerical
studies of insufficiently large systems mistakenly suggesting an Anderson transition in
two dimensions \cite{kramer1993localization}.

The signatures of localization in two dimensions manifest as quantum corrections to the classical Drude conductivity, known as weak localization \cite{abrahams1979scaling, abrahams1980critical}. These corrections arise from constructive interference between time-reversed scattering paths, which enhances the probability of backscattering.

In contrast to one and two dimensions, for $d > 2$ a genuine metal-insulator transition occurs at a finite critical disorder $W_c$. The scaling arguments that identified $d=2$ as the lower critical dimension were later substantiated by a renormalization group analysis in $d=2+\varepsilon$ dimensions \cite{hikami1992localization, foster2009termination} of the non-linear $\sigma$ model (NL$\sigma$M)~\cite{wegner1979mobility, schafer1980disordered, efetov1983supersymmetry}. These analyses confirmed that for $d>2$, a fixed point appears at a finite $g_c$, corresponding to a true metal-insulator transition.

A key feature of the transition in $d>2$ is the emergence of a mobility edge in the $W$-$E$ plane: for a fixed disorder strength $W < W_c$, states near the band center remain extended while those in the band tails become localized. Approaching the transition from the localized side, the localization length diverges as
\begin{equation}
    \xi(W) \sim |W - W_c|^{-\nu},
\end{equation}
where $\nu$ is the critical exponent characterizing the transition. This divergence mirrors the behavior of the correlation length in conventional continuous phase transitions.

Numerical studies in dimensions $d=3$ through $6$ show that $\nu$ decreases smoothly with increasing dimension, from $\nu \approx 1.57$ in $d=3$ to $\nu \approx 0.84$ in $d=6$ \cite{tarquini2017critical, mirlin1994distribution, mirlin1994statistical}. In particular, the exponent shows no sign of saturation at a finite-dimensional upper critical value, continuing to decrease as the dimension increases. This behavior has led several authors to propose that the upper critical dimension for Anderson localization is effectively infinite \cite{GarcaGarca2007, Castellani1986}.

From a rigorous mathematical perspective, the localized regime at large disorder or near spectral edges is well understood: it has been proven that eigenfunctions are exponentially localized \cite{frohlich1983absence, aizenman1993localization} and that level statistics are Poissonian \cite{minami1996local, elgart2009lifshitz}. Delocalization, however, is more challenging. Rigorous proofs of extended states exist only on the Bethe lattice
\cite{klein1994absolutely, aizenman2006absolutely, froese2006transfer}, a tree-like graph on which the Anderson model provides an analytically tractable version of the infinite-dimensional limit. In this setting, recent works have also established the existence of a mobility edge separating localized and extended states at sufficiently large connectivity \cite{aggarwal2025mobility, liu2026mobility}. In finite dimensions, no rigorous proof of delocalization has been achieved, but extensive numerical evidence supports the conjecture that eigenfunctions are extended and level statistics follow GOE in the thermodynamic limit.

The Bethe lattice (see Sec.~\ref{sec:graph} for a precise definition), which will be our focus in the following sections, provides the natural setting to study the mean-field (infinite-dimensional) limit of the problem. On this tree-like structure, where each node has fixed coordination number $k+1$ and loops are absent, the Anderson localization transition can be solved exactly.
The self-consistent equations derived by Abou-Chacra, Anderson and Thouless \cite{abou1973selfconsistent} provide the starting point for a wealth of analytical investigations ~\cite{evers2008anderson, biroli2010anderson, rizzo2024localized, tikhonov2019statistics, tikhonov2019critical, efetov1985anderson,efetov1987density,efetov1987anderson,zirnbauer1986localization,zirnbauer1986anderson,verbaarschot1988graded,mirlin1991localization,mirlin1991universality,fyodorov1991localization,fyodorov1992novel,mirlin1994statistical,mirlin1994distribution} that yield the critical point and the corresponding critical behavior.

Beyond its intrinsic importance for single-particle Anderson localization, the Bethe lattice has recently attracted renewed interest because of its deep connection with many-body localization (MBL). MBL is a dynamical quantum phase transition occurring in interacting disordered systems, where the combined effect of disorder and interactions prevents thermalization, and the system retains memory of its initial conditions even at high energy densities \cite{gornyi2005interacting, basko2006metal, Nandkishore2015, abanin2019MBLcolloquium, alet2018many, sierant2025many, Huse2014}.

To understand this connection, consider a specific paradigmatic model for MBL, namely a one-dimensional chain of $L$ spins with random-fields and random-exchanges Ising model in presence of a transverse field, studied in Ref.~\cite{imbrie2016diagonalization} (with open boundary conditions):
\begin{equation}
\label{eq:imbrie}
H_I = \sum_{i=1}^{L-1} \Delta_i \sigma_i^z \sigma_{i+1}^z + \sum_{i=1}^{L} h_i \sigma_i^z + \sum_{i=1}^{L} \Gamma_i \sigma_i^x,
\end{equation}
where $\sigma_i^x$ and $\sigma_i^z$ are the standard Pauli matrices acting on site $i$, and the couplings $\Delta_i$, $h_i$, and $\Gamma_i$ are independent random variables. The first term describes nearest-neighbor interactions, the second introduces on-site disorder via a longitudinal field, and the third (transverse field) generates quantum fluctuations by inducing spin flips.

The computational basis formed by the eigenstates of all $\sigma_i^z$ operators,
\begin{equation}
|\alpha\rangle = |s_1, s_2, \dots, s_L\rangle, \qquad s_i = \pm 1,
\end{equation}
spans the $2^L$-dimensional Hilbert space. In this basis, the first two terms of Eq.~\eqref{eq:imbrie} are diagonal, giving configuration energies
\begin{equation}\label{eq:config_E}
E_\alpha = \sum_{i=1}^{L-1} \Delta_i s_i s_{i+1} + \sum_{i=1}^{L} h_i s_i.
\end{equation}
The transverse field term $\sigma_i^x$ flips the spin on site $i$,
\begin{equation}
\sigma_i^x |s_1,\dots,s_i,\dots,s_L\rangle = |s_1,\dots,-s_i,\dots,s_L\rangle,
\end{equation}
and is therefore off‑diagonal. The full Hamiltonian connects basis states that differ by a single spin flip:
\begin{equation}
\langle \alpha | H_I | \beta \rangle = 
\begin{cases}
E_\alpha, & \text{if } \alpha = \beta,\\
\Gamma_i, & \text{if } |\alpha\rangle \text{ and } |\beta\rangle \text{ differ only at spin } i,\\
0, & \text{otherwise}.
\end{cases}
\end{equation}

\begin{figure}
    \centering
    \includegraphics[width=0.45\linewidth]{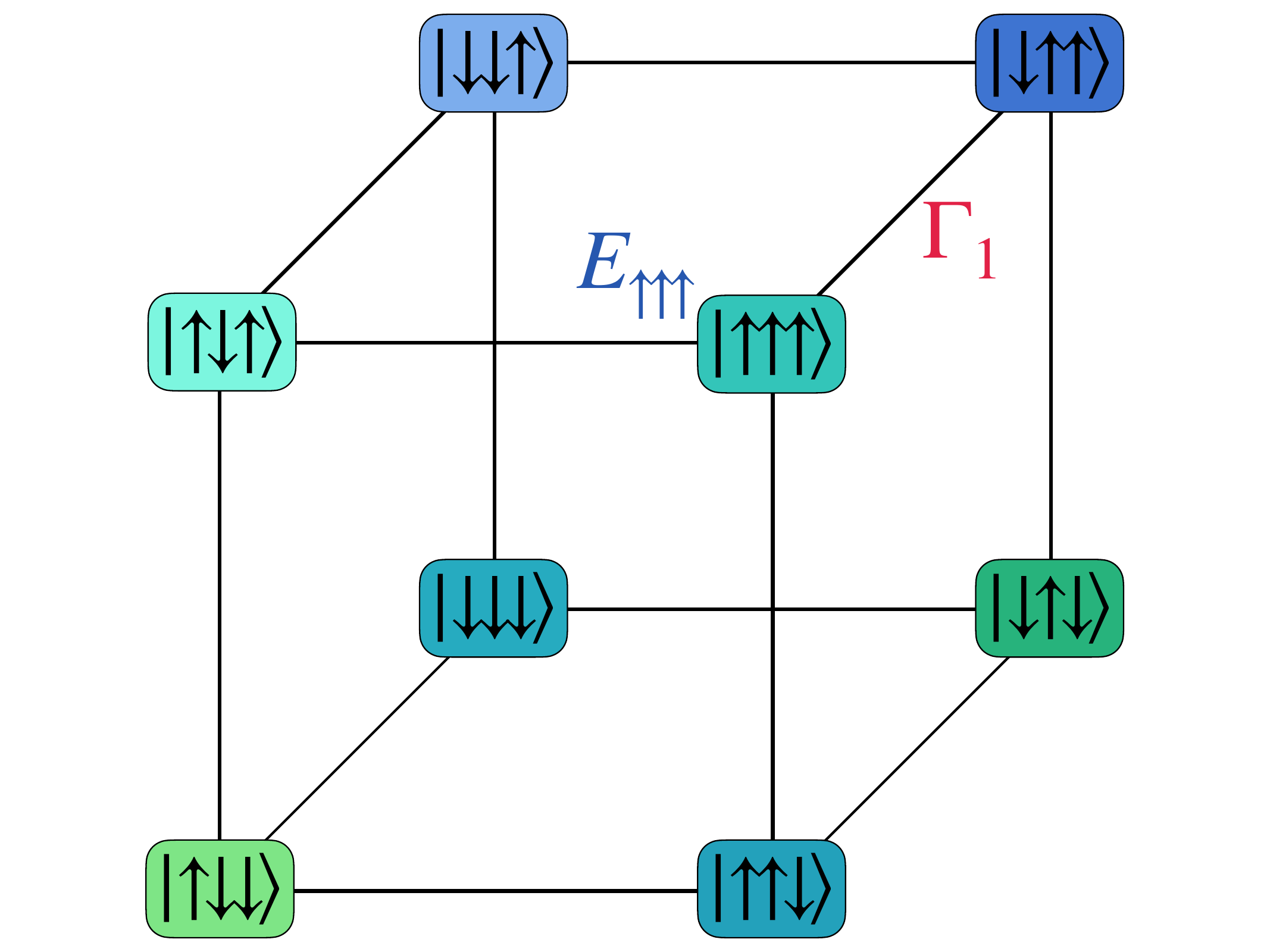}
    \caption{Schematic representation of the configuration space of the spin chain (Eq.~\eqref{eq:imbrie}) for $L=3$ in the basis of the simultaneous eigenstates of the $\{\sigma_i^z \}$ operators. Each vertex corresponds to one of the $2^3=8$ spin configurations. To each vertex, a random energy is assigned (indicated by color) containing linear combinations of the random fields $h_i$, and interaction terms $\Delta_i$, weighted by the spin values of that specific configuration. Edges connect configurations that differ by a single spin flip, forming a 3-dimensional hypercube structure. }
    \label{fig:MBLconfig}
\end{figure}

The Hilbert space can thus be viewed as a graph isomorphic to the $L$-dimensional hypercube. Each vertex represents a spin configuration (a binary string), and each edge connects two configurations that differ by a single spin flip. Consequently, each vertex has degree $L$, corresponding to the $L$ possible single-spin transitions generated by the transverse field terms.
Figure~\ref{fig:MBLconfig} illustrates  this configuration space for a chain of length $L=3$. 

This representation casts the Hamiltonian as an effective tight-binding model analogous to Eq.~\eqref{eq:H_AL_1st}, now defined on an $L$-dimensional hypercube. The diagonal terms $E_\alpha$ serve as on‑site energies, and the transverse fields $\Gamma_i$ provide hopping amplitudes between configurations connected by single spin flips. In other words, the unitary evolution of the $L$-body problem described by Eq.~\eqref{eq:imbrie} is formally mapped into a problem of single ``particle'' diffusion on a large-dimensional graph in presence of a random potential~\cite{altshuler1997quasiparticle}. 

A key difference from the Anderson model lies in the structure of the on‑site energies. The $2^L$ random energies $E_\alpha$ are strongly correlated across configurations because they are all built from the same $L$ random fields $h_i$ and $L-1$ couplings $\Delta_i$, weighted by the spin values of the specific configuration (see Eq.~\eqref{eq:config_E}). This results in very strong long-ranged ``ultrametric'' correlations that contrast with the uncorrelated disorder typical of the standard Anderson setting. 

A first pictorial approximation to the complex connectivity of many-body Hilbert spaces consists in neglecting both energy correlations and loops in the Hilbert space graph. Within this framework, the MBL problem of the eigenstates of $H_I$ is mapped onto Anderson localization on the Bethe lattice~\cite{basko2006metal, gornyi2005interacting, altshuler1997quasiparticle, tikhonov2021anderson}---a hierarchical, tree-like graph. 

Although loops are entirely absent in the Bethe lattice, this structure generically shares a key property with MBL Hilbert space graphs: both possess an effectively infinite dimensionality, since the number of nodes at a given distance from a central node grows exponentially with that distance. 

As discussed in several recent works, this toy-model representation---while highly idealized and omitting certain crucial features---lies at the heart of the earliest studies of MBL~\cite{altshuler1997quasiparticle} and continues to provide a useful and insightful starting point~\cite{tikhonov2021anderson,tikhonov2021eigenstate,tarzia2020many,de2013ergodicity,biroli2017delocalized,biroli2020anomalous,logan2019many,garcia2022critical,herre2023ergodicity, prado2026anderson}. While variations of the model~\eqref{eq:imbrie} with alternative microscopic interactions may lead to different matrix elements (note also that the Hilbert space graph itself is inherently basis-dependent) the core ideas and arguments presented below remain completely unchanged, and many features of the Hilbert space graph---the exponential growth of its volume, its effectively infinite‑dimensional character, the strong correlations among the matrix elements, and intricate loop structure---are not unique to this Hamiltonian. Rather, they are generic properties of many‑body configuration spaces and can play a central role in the phenomenology of MBL (for a comprehensive review, see Ref.~\cite{tikhonov2021anderson}).

To summarize, transport, eigenfunction moments, spectral type, and level statistics provide complementary ways of distinguishing localized from extended states. In the next section, we show that all of them are encoded in a single object, the resolvent, whose self-consistent structure becomes exact on the Bethe lattice---the tractable infinite-dimensional setting, and toy model of many-body configuration spaces, motivated above.

%=== RESOLVENT ===
\section{The resolvent}\label{sec:resolvent}

Having established the diagnostics of localization---related to transport, spectral
statistics, and inverse participation ratios---we now introduce the object that
unifies these perspectives and provides the foundation for analytical approaches:
the resolvent. The central idea is that these apparently different diagnostics can
all be expressed in terms of the resolvent, an analytic function of a complex energy
variable. Its singularities identify the spectrum, its imaginary part close to the
real axis determines the local density of states, and its analytic structure in the
thermodynamic limit distinguishes pure-point from absolutely continuous spectra.
This formulation will be particularly useful on the Bethe lattice, where the diagonal
matrix elements of the resolvent satisfy closed self-consistent equations, allowing
the localization problem to be studied directly through their statistical
properties.

\subsection{Definition and analytic properties}

\begin{figure}[h] 
    \begin{minipage}[b]{7cm}
    \centering
    \includegraphics[width=5cm]{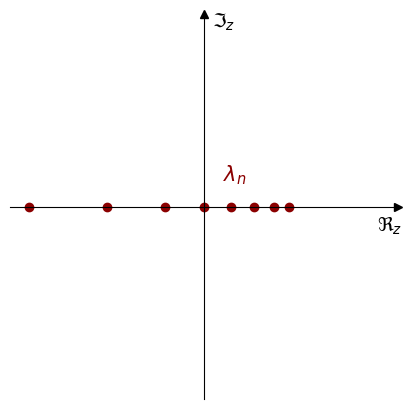}
    \end{minipage}
    \begin{minipage}[b]{7cm}
    \centering
    \includegraphics[width=6.5cm]{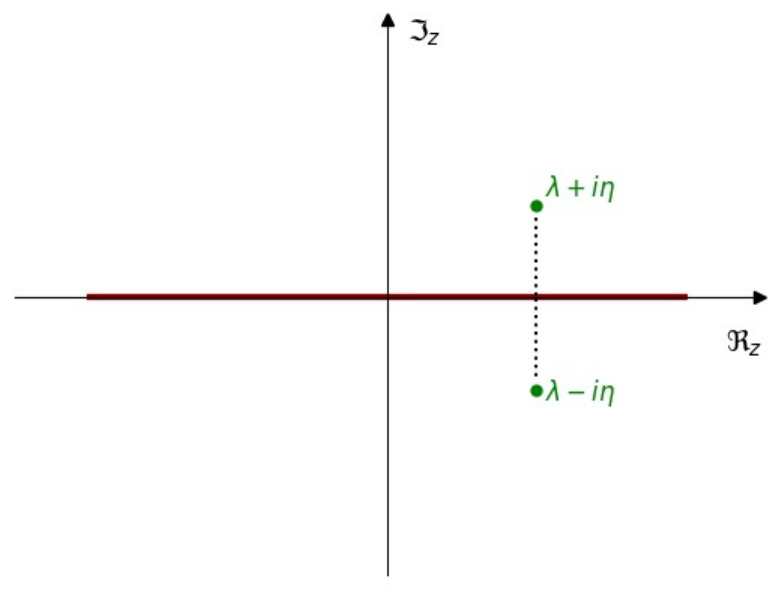}
    \end{minipage}
    \caption{(\textbf{Left}) For a finite system, $G(z)$ is analytic except at a discrete set of points on the real axis corresponding to the eigenvalues $\lambda_n$, where it exhibits simple poles. \label{fig:poles}
    (\textbf{Right}) In the thermodynamic limit ($N \to \infty$), these discrete poles
    either merge into a continuous branch cut along the real axis, if the spectrum
    becomes absolutely continuous (as in the extended phase), or remain a dense but discrete set of poles, if it
    remains pure point (as in the localized phase). The physical Green's functions are recovered by approaching
    the real axis from the complex plane through the regularization $z = \lambda \pm
    i\eta$ (green dots) as $\eta \to 0^+$. \label{fig:branch_cut}    }
\end{figure}

The resolvent of the Hamiltonian $H$ is defined as 
\begin{equation}
    G(z) = \frac{1}{z - H}, \qquad z \in \mathbb{C}\setminus\sigma(H),
\end{equation}
where $\sigma(H)$ denotes the spectrum of $H$. The matrix elements of $G$ are analytic functions of $z$ everywhere except on the spectrum. For a finite system, $G(z)$ is meromorphic with simple poles at the eigenvalues $\lambda_n$ (Fig.~\ref{fig:poles}). In the eigenbasis,
\begin{equation}
    G(z) = \sum_n \frac{|\psi_n\rangle\langle\psi_n|}{z - \lambda_n},
\end{equation}
and its matrix elements in the position basis are
\begin{equation}\label{eq:Gij_wavefunc}
    G_{ij}(z) = \langle i|G(z)|j\rangle = \sum_n \frac{\psi_n^*(i)\psi_n(j)}{z - \lambda_n}.
\end{equation}

The limits of $G(z)$ as $z$ approaches the real axis from above and below define the retarded and advanced Green's functions:
\begin{equation}
    G^R(\lambda) = \lim_{\eta\to 0^+} G(\lambda + i\eta), \qquad
    G^A(\lambda) = \lim_{\eta\to 0^+} G(\lambda - i\eta).
\end{equation}
As we will discuss below (Sec.~\ref{sec:uni_dyn}), the retarded and advanced Green's functions also admit a natural interpretation in the time domain. Since $H$ is real symmetric, $G(\lambda-i\eta)=[G(\lambda+i\eta)]^*$, and with the definition $G(z)=(z-H)^{-1}$ one has $\operatorname{Im} G_{ii}(\lambda+i\eta)<0$. For this reason, throughout these notes we work with $G(\lambda - i\eta)$, i.e., with the advanced Green's function, whose diagonal elements have a positive imaginary part that directly gives the local density of states (see Eq.~\eqref{eq:def_LDOS} below). Of course, the retarded Green's function contains exactly the same information, since $\operatorname{Im} G^R_{ii} = - \operatorname{Im} G^A_{ii}$.

Their difference reads,
\begin{align}
    G^A(\lambda) - G^R(\lambda) &= \lim_{\eta\to 0^+} \left[ G(\lambda - i\eta) - G(\lambda + i\eta) \right] \nonumber \\
    &= 2\pi i \sum_n \delta(\lambda - \lambda_n) |\psi_n\rangle\langle\psi_n| ,
\end{align}
where the last equality follows from the Sokhotski--Plemelj formula:
\begin{equation}\label{eq:principal_val}
    \lim_{\eta\to 0^+} \frac{1}{x \pm i\eta} = \mathcal{P}\frac{1}{x} \mp i\pi \delta(x),
\end{equation}
where $\mathcal{P}$ denotes the principal value. Taking the diagonal elements yields the local density of states (LDOS),
\begin{equation}\label{eq:def_LDOS}
    \rho_i(\lambda) \equiv \sum_n |\psi_n(i)|^2 \delta(\lambda - \lambda_n) = \frac{1}{\pi} \lim_{\eta\to 0^+} \operatorname{Im} G_{ii}(\lambda - i\eta).
\end{equation}
The total density of states follows by averaging:
\begin{equation}\label{eq:def_DOS}
    \rho(\lambda) = \frac{1}{N} \sum_i \rho_i(\lambda) = \lim_{\eta\to 0^+} \frac{1}{\pi N}\operatorname{Im} \operatorname{Tr} G(\lambda - i\eta).
\end{equation}

For finite $\eta>0$, the expression in Eq.~\eqref{eq:def_LDOS} provides a regularized version of the local density of states. Using Eq.~\eqref{eq:Gij_wavefunc}, one obtains
\begin{equation}\label{eq:lorentzian_eta}
    \operatorname{Im} G_{ii}(\lambda - i\eta)
    =
    \sum_n |\psi_n(i)|^2 \, \frac{\eta}{(\lambda - \lambda_n)^2 + \eta^2}.
\end{equation}
Thus, each eigenvalue contributes through a Lorentzian of width $\eta$ centered at $\lambda_n$. In the limit $\eta\to0^+$, these Lorentzians converge (in distribution) to delta functions, recovering the definition of $\rho_i(\lambda)$.

In the thermodynamic limit, the behavior of $G(z)$ near the real axis depends on the nature of the spectrum. If the spectrum is absolutely continuous (delocalized phase), the poles coalesce into a branch cut along the real axis (Fig.~\ref{fig:branch_cut}); the resolvent cannot be analytically continued across the cut. If instead the spectrum remains pure point (localized phase), the poles remain discrete even in the thermodynamic limit and form a dense set on the real axis. In this case the limits from above and below coincide almost everywhere on the real axis ($\operatorname{Im} G_{ii} \to 0$ for almost every $\lambda$), so that there is no branch cut; yet, because the poles are dense, the resolvent cannot be analytically continued across any interval of the real axis, which is a natural boundary (a line of singularities). The absence or presence of a branch cut is thus a direct signature of the spectral type and, consequently, of the phase of the system.

Note that Eq.~\eqref{eq:def_DOS} defines the DOS for a finite system of size $N$. Any finite system has a purely discrete spectrum, independently of the phase, so the distinction between localized and delocalized regimes only emerges after taking the thermodynamic limit. Accordingly, in localization problems one must carefully specify the order of limits: the limit $N\to\infty$ should be taken before $\eta \to 0^+$, so that the regulator probes the infinite-volume spectral structure rather than the discrete finite-size spectrum.

This can be understood by estimating the number of eigenvalues contributing to $\operatorname{Im} G(\lambda - i\eta)$. Since the regulator effectively averages over an energy window of width $\sim \eta$, the number of levels involved scales as $N\,\rho(\lambda)\,\eta$. For fixed $N$, this number vanishes as $\eta \to 0^+$, and the resolvent resolves individual eigenvalues. In contrast, if the thermodynamic limit is taken first, the level spacing collapses and infinitely many states contribute within any finite $\eta$, allowing the emergence of a smooth spectral density and, in the delocalized phase, of a branch cut.

\subsection{Probing the Anderson transition via the resolvent}

At first glance, one might expect the density of states $\rho(\lambda)$ itself to
provide a direct signature of the Anderson transition, since for a given disorder
realization the spectrum may remain pure point (localized phase) or develop an
absolutely continuous component (delocalized phase) in the thermodynamic limit.
However, the averaging procedure suppresses this distinction: $\rho(\lambda)$
converges to a smooth, continuous-looking function in both phases, so the
transition is encoded not in the average DOS but in the whole distribution of the
local density of states \cite{mirlin1994distribution}.

The shape of this distribution differs sharply between the two phases. In the
delocalized phase, the LDOS is approximately uniform across the lattice, resulting
in a self-averaging distribution with finite fluctuations around its mean value. In
the localized phase, by contrast, the LDOS becomes highly inhomogeneous, with most
sites carrying an exponentially small spectral weight, while a small number of
resonant sites concentrate most of the amplitude, resulting in a singular
distribution.

The resolvent gives us direct access to this distribution through its imaginary
part, whose analytic structure encodes the nature of the spectrum, as noted above.
We now make this concrete in each phase separately, starting with the extended case.

\subsubsection{Extended phase}

\begin{figure}
    \centering
    \includegraphics[width=0.5\textwidth]{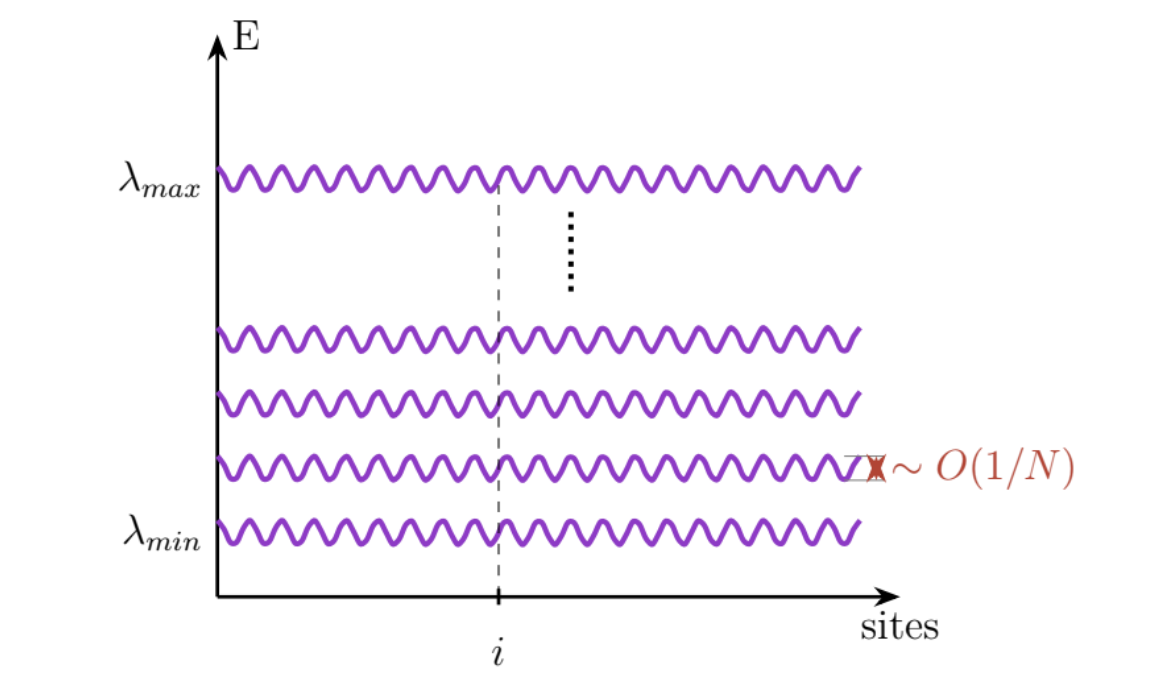}
    \caption{Schematic representation of eigenstates in the extended phase. Each horizontal line represents an eigenstate at energy $\lambda_n$, with its oscillatory pattern indicating the spatial structure. The red marker shows that the amplitude of the wavefunction is uniformly of order $O(1/N)$.}
    \label{fig:extended_spectrum}
\end{figure}

In this regime, each eigenstate spreads roughly uniformly over the entire system, with amplitudes scaling as $|\psi_n(i)|^2 \sim 1/N$, where $N$ is the number of sites (Fig.~\ref{fig:extended_spectrum}). The local density of states at site $i$ therefore receives comparable contributions from all eigenstates with energies near $\lambda$.

\begin{figure}
    \centering
    \includegraphics[width=0.5\textwidth]{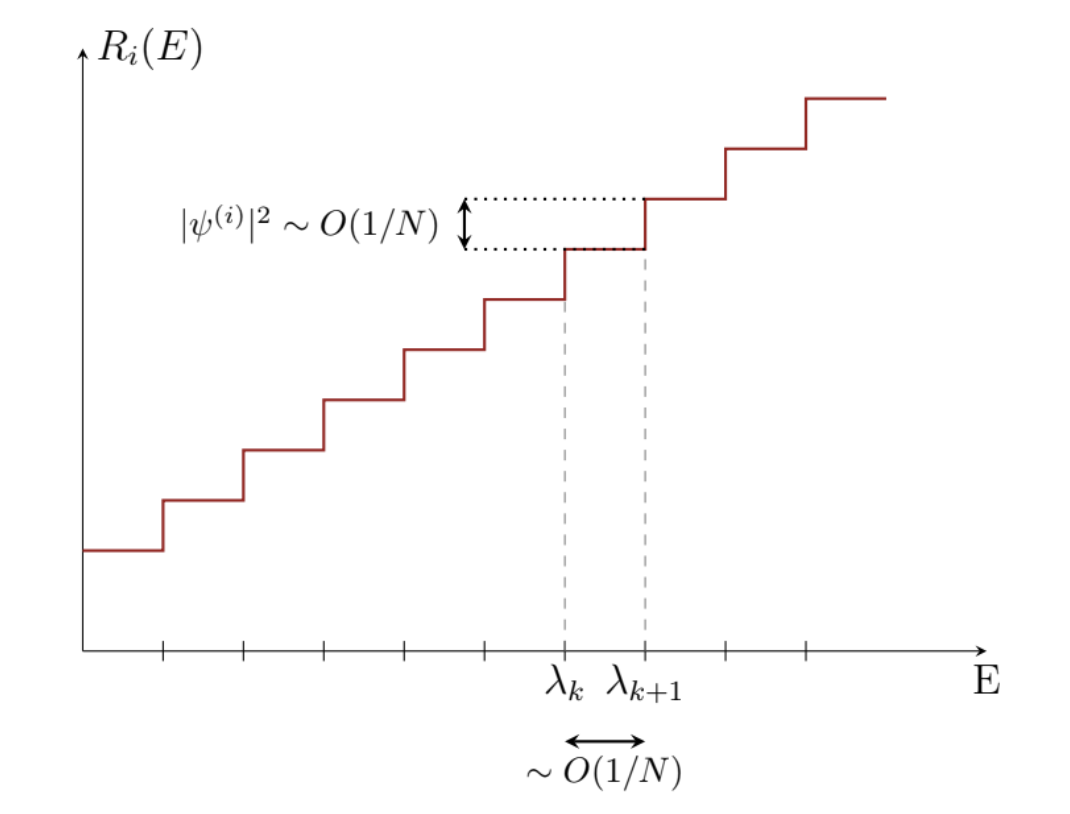}
    \caption{Integrated local density of states $R_i(E)$ in the extended phase. Both the step height (proportional to $|\psi_n(i)|^2$) and the level spacing are of order $O(1/N)$. In the thermodynamic limit, these discrete steps merge into a smooth, continuous function.}
    \label{fig:extended_integrated}
\end{figure}

To see this more clearly, let us consider the integrated local density of states,
\begin{equation}\label{eq:integrated_LDOS}
    R_i(E) = \int_{-\infty}^{E} \rho_i(\lambda) \, d\lambda = \sum_{n} |\psi_n(i)|^2 \theta(E - \lambda_n) ,
\end{equation}
which counts the cumulative weight of eigenstates up to energy $E$. As shown in Fig.~\ref{fig:extended_integrated}, $R_i(E)$ increases stepwise: each jump of height $|\psi_n(i)|^2 \sim 1/N$ occurs at an eigenvalue $\lambda_n$, and the typical spacing between successive eigenvalues is also $\Delta \lambda \sim 1/N$.

In the thermodynamic limit $N\to\infty$, the discrete steps in $R_i(E)$ become
infinitely fine, and the function converges to a smooth, continuously increasing
function of energy.

A powerful consequence of this analytic structure is that the resolvent anywhere in the complex plane can be reconstructed from its discontinuity across the cut. Starting from the spectral representation,
\begin{equation}
    G_{ij}(z) = \sum_n \frac{\psi_n^*(i)\psi_n(j)}{z - \lambda_n}
              = \int d\lambda \, \frac{ \sum_n \psi_n^*(i)\psi_n(j) \delta(\lambda - \lambda_n) }{z - \lambda},
\end{equation}
and using the Sokhotski–Plemelj formula, one finds
\begin{equation}
    G_{ij}(\lambda - i0^+) - G_{ij}(\lambda + i0^+) = 2\pi i \sum_n \psi_n^*(i)\psi_n(j) \delta(\lambda - \lambda_n).
\end{equation}
Combining these expressions yields
\begin{equation}
    G_{ij}(z) = \frac{1}{\pi} \int d\lambda \, \frac{\operatorname{Im} G_{ij}(\lambda - i0^+)}{z - \lambda},
\end{equation}
and for the diagonal elements, in particular,
\begin{equation}
    G_{ii}(z) = \int d\lambda \, \frac{\rho_i(\lambda)}{z - \lambda}.
\end{equation}
This relation shows that all the physical information about the system is encoded in the imaginary part of the Green's function on the real axis, $\operatorname{Im} G_{ij}(\lambda - i0^+)$~\cite{economou2006green}.

\subsubsection{Localized eigenstates}\label{sec:resolvent:localized}

\begin{figure}
    \centering
    \includegraphics[width= 0.5\textwidth]{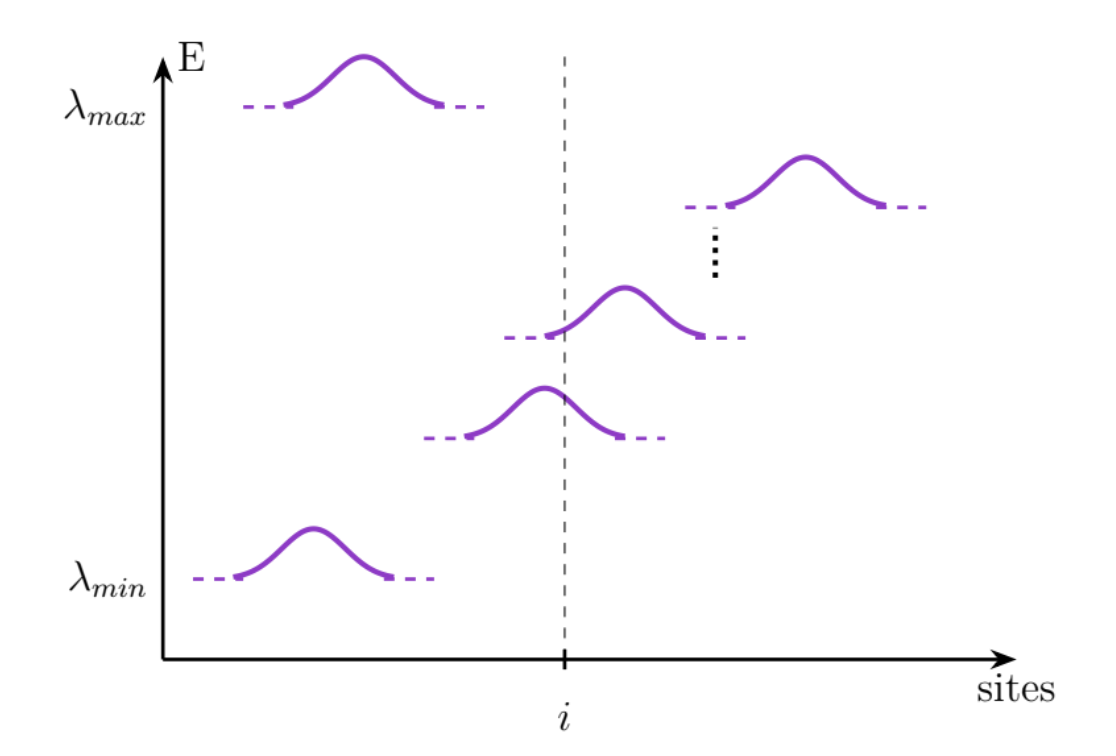}
    \caption{Schematic representation of eigenstates in the localized phase. Each eigenstate is confined to a region around its localization center. At site $i$, the local density of states receives significant contributions only from eigenstates whose centers lie within $\sim \xi$ of $i$; distant states contribute negligibly.}
    \label{fig:localized_spectrum}
\end{figure}

The localized phase presents a notably different picture. Here, eigenstates are exponentially confined to a region of size $\xi$, the localization length:
\begin{equation}
    |\psi_n(\mathbf{r})|^2 \sim e^{-|\mathbf{r} - \mathbf{r}_n|/\xi},
\end{equation}
where $\mathbf{r}_n$ denotes the localization center of the $n$-th eigenstate (Fig.~\ref{fig:localized_spectrum}).

The local density of states at site $i$ now receives contributions from a very restricted set of eigenstates. Only those whose localization centers lie within a distance $\sim \xi$ of $i$ contribute with an amplitude of order one. All other eigenstates, whose centers are far away, contribute terms that are exponentially small. Consequently, the effective number of eigenstates contributing significantly to $\rho_i(\lambda)$ at any given energy is $O(1)$, in sharp contrast to the extended case where every eigenstate contributes.

\begin{figure}
    \centering
    \includegraphics[width= 0.5\textwidth]{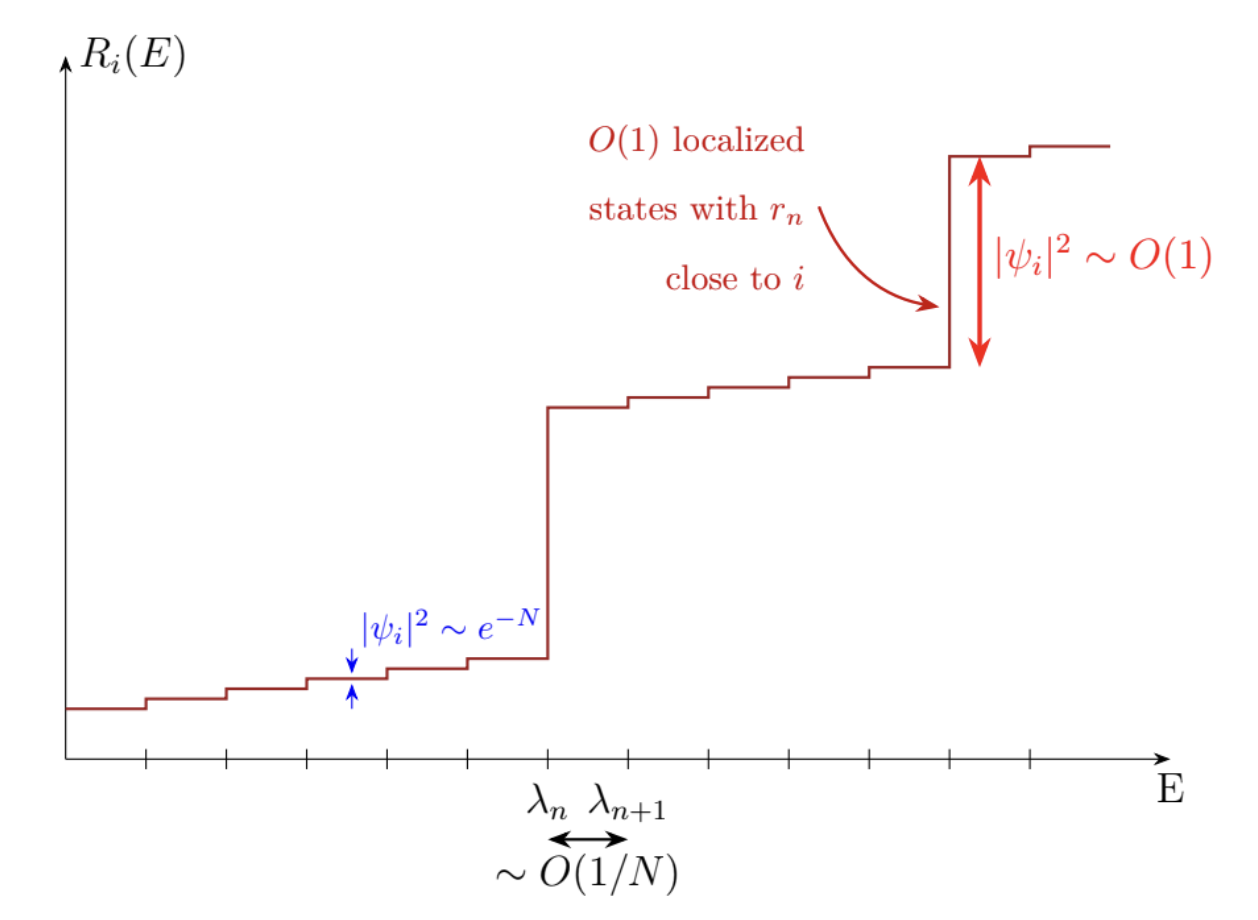}
    \caption{Integrated local density of states $R_i(E)$ in the localized phase. Most steps are exponentially small ($|\psi_n(i)|^2 \sim e^{-L/\xi}$), corresponding to eigenstates localized far (at a distance $L$) from site $i$. Rare large jumps of order one occur when an eigenstate is centered near $i$. This irregular structure persists even in the thermodynamic limit.}
    \label{fig:localized_integrated}
\end{figure}

This behavior is reflected in the integrated local density of states $R_i(E)$ (Eq.~\eqref{eq:integrated_LDOS}), shown in Fig.~\ref{fig:localized_integrated}. The function exhibits a characteristic Cantor-like or ``devil's staircase'' structure \cite{CraigSimon1983}: it remains flat over large energy intervals, punctuated by sudden jumps of order one whenever an eigenvalue $\lambda_n$ corresponds to an eigenstate localized near site $i$. The height of these jumps is irregular and does not vanish in the thermodynamic limit.

This irregular jump structure reflects a deeper fact about the local density of
states itself: $\rho_i(\lambda)$ remains non-self-averaging in the thermodynamic
limit, dominated by rare resonant eigenstates rather than by a smooth average over
many contributions. At a given site, most eigenstates contribute exponentially
small weights, while rare eigenstates localized nearby produce peaks of order one.
As a result, $\rho_i(\lambda)$ retains a discrete structure, consisting of isolated
$\delta$-peaks even as $N\to\infty$.

To understand this distribution, it is convenient to examine the imaginary part of the diagonal resolvent (Eq.~\eqref{eq:lorentzian_eta}) directly. In the extended phase, where $|\psi_n(i)|^2 \sim 1/N$, $\operatorname{Im} G_{ii}(\lambda - i\eta)$ receives contributions from the $\sim \eta N \rho(\lambda)$ eigenstates within the energy window of width $\eta$ around $\lambda$. The factor $1/N$ from the eigenfunction amplitudes compensates the number of contributing states, yielding a finite and self-averaging quantity as $\eta \to 0^+$.

In the localized phase, the situation is qualitatively different. For a typical site $i$, the amplitudes $|\psi_n(i)|^2$ associated with eigenenergies satisfying $|\lambda - \lambda_n| \lesssim \eta$ are exponentially small in the distance between the localization center of the eigenstate and the site $i$, and therefore give a negligible contribution in the thermodynamic limit. The few eigenstates that have $O(1)$ weight on site $i$ typically lie far in energy from $\lambda$, with $|\lambda - \lambda_n| \gg \eta$, and thus contribute only of order $\eta$ to $\rho_i$. By contrast, on the rare sites where an eigenstate is localized within a distance $\sim \xi$ from $i$, the corresponding amplitude is $O(1)$. For such sites, the Lorentzian peak associated with that eigenstate has a height of order $1/\eta$. As a result, the limit $\lim_{\eta \to 0^+} \operatorname{Im} G_{ii}(\lambda - i\eta)$ does not define a regular function, but rather a singular measure supported on the eigenvalues.
This implies that, in the localized phase, the distribution of $\rho_i(\lambda)$ is extremely broad and singular. The typical value---defined as $\exp\langle \ln \rho_i \rangle$---vanishes in the thermodynamic limit, even though the average $\rho=\langle \rho_i \rangle$ remains finite. This strong discrepancy between typical and average values is a hallmark of localization and explains why the full distribution of the local density of states, rather than its mean, provides the appropriate order parameter for the Anderson transition \cite{mirlin1994distribution}. Within the supersymmetry formalism, this distribution naturally appears as the order parameter function, whose nontrivial structure reflects the spontaneous breaking of the underlying graded symmetry at the transition \cite{mirlin1991localization, mirlin1994statistical, efetov1983supersymmetry}.

Combining the localization diagnostics introduced in Sec.~\ref{sec:diagnostics} with the properties of the resolvent discussed above, the contrasting behaviors of the extended and insulating phases are summarized in Table~\ref{tab:phase_comparison}.

\begin{table}[htbp]
\centering
\begin{tabular}{|p{5cm}|p{5cm}|p{5cm}|}
\hline
 & \textbf{Extended phase} & \textbf{Localized phase} \\
\hline
Transport & Diffusive (ballistic only for $W=0$) & No transport \\
\hline

Wavefunction amplitudes & $|\psi_n(i)|^2 \sim 1/N$ & $|\psi_n(i)|^2 \sim e^{-|i-i_0|/\xi}$  \\
\hline
Inverse participation ratio & $I_2 \sim 1/N$ & $I_2 = O(1)$ \\
\hline
Integrated LDOS $R_i(E)$ & Smooth, continuous & Irregular ``devil's staircase'' \\
\hline
Green’s function $G_{ii}(z)$ & Branch cut on real axis & Natural boundary\\ 
&& (line of singularities) \\
\hline
$\displaystyle \pi \rho_i (\lambda) = \lim_{\eta\to0^+} \operatorname{Im} G_{ii}(\lambda-i\eta)$ & Finite & Undefined (typically zero) \\
\hline
Spectral type & Absolutely continuous & Pure point \\
\hline
Spectral statistics & Universal RMT correlations & Poisson statistics \\
\hline
\end{tabular}
\caption{Contrasting properties of the extended (delocalized) and localized phases in Anderson localization.}
\label{tab:phase_comparison}
\end{table}

\subsection{Eigenfunction statistics}

As explained above, the probability distribution of the local density of states serves as the order parameter for Anderson localization. Below, we show that this distribution also provides a complete characterization of the moments of the eigenfunction amplitudes.

\subsubsection{Higher moments of eigenfunction amplitudes}

As discussed in Sec.~\ref{sec:diagnostics}, for a normalized eigenstate $|\psi_{n}\rangle$ with energy $\lambda_n$, the $q$-th moment of its amplitude is defined as $I_q^{(n)} = \sum_i |\psi_n(i)|^{2q}$. The scaling behavior $I_q^{(n)} \sim N^{-\zeta(q)}$ can be used to distinguish between different phases. Averaging over all eigenstates at a fixed energy $\lambda$ yields
\begin{equation}\label{eq:Iq_def}
    I_q(\lambda) = \frac{\sum_n I_q^{(n)} \delta(\lambda - \lambda_n)}{\sum_n \delta(\lambda - \lambda_n)},
\end{equation}
where the denominator corresponds to the density of states, normalizing the expression by the number of states at that energy. As explained above, the special case $q=2$ is commonly referred to as the Inverse Participation Ratio, or IPR in short, which provides a measure of the inverse volume occupied by the eigenstates at energy $\lambda$.

This definition is understood in a regularized sense, replacing the delta functions by Lorentzians of width $\eta$, consistently with the definition of the local density of states in Eq.~\eqref{eq:def_LDOS}. In this way, $I_q(\lambda)$ corresponds to an average over eigenstates within an energy window of width $\sim \eta$ around $\lambda$.

These moments can be related to the moments of the imaginary part of the resolvent. For finite $\eta>0$, taking the $q$-th power of the imaginary part of the resolvent (Eq.~\eqref{eq:lorentzian_eta}) yields
\begin{equation}
\begin{aligned}
    \left [ \operatorname{Im}G_{ii}(\lambda - i\eta) \right]^q &= \left(\sum_n |\psi_n(i)|^2 \frac{\eta}{(\lambda-\lambda_n)^2+\eta^2}\right)^q\\
    &= \sum_{n_1,\dots,n_q}
    \prod_{k=1}^q \frac{\eta |\psi_{n_k}(i)|^{2}}{(\lambda-\lambda_{n_k})^2+\eta^2}.
\end{aligned}
\end{equation}
In the limit $\eta \to 0^+$, the Lorentzian functions become sharply peaked, so that the dominant contribution comes from terms where all indices coincide, $n_1=\cdots=n_q$. Neglecting subleading contributions from distinct indices, one obtains
\begin{equation}
\begin{aligned}
    \left[ \operatorname{Im}G_{ii}(\lambda - i\eta) \right]^q &\simeq  \sum_n
    \frac{\eta^q |\psi_n(i)|^{2q}}{\big[(\lambda-\lambda_n)^2+\eta^2\big]^q}\\
    &\simeq \int dx\,
    \frac{\eta^q}{(x^2+\eta^2)^q}
    \sum_n |\psi_n(i)|^{2q}
    \delta(x-\lambda+\lambda_n)\\
    &\simeq \int dx\,\frac{\eta^q}{(x^2+\eta^2)^q}
    \sum_n |\psi_n(i)|^{2q}\delta(\lambda-\lambda_n).
\end{aligned}
\end{equation}
The last approximation comes from the fact that for small $\eta$, the kernel is sharply peaked around $x=0$, so if the energy-resolved sum varies slowly on the scale $\eta$, it can be evaluated at this point only.

Evaluating the integral for $q>1/2$, one obtains
\begin{equation}\label{eq:moment_ImG}
    \left [ \operatorname{Im}G_{ii}(\lambda-i\eta) \right]^q\simeq \frac{\eta^{1-q}\sqrt{\pi}\,\Gamma(q-\tfrac12)}{\Gamma(q)}\sum_n|\psi_n(i)|^{2q}\delta(\lambda-\lambda_n).
\end{equation}
Summing Eq.~\eqref{eq:moment_ImG} over sites and using Eq.~\eqref{eq:Iq_def}, one obtains
\begin{equation}\label{eq:ipr_from_ImG}
    I_q(\lambda) = \frac{\sqrt{\pi}\,\Gamma\left(q\right)}
    {\Gamma\left(q-\tfrac{1}{2}\right)}
    \lim_{\eta \to 0^{+}}
    \frac{\eta^{q-1}\sum_i \left[ \operatorname{Im}G_{ii}(\lambda - i\eta)\right]^{q}}
    {\sum_i \operatorname{Im} G_{ii}(\lambda - i\eta)}.
\end{equation}
Likewise, a similar expression for $I_q(\lambda)$ can be obtained from the moments of the modulus of the full resolvent,
\begin{equation}\label{eq:ipr_from_G}
    I_q(\lambda) = \frac{\sqrt{\pi}\,\Gamma\left(\tfrac{q}{2}\right)}
    {\Gamma\left(\tfrac{q-1}{2}\right)}
    \lim_{\eta \to 0^{+}}
    \frac{\eta^{q-1}\sum_i  \left| G_{ii}(\lambda - i\eta)\right|^{q}}
    {\sum_i \operatorname{Im} G_{ii}(\lambda - i\eta)}.
\end{equation}

These relations express $I_q(\lambda)$ entirely in terms of moments of the resolvent
evaluated at the complex energy $\lambda-i\eta$. This becomes particularly useful
on lattices where the diagonal resolvent $G_{ii}(z)$ satisfies a closed set of self-consistent equations, as on the Bethe lattice considered in Sec.~\ref{sec:cavity}. In that case, Eqs.~\eqref{eq:ipr_from_ImG} and \eqref{eq:ipr_from_G} provide direct access to
eigenfunction moments without requiring an explicit computation of the eigenstates.

\subsubsection{Two point correlation functions}

The single-site quantities discussed so far characterize the statistics of
eigenfunction amplitudes, but do not reveal how amplitudes at different sites are
spatially correlated. For instance, they do not tell us whether regions of large
wavefunction amplitude tend to occur close to one another or how such correlations
decay with the distance between sites. These spatial correlations provide additional
information about the structure of the eigenstates and are particularly useful near
the localization transition, where eigenfunctions exhibit multifractal fluctuations.
The resolvent also provides direct access to these spatial correlations. A natural
object is the two-point correlation function of the resolvent between two sites
separated by distance $r$, $\langle |G_{0r}(\lambda - i\eta)|^{2}\rangle$, where the
average combines spatial averaging over pairs of sites at distance $r$ and disorder
averaging. This correlation function is resolved at energy $\lambda$, while the finite
regulator $\eta$ introduces an energy broadening. Using the spectral decomposition
of the resolvent, one finds for finite $\eta>0$,
\begin{align}
    \eta\, |G_{0r}(\lambda - i\eta)|^2 
    &= \sum_n |\psi_n(0)|^2 |\psi_n(r)|^2 
    \frac{\eta}{(\lambda - \lambda_n)^2 + \eta^2}.
\end{align}
In the limit $\eta \to 0^+$, the Lorentzian converges (in the sense of distributions) to a delta function, yielding
\begin{align}
    \eta\, |G_{0r}(\lambda - i\eta)|^2 
    \;\xrightarrow{\eta \to 0^+}\;
    \pi \sum_n |\psi_n(0)|^2 |\psi_n(r)|^2 \delta(\lambda - \lambda_n).
\end{align}
This shows that the squared resolvent probes correlations of eigenfunction amplitudes at fixed energy. Introducing the normalized amplitude correlation function, one obtains the relation
\begin{align}\label{eq:corr_def}
    C(r;\lambda)
    &= \Bigg\langle 
    \frac{\sum_n |\psi_n(0)|^2 |\psi_n(r)|^2 \delta(\lambda - \lambda_n)}
    {\rho(\lambda)}
    \Bigg\rangle = \lim_{\eta \to 0^+}
    \frac{\eta\, \langle |G_{0r}(\lambda - i\eta)|^2 \rangle}
    {\langle \mathrm{Im}\, G_{00}(\lambda - i\eta) \rangle},
\end{align}
where $\rho(\lambda)$ is the density of states.

Equivalently, using the relation between the resolvent and the local density of states, this correlator can be expressed as the normalized two-point correlation function of the LDOS~\cite{tikhonov2019statistics},
\begin{align}
    C(r;\lambda)
    =  \lim_{\eta \to 0^+} \frac{2\,\eta \,  \langle \text{Im} \, G_{00} (\lambda - i\eta) \, \text{Im} \, G_{rr} (\lambda - i\eta) \rangle}{\langle \text{Im} \, G_{00} (\lambda - i \eta) \rangle }.
\end{align}

\subsection{Relationship with unitary quantum dynamics}\label{sec:uni_dyn}

Below, we finally show that transport and dissipation are also fully encoded in the statistics of the Green’s functions. To show this, we focus on the specific example of the return probability for a particle prepared on a given node $|0\rangle$ at time $0$.
In particular, the return probability $R(\tau)=|\langle 0|e^{-iH\tau}|0\rangle|^2$ can be expressed in terms of the resolvent using the integral representation of the time evolution operator. Starting from the spectral decomposition,
\begin{equation}
\langle 0|e^{-iH\tau}|0\rangle = \sum_n e^{-i\lambda_n \tau}\,|\langle \psi_n|0\rangle|^2,
\end{equation}
and using Cauchy's integral formula to represent the exponential,
\begin{equation}
e^{-i\lambda_n \tau} = \frac{1}{2\pi i}\oint_{\mathcal{C}} dz\, \frac{e^{-iz\tau}}{z-\lambda_n},
\end{equation}
where the contour $\mathcal{C}$ encircles the spectrum counterclockwise, we obtain
\begin{equation}\label{eq:return_int}
\langle 0|e^{-iH\tau}|0\rangle = \frac{1}{2\pi i}\oint_{\mathcal{C}} dz\, e^{-iz\tau} G_{00}(z).
\end{equation}
This representation is essentially the inverse Laplace transform. More precisely, for $\operatorname{Im} z > 0$, one can write
\begin{equation}\label{eq:Green_Laplace}
    G(\lambda + i\eta) = -i \int_0^\infty d\tau\, e^{i\lambda \tau} e^{-\eta \tau} e^{-iH\tau},
\end{equation}
which corresponds to the retarded Green's function (the advanced one is obtained
analogously for $\operatorname{Im} z < 0$). Indeed, the time-domain retarded
Green's function is $G^R(\tau) = -i\Theta(\tau)e^{-iH\tau}$, so that
$G(z) = \int_0^\infty d\tau\, e^{iz\tau}G^R(\tau)$ for $\operatorname{Im} z > 0$; conversely,
the time-domain advanced Green's function is $G^A(\tau) = i\Theta(-\tau)e^{-iH\tau}$, and
$G(z) = \int_{-\infty}^{0} d\tau\, e^{iz\tau}G^A(\tau)$ for $\operatorname{Im} z < 0$.

For $\tau>0$, the contour in Eq.\eqref{eq:return_int} can be deformed to run just below the real axis, picking up the contribution from the branch cut (or, equivalently, the dense set of poles) along the real line. Using the identity $G_{00}(\lambda - i\eta) - G_{00}(\lambda + i\eta) = 2i\,\Im G_{00}(\lambda - i\eta)$ together with the Sokhotski--Plemelj formula, one finds
\begin{equation}
\langle 0|e^{-iH\tau}|0\rangle = \frac{1}{\pi}\int_{-\infty}^{\infty} d\lambda\, e^{-i\lambda \tau}\,\Im G_{00}(\lambda - i0^+).
\end{equation}
This shows that the return amplitude is the Fourier transform of the local density of states.

In the delocalized phase, $\Im G_{00}(\lambda - i0^+)$ converges to a smooth function that vanishes fast enough as $\lambda \to \pm \infty$ (recall that the average of $\Im G_{00}(\lambda)$ over all sites is the total density of states at energy $\lambda$, whose integral over $\lambda$ is normalized to $1$). Its Fourier transform therefore decays to zero as $\tau \to \infty$, implying that the return probability vanishes at long times.
This provides a transparent physical interpretation of $\Im G_{00}$ in the delocalized phase. The convergence of the LDOS $\rho_0(\lambda)=\frac{1}{\pi}\Im G_{00}(\lambda-i0^+)$ to a continuous function means that the local spectral weight is spread over a finite energy
interval rather than concentrated in isolated delta peaks. Its Fourier transform therefore decays in time, reflecting the escape of a particle initially localized at site $0$ into the rest of the system.

Conversely, in the localized phase, $\Im G_{00}(\lambda - i0^+)$ does not admit a well-defined smooth limit and can be viewed as a sum of $\delta$-peaks. As a result, $\langle 0|e^{-iH\tau}|0\rangle$ becomes a sum of oscillatory terms $e^{-i\lambda_n \tau}$ that do not decay, and the return probability remains finite at all times.

Let us consider a toy model that shows the emergence of the branch-cut and its concrete relationship with the particle lifetime. Assume that all eigenvalues $\lambda_n, \quad n=1,\dots,N$ are equally spaced on the whole real axis. We assume that the broadening of the eigenstates' amplitudes can be modeled by a Lorentzian,
\begin{equation}
    |\psi_n(0)|^2=\frac{1}{\pi}\frac{\gamma}{(\lambda_n-\varepsilon_0)^2+\gamma^2},
\end{equation}
with $\gamma\sim O(1)$ controlling the broadening. For finite $N$, the resolvent has real poles corresponding to the eigenvalues $z=\lambda_n$. However, in the thermodynamic limit $N\to \infty$ the spectrum becomes continuous and,
\begin{equation}
    G_{00}(z)=\int d\lambda \frac{1}{\pi}\frac{\gamma}{(\lambda-\varepsilon_0)^2+\gamma^2}\frac{1}{z-\lambda}.
\end{equation}
Using the residue theorem in the $\lambda$-plane, and closing the contour in the
half-plane that avoids the pole at $\lambda=z$ (upper half if $\operatorname{Im}
z<0$, lower half if $\operatorname{Im}z>0$), one obtains
\begin{equation}
    G_{00}(z)=
    \begin{cases}
        \dfrac{1}{z-\varepsilon_0 + i\gamma}, & \operatorname{Im}z>0\\[8pt]
        \dfrac{1}{z-\varepsilon_0 - i\gamma}, & \operatorname{Im}z<0
    \end{cases}\quad.
\end{equation}
This expression shows that the resolvent is not a single analytic function in the whole complex plane. Instead, it consists of two different analytic branches: one valid in the upper half‑plane (the retarded Green’s function $G^R(z)$) and one in the lower half‑plane (the advanced Green’s function $G^A(z)$). The real axis is a branch cut, and therefore the limiting values from above and below differ,
\begin{equation}
    \lim_{\eta\to0^+} G_{00}(\lambda+i\eta) = \frac{1}{\lambda-\varepsilon_0+i\gamma},\qquad
    \lim_{\eta\to0^+} G_{00}(\lambda-i\eta) = \frac{1}{\lambda-\varepsilon_0-i\gamma},
\end{equation}
and their difference gives the local density of states,
\begin{equation}
    \rho_0(\lambda)=\frac{G_{00}(\lambda-i0^+)-G_{00}(\lambda+i0^+)}{2\pi i}=\frac{1}{\pi}\frac{\gamma}{(\lambda-\varepsilon_0)^2+\gamma^2}.
\end{equation}
Then the return amplitude becomes
\begin{equation}
\langle 0|e^{-iH\tau}|0\rangle = \int_{-\infty}^{\infty} d\lambda\, e^{-i\lambda \tau}\,\frac{1}{\pi}\frac{\gamma}{(\lambda-\varepsilon_0)^2+\gamma^2} = e^{-i\varepsilon_0 \tau}\,e^{-\gamma \tau},
\end{equation}
and the return probability decays exponentially:
\begin{equation}
R(\tau) = |\langle 0|e^{-iH\tau}|0\rangle|^2 = e^{-2\gamma \tau}.
\end{equation}
Thus, $\gamma$ acts as an inverse lifetime. In the Lorentzian example, the width $\gamma$ captures the broadening of the energy levels. In a more realistic and more complex setting, of course, the local density of states does not have a perfectly Lorentzian shape, but this toy model captures the essential features. More generally, such a broadening is encoded in the imaginary part of the resolvent, as we will see in the case of the Bethe lattice in Sec.~\ref{sec:critical}.

\subsection{Analogy with standard second order phase transitions}

A helpful conceptual analogy can be drawn between the Anderson localization transition and standard second-order phase transition, such as, for instance, the ferromagnetic transition described by the classical Ising model \cite{Ising1925}. Both transitions are characterized by an order parameter that is non‑zero in one phase and zero in the other, but the distinction emerges only when the thermodynamic limit is taken before an auxiliary parameter is sent to zero. In the Ising model, the order parameter is the local magnetization $m_i = \langle \sigma_i \rangle$. For any finite system, $m_i = 0$ regardless of temperature, because the average over the symmetric distribution vanishes. The ferromagnetic phase is revealed by taking the thermodynamic limit $N\to\infty$ first, and then sending the symmetry‑breaking field $h\to0^+$; this yields a non‑zero magnetization. In the paramagnetic phase, instead, the magnetization is zero regardless of the order of limits.

In Anderson localization, the analogous role is played by the typical value of $\operatorname{Im} G_{ii}(\lambda - i\eta)$, where $\eta>0$ is a small broadening. For a finite system, $\operatorname{Im} G_{ii}(\lambda - i\eta)$ is a sum of Lorentzians that are non‑zero only in the immediate vicinity of the eigenvalues. In the thermodynamic limit, the order of limits distinguishes the phases:
\begin{itemize}
    \item Delocalized phase ($W<W_c$): $\displaystyle \lim_{\eta\to0^+}\lim_{N\to\infty} \operatorname{Im} G_{ii}(\lambda - i\eta) > 0$,
    \item Localized phase ($W>W_c$): $\displaystyle \lim_{\eta\to0^+}\lim_{N\to\infty} \operatorname{Im} G_{ii}(\lambda - i\eta) = 0$ (for most sites).
\end{itemize}
Taking the limits in the opposite order ($\lim_{N\to\infty}\lim_{\eta\to0^+}$) yields zero in both phases, because for any finite $N$ the spectrum is discrete and the resolvent has isolated poles. Thus, the delocalized phase is accessed only when the thermodynamic limit is taken before the broadening is removed, mirroring the Ising case.

The analogy is summarised in Table~\ref{tab:ising_anderson}. It highlights that both transitions share a common structure: a non‑commutativity of limits that defines the ordered phase. A crucial difference, however, is that in Anderson localization the local density of states is a random variable; the true order parameter is its probability distribution rather than its mean, which remains continuous across the transition \cite{mirlin1994distribution}.

As a matter of fact, the Anderson localization transition can also be described within a field‑theoretic framework, the supersymmetric non‑linear sigma model (SUSY NL$\sigma$M) \cite{wegner1979mobility, schafer1980disordered, efetov1983supersymmetry}. In this formulation, the transition can be interpreted as a spontaneous breaking of a graded symmetry, and the relevant order parameter is not a simple expectation value but the probability distribution of the local density of states (LDOS) \cite{mirlin1991localization, mirlin1994distribution, evers2008anderson, hikami1992localization}. This order parameter becomes non‑trivial in the localized phase, signaling the breakdown of the symmetry that is present in the metallic phase.

\begin{table}[h]
\centering
\begin{tabular}{|p{0.48\textwidth}|p{0.48\textwidth}|}
\hline
\textbf{Ising model} & \textbf{Anderson localization} \\
\hline
Local magnetization: $m_i = \langle \sigma_i \rangle$ &
Distribution of the LDOS: characterized by the typical value of $\operatorname{Im} G_{ii}(\lambda - i\eta)$ \\
\hline
$m_i = 0$ both above and below $T_c$ in finite-size systems&
$\operatorname{Im} G_{ii}(\lambda - i0^+) = 0$ in finite-size systems  (except if $\lambda$ coincides with an eigenvalue) \\
\hline
Ferromagnetic phase ($T<T_c$): &
Delocalized phase ($W<W_c$): \\
$\displaystyle \lim_{h\to0^+}\lim_{N\to\infty} m_i(h) = \pm m > 0$ &
$\displaystyle \lim_{\eta\to0^+}\lim_{N\to\infty} \operatorname{Im} G_{ii}(\lambda - i\eta) > 0$ \\
\hline
Paramagnetic phase ($T>T_c$): &
Localized phase ($W>W_c$): \\
$\displaystyle \lim_{h\to0^+}\lim_{N\to\infty} m_i(h) = 0$ &
$\displaystyle \lim_{\eta\to0^+}\lim_{N\to\infty} \operatorname{Im} G_{ii}(\lambda - i\eta) = 0$ (for most sites) \\
\hline
\end{tabular}
\caption{Comparison between the transition in the Ising model and in Anderson localization.}
\label{tab:ising_anderson}
\end{table}

In summary, the resolvent encodes all the diagnostics of localization: the spectral type is reflected in its analytic structure close to the real axis, the eigenfunction moments, spatial correlations, and return probability are all expressed through its matrix elements, and the full distribution of $\operatorname{Im} G_{ii}$---taking $N\to\infty$ before $\eta\to0^+$---is the order parameter of the transition. We now turn to the lattice on which this distribution can be computed exactly: after defining its geometry in Sec.~\ref{sec:graph}, we derive the cavity equations in Sec.~\ref{sec:cavity}.

%=== GRAPHS ===
\section{Introduction to graph theory}\label{sec:graph}

The resolvent formulation introduced in the previous section becomes particularly
powerful on tree-like geometries, where the absence of short loops allows the
Green's functions on different branches of the graph to be related through
self-consistent equations. Before deriving these equations, we therefore introduce
the graph-theoretical notions needed to define the Bethe lattice and its natural
finite-size counterpart, the random regular graph (RRG). In particular, we will
clarify the distinction between the Bethe lattice, RRGs, and finite Cayley trees,
which becomes important when discussing Anderson localization.

\subsection{Basic definitions}

A graph is a mathematical structure consisting of a set of objects that are pairwise related. The objects are abstracted as vertices (or nodes) and their relations are represented by edges. Formally, a graph is a pair $G = (V,E)$, where $V$ is the set of vertices and $E$ is a set of pairs $(i,j)$ of vertices. In these notes we consider simple undirected graphs: there are no self-loops (an edge $(i,i)$ is not allowed), there is at most one edge between any two vertices, and edges have no orientation, so $(i,j) = (j,i)$.

A graph is called unweighted when all edges are treated equally. If a numerical
value (weight) is assigned to each edge, the graph is said to be weighted. The number of neighbors of a given vertex $i$ is its degree, denoted by $k_i$.

The adjacency matrix $A$ of an undirected simple graph with $N$ vertices is an $N\times N$ symmetric matrix with entries
\begin{equation}
    A_{ij} = 
    \begin{cases}
    1 & \text{if } \{i,j\}\in E,\\
    0 & \text{otherwise},
    \end{cases}
\end{equation}
and $A_{ii}=0$ because there are no self-loops. The degree of vertex $i$ is then
\begin{equation}
    k_i = \sum_{j=1}^N A_{ij},
\end{equation}
and the number of edges is
\begin{equation}
    |E| = \frac12\sum_{i,j=1}^N A_{ij}.
\end{equation}

A walk is a sequence of vertices such that consecutive vertices are connected by an edge. A path is a walk with no repeated vertices. A cycle (or loop) is a closed path, \ie, a path that starts and ends at the same vertex, with all other vertices distinct. A graph without cycles is called acyclic. A connected acyclic graph is a tree. For a tree with $N$ vertices,
there exists exactly one simple path between any two vertices. Graphs with $|E| = O(N)$ (\ie, the number of edges scales at most linearly with the number of vertices) are called sparse. In such graphs, the average degree $\langle k \rangle = 2|E|/N$ remains finite as $N\to\infty$.

The adjacency matrix also encodes the number of walks. The number of walks of length $\ell$ between vertices $i$ and $j$ is given by
\begin{equation}
    n_{\text{walks}}^{(\ell)}(i,j) = (A^{\ell})_{ij}.
\end{equation}
We can also write the degree of a node as $k_i=(A^2)_{ii}$ and the number of edges as $|E|=\tfrac12\text{Tr}(A^2)$.

If we assign a weight $t$ to each step, the generating function for the number of
walks between vertices is
\begin{equation}
\mathcal{G}(t) = I + \sum_{\ell=1}^{\infty} t^{\ell} A^{\ell} = (I - tA)^{-1},
\end{equation}
where $I$ is the identity matrix (which corresponds to walks of length zero). This
generating function is closely related to the resolvent $G(z)=(z-H)^{-1}$
introduced in Sec.~\ref{sec:resolvent}: setting $H=A$ (the adjacency matrix with no
on-site disorder, and with the opposite overall sign convention used for the
hopping term in Eq.~\eqref{eq:H_AL_1st}) and $z=1/t$ gives
$\mathcal{G}(t) = t^{-1}\,G(1/t)$, and is useful when studying walks and spectral
properties of graphs.

\subsection{Random tree-like graphs}

A random graph is a statistical ensemble of individual graphs, each occurring with a certain probability. Observables are then defined as averages over this ensemble.

We focus on a special class of random graphs that are locally tree-like. A locally tree-like graph is one in which, for any fixed distance, the neighborhood of a typical node is a tree with probability tending to one as $N\to\infty$.
This does not imply that the graph contains no cycles: rather, a typical node lies at a distance that grows with system size from the nearest loop. The random graph ensembles considered below are sparse, with finite average degree as $N\to\infty$.

Locally tree-like graphs are used as a mean-field version of $d$-dimensional Euclidean lattices, since the number of nodes within a given distance from a given node grows exponentially with the distance. Their hierarchical structure not only renders them analytically tractable, but also preserves a notion of distance and finite local connectivity. They are therefore widely used as approximations of more realistic networks.

The systematic study of tree-like random graphs began in the second half of the 1950s with the work of Erdős and Rényi \cite{erdos1959random1, erdos1960evolutionRandomGraph, erdos1961connectedness}. They introduced two equivalent models that are now collectively known as the Erdős–Rényi random graph.

\begin{itemize}
    \item \textbf{The $G(N,p)$ model:}  
    Consider a set of $N$ vertices and connect each pair of vertices independently with probability $p$. The ensemble contains all $2^{N(N-1)/2}$ labeled simple graphs on $N$ vertices; a graph $G$ with $|E|$ edges has probability  
    \begin{equation}
        \mathcal{P}(G) = p^{|E|}(1-p)^{N(N-1)/2 - |E|}.
    \end{equation}
    The degree of a vertex follows the binomial distribution  
    \begin{equation}
        P(k) = \binom{N-1}{k} p^{k} (1-p)^{N-1-k},
    \end{equation}
    so the average degree is $\langle k \rangle = p(N-1)$. In the limit $N\to\infty$ with $\langle k \rangle = c$ fixed (the sparse regime), this distribution approaches a Poisson distribution  
    \begin{equation}
        P(k) = e^{-c} \frac{c^{k}}{k!},
    \end{equation}
    hence the $G(N,p)$ model is often called a \textit{Poisson random graph}.

    \item \textbf{The $G(N,M)$ model:}  
    Consider now the uniform ensemble of all labeled simple graphs with exactly $N$ vertices and $M$ edges. The total number of such graphs is $\binom{N(N-1)/2}{M}$, and each graph has the same probability $\mathcal{P}(G)=1/Z(N,M)$. In the sparse limit $N\to\infty$, $M = cN/2$ (so that $\langle k\rangle = 2M/N = c$), the two ensembles become equivalent: the degree distribution is again Poisson \cite{Bogacz2006}.
\end{itemize}

In the $G(N,p)$ model the number of edges fluctuates; therefore, it is a grand canonical ensemble. In contrast, the $G(N,M)$ model fixes the number of edges exactly and is a microcanonical ensemble. 
In both cases, the presence of an edge between any two vertices is independent of all others, so the networks are uncorrelated: the degree of a vertex carries no information about the degrees of its neighbors.

\subsubsection{The configuration model}\label{sec:config_model}

The Erdős–Rényi model yields random graphs with a Poisson degree distribution. To generate random graphs with an arbitrary prescribed degree distribution, one can use the configuration model. Given a degree sequence $(k_1,\dots,k_N)$ with $\sum_i k_i$ even, we attach $k_i$ “stubs” (or half-edges) to each vertex $i$. The graph is built by repeatedly choosing two stubs uniformly at random from the remaining pool and joining them to form an edge, until all stubs are paired.

This construction produces a random multigraph: every pairing of stubs that respects the given degrees is equally likely. The resulting graph may contain self-loops (a stub of a vertex paired with another stub of the same vertex), multiple edges between the same pair of vertices, and cycles of various lengths. Moreover, the graph may consist of several connected components.

For large $N$ and degree sequences that are sparse (i.e., $\langle k\rangle$ finite) and have finite second moment $\langle k^2\rangle$, the expected numbers of self-loops, multiple edges, and short cycles are finite (i.e., they do not grow with $N$); therefore, they constitute a negligible fraction of the graph in the thermodynamic limit. Under the same conditions, the graph has a unique giant connected component with probability tending to $1$ as $N\to\infty$ whenever the Molloy–Reed criterion $\langle k^2\rangle > 2\langle k\rangle$ is satisfied \cite{molloy1995critical}. For degree sequences that do not satisfy this condition, no giant component exists. In all cases, the configuration model provides a versatile tool to generate random locally tree-like graphs with a prescribed degree distribution \cite{newman2018networks}.

\subsection{Random regular graphs and the Bethe lattice}

A particularly important special case of the configuration model is the random regular graph (RRG)~\cite{wormald1999models}. Here, all vertices have the same degree $k+1$ with $k\ge 2$. The RRG ensemble consists of all simple graphs with $N$ vertices where each vertex has exactly $k+1$ neighbors, each graph taken with equal probability. In practice, one often generates such graphs via the configuration model with a uniform degree sequence.

\begin{figure}
    \centering
\includegraphics[width=0.35\linewidth]{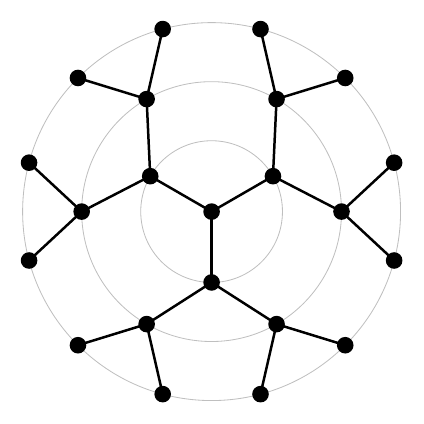}
    \caption{Cayley tree with branching number $k=2$ (degree $k+1=3$) up to the third generation. Locally, it is
    identical to the neighborhood of the RRG and of the Bethe lattice of the same
    degree. Each concentric circle shows a successive generation of nodes, where the distance $\ell$ from the central root increases by one at each layer.}
    \label{fig:BL_rings}
\end{figure}

The RRG is locally tree-like: for any fixed distance $\ell$, the $\ell$-neighborhood of a typical vertex is a tree with probability tending to $1$ as $N\to\infty$. The characteristic distance from a typical vertex at which loops are first encountered scales as $O(\ln N/\ln k)$ (see \cite{dorogovtsev2022nature,Marinari2004,wormald1999models,MarinariMonassonSemerjian2004} for a review of this result),  so any fixed neighborhood becomes tree-like as $N$ grows. Fig.~\ref{fig:BL_rings} shows the tree-like neighborhood of a node of the RRG.

A frequent source of confusion lies in identifying the infinite Bethe lattice with the thermodynamic limit of a finite Cayley tree.  A Cayley tree of $n$ generations is a loopless hierarchical structure whose root node is connected to $k+1$ offspring in the first generation, while all subsequent nodes up to generation $n-1$ branch into $k$ offspring in the next generation. This results in a number of vertices accessible at distance $\ell$ from the center that grows exponentially as $(k+1)k^{\ell-1}$. The last generation ($n$) is composed of leaves and forms a boundary that comprises a finite fraction of the total number of nodes, approaching $(k-1)/k$ as $n\to\infty$. Therefore, even at the thermodynamic limit $n\to\infty$, the boundary vertices remain non‑negligible, and the nodes belonging to different generations have different statistical properties. In contrast, the Bethe lattice is an infinite tree which has no boundaries and is translationally invariant, \ie, all vertices are equivalent. Recent studies have demonstrated that these boundary effects cause Anderson localization on Cayley trees to exhibit profound qualitative and quantitative differences compared to random regular graphs --particularly within the delocalized metallic phase --discrepancies that persist robustly in the thermodynamic limit~\cite{biroli2020anomalous,sonner2017multifractality,tikhonov2016fractality}.

RRGs circumvent this problem because they have no boundary: each vertex is statistically equivalent, and the graph is homogeneous. In the thermodynamic limit $N\to\infty$, an RRG converges locally to the Bethe lattice, meaning that any finite neighborhood of a randomly chosen vertex is, with high probability, isomorphic to the corresponding neighborhood in the Bethe lattice \cite{wormald1999models, Bollobas2001}. Hence, RRGs provide the most natural finite-size realization of the Bethe lattice for numerical studies.

For a large enough system size $N$ and sufficiently small distances $\ell$, the
number of vertices at distance $\ell$ from a given vertex in an RRG grows
exponentially as $\sim k^\ell$, just as in a Cayley tree. However, unlike the
Cayley tree, the RRG contains loops: a vertex reached at distance $\ell$ may
actually coincide with an already visited vertex, forming a cycle. The probability
that a given vertex at distance $\ell$ is not new (\ie, creates a loop) is of order
$k^\ell/N$, which is negligible for small enough $\ell$. More precisely, when
$\ell \sim O(\ln N/\ln k)$, this probability becomes relevant and the exponential
growth predicted by the tree approximation breaks down, so that $\ln N/\ln k$ is
the typical value of loop size and can be understood as the diameter of the graph.

In the thermodynamic limit $N\to\infty$, an RRG is therefore locally a tree up to
arbitrarily large distances, while globally it is a compact structure with a
diameter that grows logarithmically with $N$ --- making it the ideal finite-$N$
approximation to the Bethe lattice.

A clear and direct manifestation of the pathological nature of Cayley trees, as compared to RRGs, is provided by the spectral statistics of their adjacency matrices. In the absence of disorder, the eigenvalue statistics of RRG adjacency matrices obey GOE statistics (indeed, RRGs behave in many respects as a sparse discrete counterpart of GOE random matrices). In contrast, the eigenvalue statistics of finite Cayley trees are non-universal: they are neither described by random matrix theory nor by Poisson statistics for any finite generation $n$, and they do not converge to the RMT prediction in the large-$n$ limit.

The distinction established above will be crucial in what follows. Locally, an RRG becomes indistinguishable from the Bethe lattice as $N\to\infty$, while being, at any system size, free of the extensive boundary that characterizes
finite Cayley trees. This local tree structure is precisely what allows the resolvent equations to close recursively. In the next section, we exploit this property to derive the cavity
equations for the Anderson model on the Bethe lattice.

%=== CAVITY ===
\section{Cavity equations}\label{sec:cavity}

As explained above, the central object for describing the spectral properties of the Anderson model and the localization transition is the resolvent matrix $G(z)=(z-H)^{-1}$, whose diagonal elements give the local density of states via $\rho_i(\lambda) = \frac{1}{\pi}\lim_{\eta\to0^+}\Im G_{ii}(\lambda-i\eta)$. On the Bethe lattice, the absence of loops allows us to derive exact recursive equations for the diagonal elements of the resolvent.
In this section, we explain how to derive these equations using the so-called cavity method (see \cite{braunstein2023cavity, krzakala2011belief, mezard2009information} for comprehensive reviews in the broader context of statistical physics and optimization problems, and \cite{Susca_Vivo_Kuhn_2021, potters2020first} for applications to the spectral properties of random matrices). 

\subsection{Derivation of the cavity equations through a Gaussian representation}

A convenient starting point is a Gaussian representation of the resolvent \cite{Susca_Vivo_Kuhn_2021} in the presence of an external field. For a real symmetric matrix $M$ and a source vector $J$, the identity
\begin{equation}
\begin{aligned}
    & \int \prod_{i=1}^N d\phi_i \; \phi_a\phi_b \; e^{-\frac12\sum_{i,j}\phi_i M_{ij}\phi_j + \sum_i J_i\phi_i} \\
    & \qquad = \frac{(2\pi)^{N/2}}{\sqrt{\det M}}\; e^{\frac12\sum_{i,j} J_i (M^{-1})_{ij} J_j} \Bigl[ (M^{-1})_{ab} + \sum_{i,j} (M^{-1})_{ai}J_i\,(M^{-1})_{bj}J_j \Bigr]
    \end{aligned}
\end{equation}
holds provided the integrals converge. For $M = zI - H$, the quadratic form is not positive definite when $z$ lies in the complex plane. To ensure convergence, one works with the complex Gaussian representation
\begin{equation}
    \int \prod_{i=1}^N d\phi_i \; \phi_a\phi_b \;
    e^{-\frac{i}{2}\sum_{i,j}\phi_i (zI-H)_{ij}\phi_j}
    =
    -i\,\frac{(2\pi)^{N/2}}{\sqrt{\det[i(zI-H)]}}\,
    \bigl[(zI-H)^{-1}\bigr]_{ab},
\end{equation}
where the integration contour is taken such that the quadratic form has a positive real part. For $z$ in the lower half-plane ($\Im z<0$), the matrix $i(zI-H)$ has a positive definite real part, ensuring convergence. Thus, the matrix elements of the resolvent can be expressed as correlation functions of the Gaussian fields $\phi_i$ with the action
\begin{equation}
    S[\phi] = -\frac{i}{2}\sum_{i=1}^N (z-\varepsilon_i)\phi_i^2 - i\sum_{\langle i,j\rangle} t\,\phi_i\phi_j ,
\end{equation}
where $i$ is a node index, summed over all nodes and $\langle i,j \rangle$ the sum over nearest-neighbor pairs on the lattice.

The elements of the resolvent can thus be obtained as
\begin{equation}\label{eq:Gint_def}
    -iG_{ab}(z) = \frac{\int D\phi\; \phi_a\phi_b \; e^{S[\phi]}}{\int D\phi\; e^{S[\phi]}}.
\end{equation}
The denominator $Z=\int D\phi\,e^{S[\phi]}$ plays the role of a partition function, and we introduce the normalized Gaussian weight $P[\phi]=e^{S[\phi]}/Z$.
 
\begin{figure}
    \centering  \includegraphics[width=0.5\linewidth]{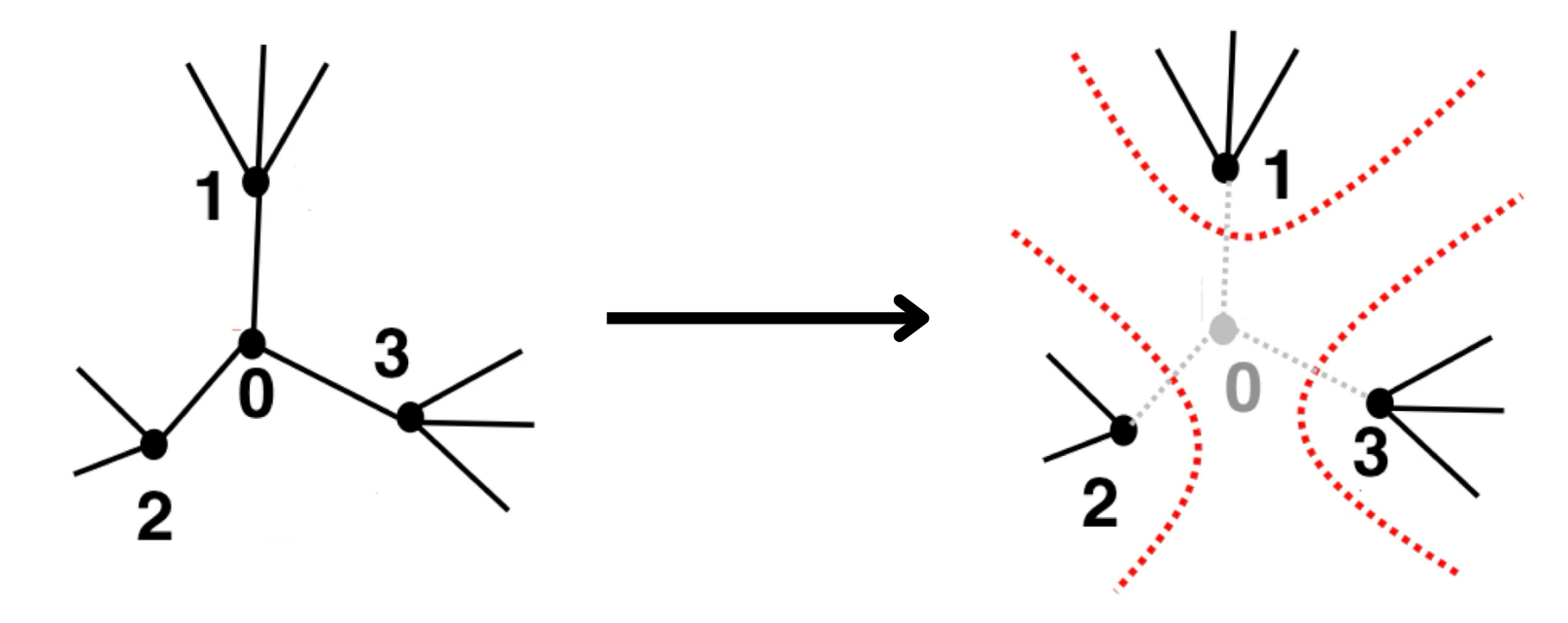}
    \caption{Schematic representation of the cavity method factorization.}
    \label{fig:cavity_graph}
\end{figure}

Since the Bethe lattice is a tree, removing a vertex $i$ disconnects the
graph into separate branches, one rooted at each of its neighbors (see
Fig.~\ref{fig:cavity_graph}). This property implies that the joint marginal
distribution of the fields on the neighbors of $i$ factorizes once vertex
$i$ is removed. For an RRG, the same factorization becomes asymptotically
exact for finite neighborhoods as $N\to\infty$, as a consequence of its
locally tree-like structure. Consider a vertex $0$ and its $k+1$ neighbors. The marginal distribution of the fields $\phi_1,\dots,\phi_{k+1}$ in a system where vertex $0$ is absent factorizes into a product of independent distributions for each branch:
\begin{equation}
    P^{(0)}(\phi_1,\dots,\phi_{k+1}) = \prod_{j=1}^{k+1} P_{j}^{(0)}(\phi_j),
\end{equation}
where the superscript $(0)$ denotes a cavity in node $0$. Therefore, $P_{j}^{(0)}(\phi_j)$ is the distribution of the field on vertex $j$ in a graph where node $0$ was removed. This factorization is exact on a tree and becomes asymptotically exact on a random regular graph as $N\to\infty$, because short loops become extremely rare \cite{BordenaveLelarge2010}.

From the Gaussian nature of the action, each $P_{j}^{(0)}(\phi)$ is itself Gaussian. We therefore parametrize it as
\begin{equation}
   P_{j}^{(0)}(\phi_j) \propto \exp\!\left(-\frac{i}{2}\,\frac{\phi_j^2}{G_{j}^{(0)}(z)}\right).
\end{equation}
From Eq.~\eqref{eq:Gint_def} we can recognize the complex number $G_{j}^{(0)}(z)$ as the diagonal element of the resolvent of the system where vertex $0$ has been removed, evaluated at vertex $j$.

We first derive the diagonal resolvent $G_{00}$ in terms of the cavity resolvents $G_j^{(0)}$ of its $k+1$ neighboring branches. The corresponding
recursion for a cavity resolvent $G_0^{(i)}$ is then obtained by excluding neighbor $i$. The joint distribution of $\phi_0$ and its neighbors is,
\begin{equation}
    P(\phi_0,\phi_1,\dots,\phi_{k+1}) \propto e^{-\frac{i}{2}(z-\varepsilon_0)\phi_0^2 - i t\phi_0\sum_{j=1}^{k+1}\phi_j}\;
    \prod_{j=1}^{k+1} e^{-\frac{i}{2}\frac{\phi_j^2}{G_{j}^{(0)}}}.
\end{equation}
Integrating out the neighbors $\phi_j$ (which are independent Gaussians) gives a marginal distribution for $\phi_0$:
\begin{equation}
    P_0(\phi_0) \propto e^{-\frac{i}{2}(z-\varepsilon_0)\phi_0^2}\;
    \prod_{j=1}^{k+1} \int d\phi_j\;
    e^{-i t\phi_0\phi_j - \frac{i}{2}\frac{\phi_j^2}{G_{j}^{(0)}}}.
\end{equation}
Each Gaussian integral over $\phi_j$ yields a factor
\begin{equation}
    \int d\phi_j\;
    e^{-i t\phi_0\phi_j - \frac{i}{2}\frac{\phi_j^2}{G_{j}^{(0)}}}
    \propto
    \exp\!\left(\frac{i}{2}\,t^2G_{j}^{(0)}\phi_0^2\right).
\end{equation}
Collecting the contributions from all neighbors, we obtain
\begin{equation}
    P_0(\phi_0) \propto \exp\!\left(-\frac{i}{2}\Bigl[ z-\varepsilon_0 - t^2\sum_{j=1}^{k+1} G_{j}^{(0)} \Bigr]\phi_0^2\right).
\end{equation}
Comparing with the definition of the Green's function (Eq.~\eqref{eq:Gint_def}) we have that,
\begin{equation}\label{eq:resolventBL}
     G_{00}(z) = \frac{1}{\,z - \varepsilon_0 - t^2\sum_{j\in\partial0} G_{j}^{(0)}(z)\,}.
\end{equation}
To close the set of equations, we repeat the procedure for the cavity resolvent, 
\begin{equation}\label{eq:cavity_bl}
    G_{0}^{(i)}(z) = \frac{1}{z - \varepsilon_0 - t^2\sum_{j\in\partial0\setminus i} G_{j}^{(0)}(z)\,}.
\end{equation}
Equation~\eqref{eq:cavity_bl} forms a closed recursion for the cavity Green's functions. For a random regular graph, this approximation becomes asymptotically exact in the thermodynamic limit because loops become rare \cite{BordenaveLelarge2010}.

The sum in the denominator of Eq.~\eqref{eq:cavity_bl} is often referred to as the cavity self-energy~\cite{abou1973selfconsistent},
\begin{equation}\label{eq:cavselfE_def}
    \Sigma_{0}^{(i)}(z) = t^2\sum_{j\in\partial0\setminus i} G_{j}^{(0)}(z).
\end{equation}
It renormalizes the on‑site energy due to the coupling to the rest of the graph. Using the cavity equation for $G_{0}^{(i)}(z) = (z - \varepsilon_0- \Sigma_{0}^{(i)}(z))^{-1}$, we obtain a closed self‑consistent relation for the cavity self-energy:
\begin{equation}\label{eq:cavselfE_recursion}
    \Sigma_{0}^{(i)}(z) = t^2 \sum_{j\in\partial0\setminus i} \frac{1}{z -\varepsilon_j- \Sigma_{j}^{(0)}(z)}.
\end{equation}

Likewise, one can define the self-energy 
\begin{equation}\label{eq:selfE_def}
    \Sigma_{0}(z) = t^2\sum_{j\in\partial0} G_{j}^{(0)}(z),
\end{equation}
and rewrite the resolvent as
\begin{equation}\label{eq:resolventBL_selfE}
     G_{00}(z) = \frac{1}{\,z - \varepsilon_0 - \Sigma_0(z)}.
\end{equation}

Since the disorder makes the cavity Green's functions random variables, Eq.~\eqref{eq:cavity_bl} induces a self-consistent recursion for their probability distribution. By statistical homogeneity, the cavity Green's functions are identically distributed after disorder averaging; let $P(G)$
denote their common distribution. In the following, $G$ and $G_j$ denote generic realizations drawn from this distribution, with site labels and cavity superscripts suppressed. Equation~\eqref{eq:cavity_bl} then becomes

\begin{equation}\label{eq:Pprob_cavity}
    P(G) =\int d\varepsilon \gamma(\varepsilon) \prod_{j=1}^k dG_j P(G_j)\delta\left(G- \frac{1}{z - \varepsilon - t^2\sum_{j=1}^{k} G_{j}(z)}\right).
\end{equation}
The fixed point of this equation can then be used to obtain the distribution $Q(G)$ of the diagonal resolvent of the original problem. Here, $G$ denotes a realization of the full diagonal resolvent, while the $G_j$ entering the recursion remain cavity Green's functions distributed according to $P(G)$,
\begin{equation}\label{eq:Qprob_G}
    Q(G) =\int d\varepsilon \gamma(\varepsilon) \prod_{j=1}^{k+1} dG_j P(G_j)\delta\left(G- \frac{1}{z - \varepsilon - t^2\sum_{j=1}^{k+1} G_{j}(z)}\right).
\end{equation}

This distribution can be solved numerically by population dynamics (which we will explain in Sec.\ref{sec:popdyn}).

\subsection{Block-matrix inversion derivation of the cavity equations}

An alternative derivation of the cavity equations is based on the Schur complement (block-matrix inversion) formula \cite{arous2008spectrum,potters2020first}. This approach is purely algebraic and directly relates the full resolvent to the resolvent of the graph with a vertex removed, for a generic graph.

Consider the Anderson Hamiltonian $H$ on a graph. For a given vertex $0$, we write the matrix $M = zI - H$ in block form, separating the row and column corresponding to vertex $0$ from the rest:
\begin{equation}\label{eq:matrix}
    M = \begin{pmatrix}
        z - \varepsilon_0 & t\,\mathbf{1}_{\partial 0}^{\mathsf{T}} \\
        t\,\mathbf{1}_{\partial 0} & M^{(0)}
    \end{pmatrix},
\end{equation}
where $\mathbf{1}_{\partial 0}$ is a column vector (of size $N-1$) with entries equal to $1$ for each neighbor $i$ of $0$ and $0$ otherwise, and $M^{(0)} = zI^{(0)} - H^{(0)}$ is the $(N-1)\times(N-1)$ matrix obtained by removing the row and column corresponding to $0$ from $zI-H$. The resolvent is given by $G(z) = M^{-1}$.

Applying the Schur complement formula~\cite{Schur1917} to Eq.~\eqref{eq:matrix}, we obtain
\begin{equation}\label{eq:G00_block}
    G_{00} = \Bigl( z - \varepsilon_0 - t^2\,\mathbf{1}_{\partial 0}^{\mathsf{T}} (M^{(0)})^{-1} \mathbf{1}_{\partial 0} \Bigr)^{-1}
    = \frac{1}{z - \varepsilon_0 - t^2 \sum_{i,j \in \partial 0} G^{(0)}_{ij}},
\end{equation}
where $G^{(0)} = (M^{(0)})^{-1}$ is the \emph{cavity} resolvent of the graph with vertex $0$ removed, and $G^{(0)}_{ij}$ are its matrix elements between neighbors $i$ and $j$ of $0$.

Equation~\eqref{eq:G00_block} is exact for any graph. On a general graph, the off-diagonal terms $G^{(0)}_{ij}$ for $i \neq j$ are nonzero, since paths connecting distinct neighbors of $0$ still exist after removing $0$. However, on a tree (or on a locally tree-like graph in the thermodynamic limit), removing vertex $0$ disconnects its neighbors. Consequently, there are no paths between $i$ and $j$ in the absence of $0$, and thus $G^{(0)}_{ij} \approx 0$ for $i \neq j$. In random regular graphs, loops become rare as $N \to \infty$, and the probability of finite off-diagonal contributions vanishes. Within the Bethe approximation, one therefore neglects the off-diagonal terms and retains only the diagonal ones, recovering Eq.~\eqref{eq:cavity_bl}.

Once the off-diagonal terms are neglected in Eq.~\eqref{eq:G00_block}, one obtains an expression for the Green's function at node $0$ in terms of the cavity Green's functions $G_j^{(0)}$ on its neighbors. To close the system of equations, one can apply the same block inversion strategy to the cavity matrix $M^{(0)}$, namely the resolvent of the Hamiltonian where node $0$ has been removed. This yields
\begin{equation}\label{eq:Gj_block}
    G_{j}^{(0)} = \Bigl( z - \varepsilon_j - t^2\,\mathbf{1}_{\partial j \setminus 0}^{\mathsf{T}} (M^{(0,j)})^{-1} \mathbf{1}_{\partial j \setminus 0} \Bigr)^{-1}
    = \frac{1}{z - \varepsilon_j - t^2 \sum_{m_1,m_2 \in \partial j \setminus 0} G^{(0,j)}_{m_1 m_2}},
\end{equation}
where $\mathbf{1}_{\partial j \setminus 0}$ is a vector (of size $N-2$) with entries equal to $1$ for neighbors of $j$ excluding $0$, and $M^{(0,j)} = zI^{(0,j)} - H^{(0,j)}$ is the $(N-2)\times(N-2)$ matrix obtained by removing both $0$ and $j$ from $zI-H$.

This equation is again exact on any graph, but it does not provide a closed system. On tree-like graphs, however, the same simplifications apply: in the absence of node $j$, its neighbors become disconnected, so $G^{(0,j)}_{m_1 m_2} \approx 0$ for $m_1 \neq m_2$. Moreover, since removing node $0$ does not affect the branches rooted on the nodes $m$ in absence of $j$, one has $G_m^{(0,j)} \approx G_m^{(j)}$. These approximations---which become asymptotically exact on infinite random regular graphs~\cite{BordenaveLelarge2010}---lead back to the self-consistent cavity equations derived above, Eq.~\eqref{eq:resolventBL}.

Finally, it is worth noting that the self-consistent equations~\eqref{eq:resolventBL} were originally derived in a different way in the seminal work of Abou-Chacra, Anderson, and Thouless, who first formulated and solved the Anderson localization problem on the Bethe lattice~\cite{abou1973selfconsistent}. In that work, the equations were obtained without resorting to either the Gaussian integral representation of the resolvent or the block-matrix inversion formula. Instead, they were derived by an exact resummation of the perturbative expansion in the hopping amplitude---known as the locator expansion and briefly introduced in Sec.~\ref{sec:model}. This resummation can be carried out exactly on loopless, tree-like structures, where any two nodes are connected by a unique path.

A detailed presentation of this approach goes beyond the scope of this review; we refer the interested reader to Refs.~\cite{abou1973selfconsistent,scardicchio2017perturbation} for comprehensive discussions.

\subsection{The off diagonal elements}

As discussed in Sec.~\ref{sec:resolvent}, the off-diagonal elements of the
resolvent are also of interest, since they provide access to spatial correlation
functions, such as the one defined in Eq.~\eqref{eq:corr_def}. They can be
computed in a way analogous to the diagonal elements, using the Gaussian
representation in Eq.~\eqref{eq:Gint_def}.

\begin{figure}
    \centering
    \includegraphics[width=0.5\linewidth]{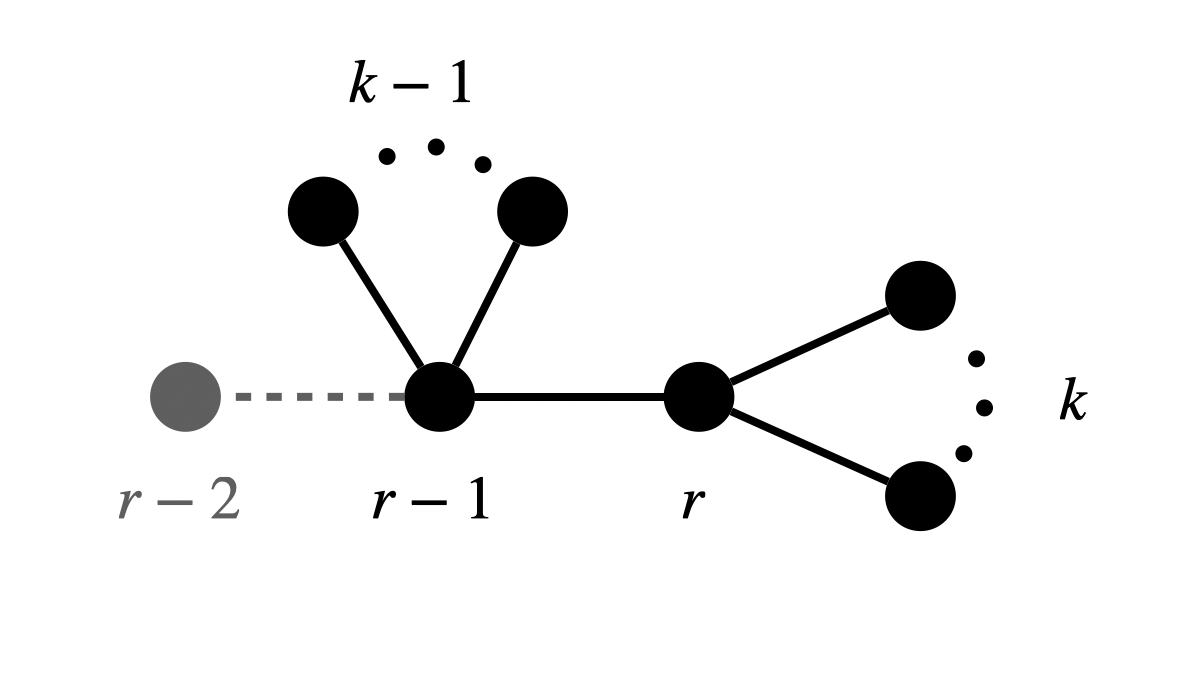}
    \caption{Branching structure entering the computation of the off-diagonal resolvent $G_{0r}(z)$ on the Bethe lattice. The final section of the unique path between nodes $0$ and $r$ is shown horizontally. The final node, $r$, has $k$ neighbors to be integrated out that are not part of the path connecting it to node $0$, while all the intermediary nodes along the path, e.g. node $r-1$, have $k-1$ such neighbors.}
    \label{fig:offdiag_path}
\end{figure}

Since the Bethe lattice is locally a tree, there is a unique path connecting any two
nodes. Consider two nodes, denoted by $0$ and $r$, separated by graph distance
$r$. We start by integrating out the first node, $0$, together with the branches
attached to the first node along the path. Let node $1$ be the neighbor of $0$
that lies on the path to $r$, and let $j_n$, with $n=1,\dots,k-1$, denote the
remaining neighbors of node $1$ that do not belong to this path. In the numerator
of Eq.~\eqref{eq:Gint_def}, this gives
\begin{equation}
\begin{aligned}
    \mathcal{I}_1
    &=\int d\phi_0 \prod_{n=1}^{k-1}d\phi_{j_n}\;
    \phi_0
    e^{-\tfrac i2\tfrac{\phi_0^2}{G_{0}^{(1)}}}
    \exp{\left(
    -\tfrac i2\sum_n\tfrac{\phi_{j_n}^2}{G_{j_n}^{(1)}}
    -it\phi_1\left(\phi_0+\sum_n\phi_{j_n}\right)
    \right)}
    e^{-\tfrac i2\phi_1^2(z-\varepsilon_1)}
    \\
    &=
    -A_1 t G_{0}^{(1)}\phi_1
    e^{-\tfrac i2\tfrac{\phi_1^2}{G_{1}^{(2)}}}.
\end{aligned}
\end{equation}
The constant $A_1$ is exactly canceled by the corresponding contribution in the
denominator. Notice that the integration has transferred the same structure to the next node along the path, node $2$, multiplied by the
factor $-tG_{0}^{(1)}$. Since node $2$ is the only remaining neighbor of node $1$
on the path to $r$, the next integration can be performed in the same way. This
recursive structure continues along the path until the final node, $r$, where
one recovers the integral corresponding to the diagonal resolvent $G_{rr}(z)$. See Fig.~\ref{fig:offdiag_path} for an illustration of the branching structure along the unique path connecting nodes $0$ and $r$.
Therefore, the off-diagonal component of the resolvent is
\begin{equation}\label{eq:G_off-d}
    G_{0r}(z)=(-t)^rG_{rr}(z)\prod_{s=0}^{r-1}G_s^{(s+1)}(z).
\end{equation}

To summarize, the cavity recursion~\eqref{eq:cavity_bl}, derived here in two complementary ways, is exact on the Bethe lattice and asymptotically exact on RRGs. After disorder averaging it becomes a self-consistent equation for the distribution of the cavity Green's functions, Eqs.~\eqref{eq:Pprob_cavity}--\eqref{eq:Qprob_G}, while off-diagonal elements follow from products of cavity propagators along the unique path connecting two sites, Eq.~\eqref{eq:G_off-d}. In the next section, we use these equations to characterize the two phases and the transition.

%=== CRITICAL ===
\section{The localization transition}\label{sec:critical}

The self-consistent integral equation~\eqref{eq:Pprob_cavity} determines the joint probability distribution $P(\operatorname{Re}G,\operatorname{Im}G)$ of the cavity resolvent, from which the distribution $Q(\operatorname{Re}G,\operatorname{Im}G)$ of the full diagonal resolvent follows through Eq.~\eqref{eq:Qprob_G}. The localization transition is reflected most directly in the behavior of the imaginary part as the regulator $\eta\to0^+$. In the localized phase, $\operatorname{Im}G$ vanishes at a typical site while developing increasingly broad fluctuations due to rare
resonances. In the delocalized phase, instead, its distribution approaches a finite limiting form as $\eta\to0^+$, although close to the transition it retains a similar resonant structure, with the cutoff now generated by the system itself rather than imposed by $\eta$. We first examine these two behaviors before turning to the critical point itself.

It is instructive to analyze the qualitative behavior of the distribution $P(\operatorname{Re} G, \operatorname{Im} G)$ in the two phases, starting from the localized regime. The simplest starting point is the fully localized limit $t=0$, where the Hamiltonian becomes diagonal, and the Green's function is purely local. Writing explicitly the real and imaginary parts (and focusing, for simplicity, on the $E=0$ case),
\begin{align}
    \operatorname{Re} G_{ii} &= \frac{-\varepsilon_i}{\varepsilon_i^2+\eta^2},\qquad 
    \operatorname{Im} G_{ii} = \frac{\eta}{\varepsilon_i^2+\eta^2}.
\end{align}
For a typical site, $|\varepsilon_i|$ is of order $W$, so that
$\operatorname{Im} G_{ii}\sim \eta$.
However, rare resonant sites satisfying
$|\varepsilon_i|\lesssim \eta$
produce anomalously large values,
$\operatorname{Im} G_{ii}\sim \eta^{-1}$.
Thus, the distribution of $\operatorname{Im} G$ is extremely broad: most sites contribute values of order $\eta$, while rare resonances generate large fluctuations extending up to $\eta^{-1}$. This structure can be obtained explicitly. Introducing $y=\operatorname{Im} G_{ii}$ for $\eta\ll 1$
one finds
\begin{equation}\label{eq:dist_ImG_loc_inf}
    P(y)\simeq \frac{\sqrt{\eta}}{W} \frac{1}{y^{3/2}} \mathbf{1}_{\left[ \frac{4 \eta}{W^2},\eta^{-1}\right]}(y),
\end{equation}
where $\mathbf{1}_{[a,b]}(y)$ denotes the indicator function.  Thus, Eq.~\eqref{eq:dist_ImG_loc_inf} has a lower cutoff at $y\simeq 4 \eta/W^2$ and upper cutoff at $y=\eta^{-1}$. Consequently, the typical value, $y_{\text{typ}}=e^{\langle\ln y\rangle}$, is of order $\eta$ but large moments of $y$ are dominated by the upper cutoff (while $\langle y \rangle$, which is proportional to the average DOS, is of order $1$). 

\begin{figure}
    \centering
    \includegraphics[width=0.65\linewidth]{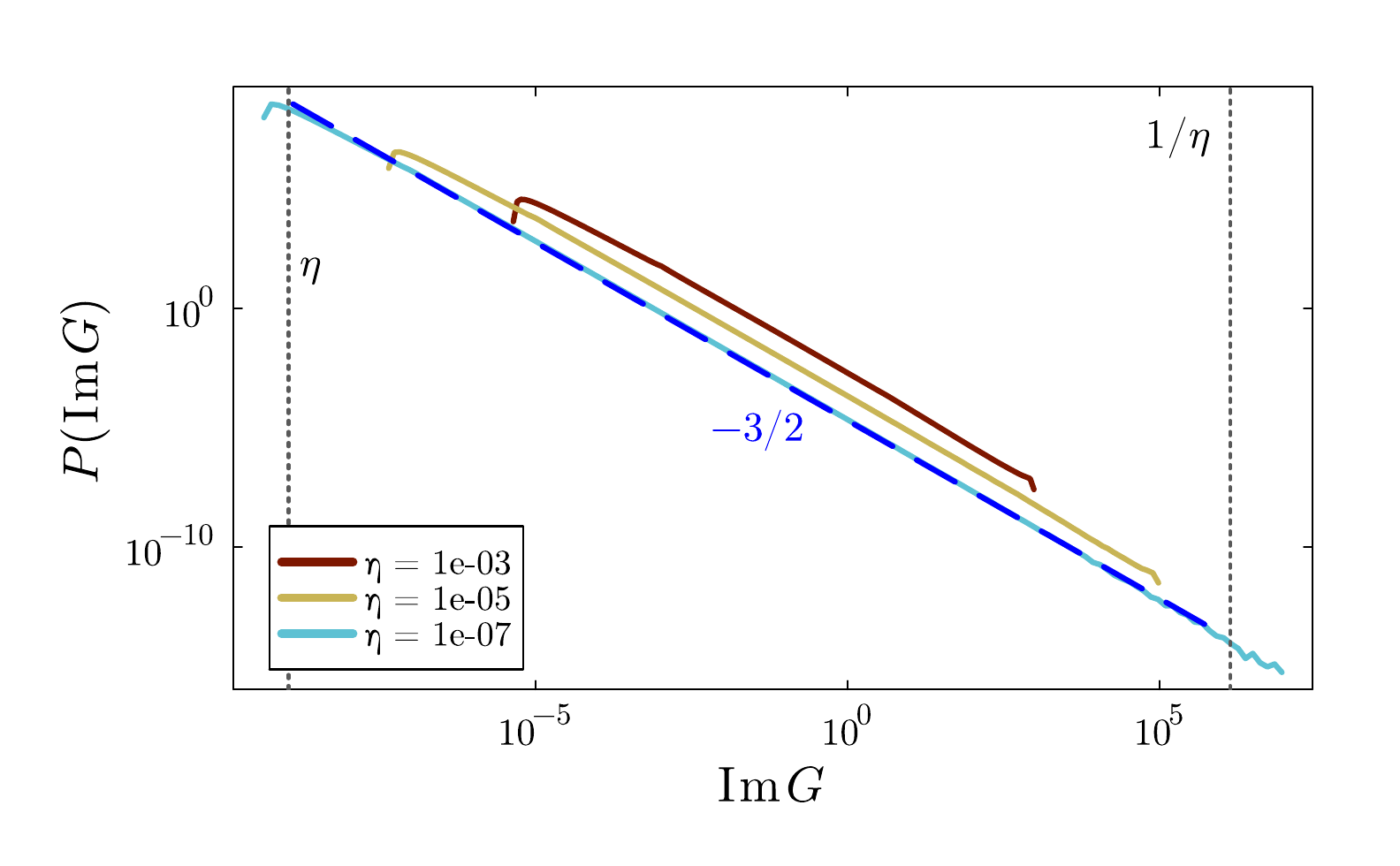}
    \caption{Probability distribution of $\operatorname{Im}G$ in the localized phase for different values of the regulator $\eta$. The distributions are obtained using population dynamics (see Sec.\ref{sec:popdyn}) for a population size $\Omega=2^{25}$, and fixed energy $E=0.0$ and disorder strength $W=30.0$, on a Bethe lattice with branching $k=2$. }
    \label{fig:dist_loc}
\end{figure}

Although finite hopping couples nearby resonances and generates spatial correlations, the qualitative picture found in the infinite-disorder limit survives throughout the localized phase. As discussed in Sec.~\ref{sec:resolvent:localized}, the local spectral measure remains singular rather than absolutely continuous as $\eta\to0^+$. In terms of the
resolvent, this singularity manifests itself in the fact that $\operatorname{Im}G_{ii}(\lambda-i\eta)$ does not approach a smooth, $\eta$-independent function. Instead, its typical value vanishes with $\eta$, while rare resonances produce anomalously large values. The main properties of the functional form of the probability distribution of $\operatorname{Im}G$ derived in the infinite-disorder limit remain valid at finite disorder throughout the localized phase:
\begin{equation}\label{eq:dist_ImG_loc}
    P(\operatorname{Im} G) \sim \frac{\sqrt{\eta}}{\left
    (\operatorname{Im} G\right)^{3/2}} , \qquad \text{for } \eta \ll \operatorname{Im} G \ll \eta^{-1}.
\end{equation}
As $\eta\to0^+$, the lower cutoff moves to zero while the upper cutoff diverges, so that the distribution remains strongly inhomogeneous and does not approach an $\eta$-independent regular form. At the same time, the typical value of $\operatorname{Im}G_{ii}$ collapses to zero proportionally to $\eta$. Note that the exponent $3/2$ in Eq.~\eqref{eq:dist_ImG_loc} refers to the marginal distribution of $\operatorname{Im}G$, integrated over $\operatorname{Re}G$. As discussed in Sec.~\ref{sec:critical_behavior} (see Eq.~\eqref{eq:tail_marginal}), the distribution of $\operatorname{Im}G$ at fixed $\operatorname{Re}G$ has instead a different, disorder-dependent tail exponent, which controls the transition. Figure~\ref{fig:dist_loc} shows the effect of reducing the regulator $\eta$ on the distribution of $\operatorname{Im}G$ on a Bethe lattice of branching $k=2$, deep in the localized phase $W=30.0>W_c$.

This behavior is directly reflected in the inverse participation ratio.
Using the relation between the generalized IPR and the resolvent, Eq.~\eqref{eq:ipr_from_ImG}, together with the distribution in Eq.~\eqref{eq:dist_ImG_loc}, one obtains $\langle \operatorname{Im} G^q \rangle \sim \eta^{1-q}$ for $q>1/2$. In particular, for $q=2$ (for which the prefactor in Eq.~\eqref{eq:ipr_from_ImG} equals $\sqrt{\pi}\,\Gamma(2)/\Gamma(3/2) = 2$),
\begin{equation}
    I_2 = \lim_{\eta\to0^+} \frac{2\,\eta \langle \operatorname{Im} G_{ii}^2 \rangle}{\langle \operatorname{Im} G_{ii} \rangle}
    = O(1).
\end{equation}

\begin{figure}
    \centering
    \includegraphics[width=0.65\linewidth]{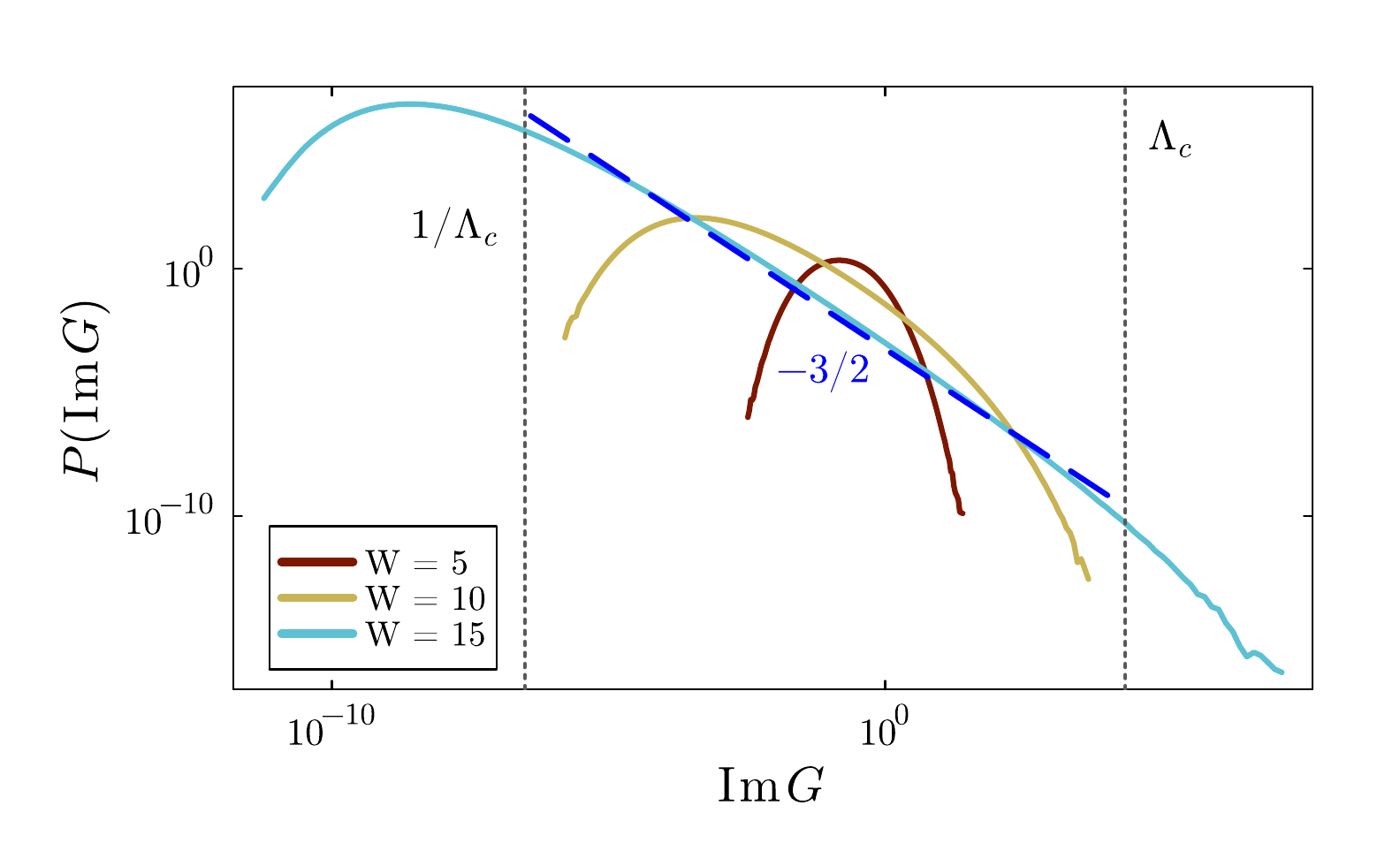}
    \caption{Probability distribution of $\operatorname{Im}G$ in the delocalized phase for different values of the disorder strength $W$. The distributions are obtained using population dynamics (see Sec.\ref{sec:popdyn}) for a population size $\Omega=2^{25}$, and fixed energy $E=0.0$ and no regulator $\eta=0.0$, on a Bethe lattice with branching $k=2$, for which the critical disorder is $W_c\approx18$. }
    \label{fig:dist_deloc}
\end{figure}

On the delocalized side, by contrast, \(P(\operatorname{Re} G, \operatorname{Im} G)\) (and therefore \(Q(\operatorname{Re} G, \operatorname{Im} G)\)) converges to a stable and well-defined probability distribution in the limit \(\eta \to 0^+\). However, as the disorder increases, the marginal probability distribution of the imaginary part becomes progressively broader and more asymmetric. Sufficiently close to the transition, the distribution of \(\operatorname{Im} G\) becomes qualitatively very similar to that found in the localized phase. As shown in Fig.~\ref{fig:dist_deloc}, it develops a broad power-law regime analogous to the one observed in the localized phase:
\begin{equation}\label{eq:dist_ImG_deloc}
    P(\operatorname{Im} G) \sim \frac{1}{\sqrt{\Lambda_c} \left(\operatorname{Im} G\right)^{3/2}} , \qquad \text{for } \Lambda_c^{-1} \ll \operatorname{Im} G \ll \Lambda_c.
\end{equation}
The crucial difference is that, in the delocalized phase, the cutoff that plays the role of the imaginary regulator is not imposed externally by the broadening \(\eta\). Instead, it is generated self-consistently by the system itself. 

As shown in Fig.~\ref{fig:dist_deloc}, this self-generated cutoff \(\Lambda_c\) strongly depends on the disorder and grows rapidly as \(W_c\) is approached. The quantity \(\Lambda_c\), which plays the role of \(\eta^{-1}\), is referred to as the correlation volume. To clarify why this identification is appropriate, it is useful to return to the comparison between \(P(\operatorname{Im} G)\) in the localized phase at finite \(\eta\) and in the delocalized phase close to \(W_c\).

In the localized phase, the eigenstates are exponentially localized around randomly distributed localization centers, and the regulator $\eta$ sets an
energy resolution: levels separated by less than $\eta$ cannot be resolved individually and contribute simultaneously to the broadened local spectrum.
In the delocalized phase, by contrast, the relevant broadening does not vanish when this external scale is removed. Concretely, the cavity recursion gives
\begin{equation}
    \operatorname{Im}G_{ii}
    =|G_{ii}|^2\left[\eta+\operatorname{Im}\Sigma_i\right],
    \qquad
    \operatorname{Im}\Sigma_i
    =t^2\sum_{j\in\partial i}\operatorname{Im}G_{j}^{(i)},
\end{equation}
so that $\eta$ and $\operatorname{Im}\Sigma_i$ play equivalent roles. There, $\operatorname{Im}\Sigma_i$ remains finite as $\eta\to0^+$, since the coupling to the rest of the lattice produces a self-consistent broadening whose typical scale is $\Lambda_c^{-1}$. This scale therefore plays, for delocalized states, the role that $\eta$ plays in the localized phase.

Close to $W_c$, typical delocalized eigenstates can then be viewed as arising from the hybridization of many localized eigenstates within an energy window of width $\Lambda_c^{-1}$. This window acts as an intrinsic energy resolution associated with the resonant structure of the eigenstates.

\begin{figure}
    \centering
    \includegraphics[width=0.65\linewidth]{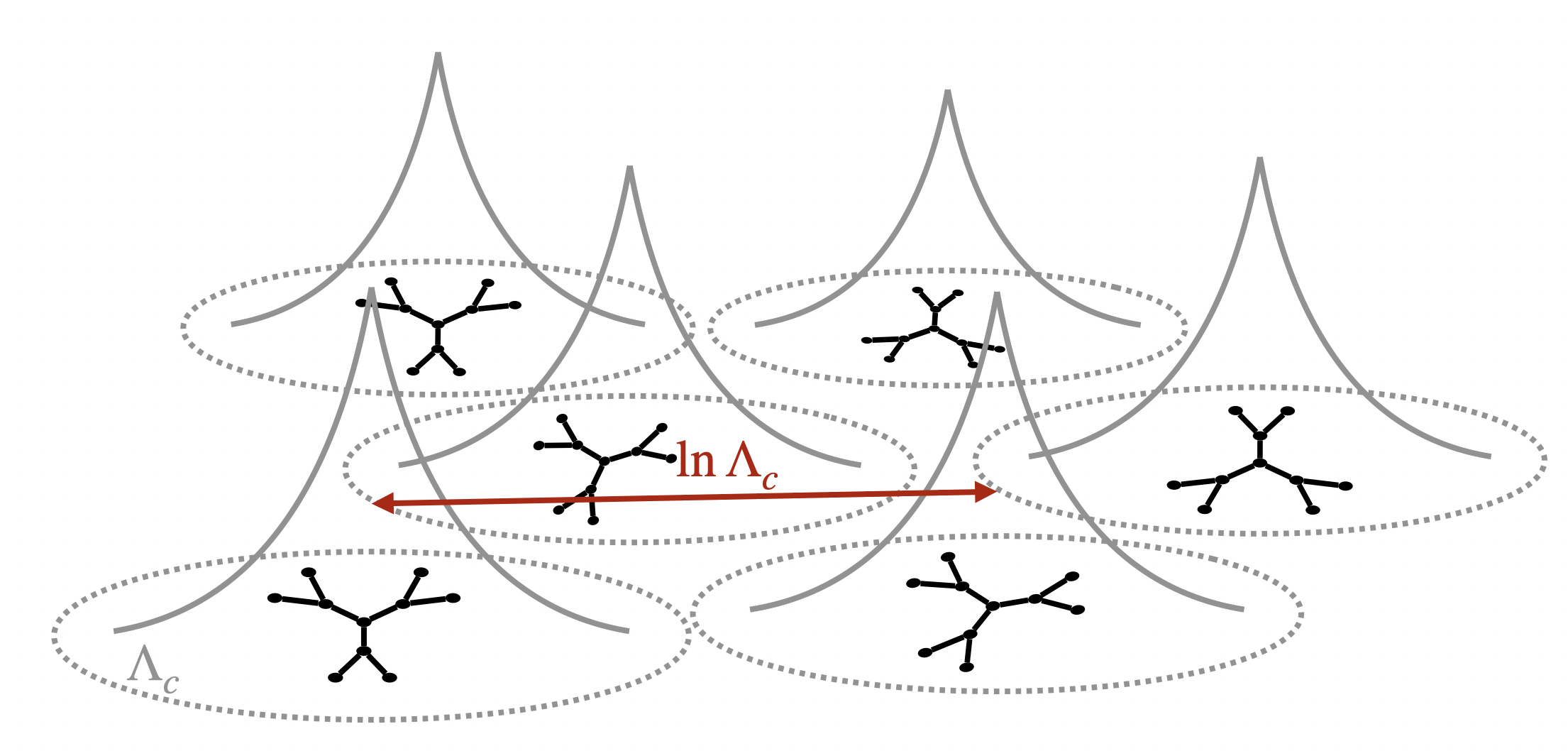}
    \caption{Schematic sketch of the spatial profile of the amplitude of typical delocalized eigenstates close to localization threshold, resulting of the hybridization of $N/\Lambda_c$ resonant peaks, each one occupying a volume $\Lambda_c$. }
    \label{fig:placeholder}
\end{figure}

For a large system of $N$ sites, the number of eigenstates within this energy window---and hence the number of hybridized localized states---is of order $N / \Lambda_c$. Consequently, the spatial profile of typical delocalized eigenstates close to the localization threshold can be schematically understood as in Fig.~\ref{fig:placeholder}. They appear as a superposition of an extensive number of resonant localized peaks within a narrow energy window.

From this perspective, the interpretation of \(\Lambda_c\) is straightforward: it corresponds to the volume associated with a single resonant peak and therefore to the spatial region over which the wavefunction amplitudes are strongly correlated. An equivalent viewpoint is the following: in exact diagonalizations at finite \(N\), one must have \(N \gg \Lambda_c\) to observe that the number of peaks in the eigenstates grows with \(N\), thus distinguishing them from localized or critical states.

For any $W<W_c$, the correlation volume is finite, although on the Bethe lattice it diverges exponentially upon approaching the transition~\cite{mirlin1991localization,mirlin1994statistical,tikhonov2019critical}, see Eq.~\eqref{eq:Lambdac} below. As a consequence, all moments of $\operatorname{Im} G$ remain finite in the delocalized phase. In particular, high-order moments ($q>1/2$) are dominated by the upper cutoff and scale as $\langle (\operatorname{Im} G)^q \rangle \sim \Lambda_c^{q-1}$.
As a result, Eq.~\eqref{eq:ipr_from_G} immediately shows that the IPR vanishes after multiplying by $\eta$ and taking the limit $\eta \to 0^+$, as expected in the delocalized phase. The typical value of $\operatorname{Im} G$ scales as $\Lambda_c^{-1}$. Since $\operatorname{Im} G_{ii}$ is directly related to the inverse escape time of a particle initially created at site $i$, it follows that $\Lambda_c^{-1}$ is proportional to the diffusion coefficient in the delocalized phase. Therefore, the diffusion coefficient vanishes exponentially fast upon approaching the critical disorder $W_c$.

\subsection{The critical behavior}\label{sec:critical_behavior}

The contrasting behaviors of $P(\operatorname{Re} G, \operatorname{Im} G)$ in the two phases suggest a practical criterion to locate the mobility edge. Instead of solving for the full distribution, one can study the linear stability of the real solution of the cavity equations against a small imaginary perturbation ~\cite{abou1973selfconsistent, mirlin1991localization, tikhonov2019critical, parisi2019anderson, bapst2014largekAL, tarquini2016anderson, Biroli_Tarzia_semerjan_2010}. The idea is to first solve the cavity equations for real $z=E$ (i.e., with $\eta=0$), obtaining the real cavity Green's functions $g_i^{(j)}$. One then adds a small imaginary part and examines whether it grows or decays under iteration of the linearized equations. In the delocalized phase, a small initial imaginary part grows, indicating that the system cannot sustain a purely real solution; in the localized phase, it decays, signaling stability. The transition point is identified when the stability changes sign.

To implement this criterion, we fix the energy $z = E - i\eta$ with a small regulator $\eta$ and parametrize the cavity resolvent as $G_i^{(j)} = g_i^{(j)} + i\eta\,\hat{g}_i^{(j)}$. Expanding the cavity recursion Eq.~\eqref{eq:cavity_bl} to first order in $\eta$ yields
\begin{align}
    g_i^{(j)} &= \frac{1}{E - \varepsilon_i - t^2 \sum_{k\in\partial i\setminus j} g_k^{(i)}}, \label{eq:Re_cavity_bl}\\
    \hat{g}_i^{(j)} &= \bigl(g_i^{(j)}\bigr)^2 \left(1 + t^2 \!\! \sum_{k\in\partial i\setminus j} \hat{g}_k^{(i)}\right). \label{eq:Im_cavity_bl}
\end{align}
The real part $g$ is independent of $\hat{g}$ and can be solved separately. The linear equation for $\hat{g}$ determines whether a small initial imaginary part grows or decays under iteration. The localization threshold $W_c$ is identified as the disorder at which the linearized equations, in the absence of the regulator, become unstable.

\begin{figure}
    \centering
    \includegraphics[width=0.6\linewidth]{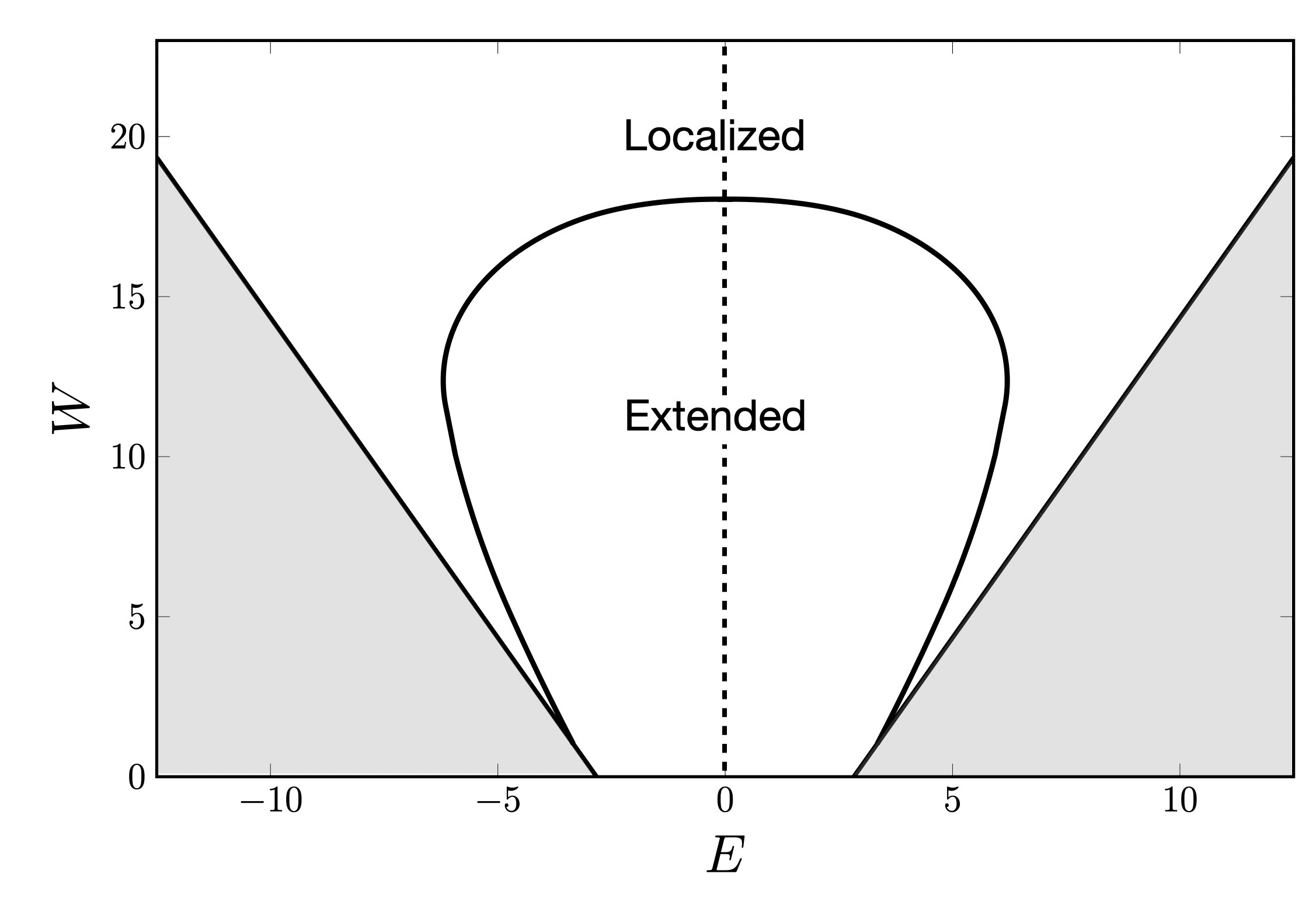}
    \caption{Schematic phase diagram of the Anderson model for the Bethe lattice for $k=2$  in the $W$-$E$ plane~\cite{Biroli_Tarzia_semerjan_2010,tonetti2025testing,tonetti2026geometrylocalizationprobinglocalization}. The inner curve marks the mobility edge separating localized and extended states. The straight lines indicate the edges of the support of the density of states, $E=\pm(2\sqrt{k}+W/2)$; the shaded outer regions therefore correspond to energies outside the spectrum of the model. The vertical dashed line at $E=0$ indicates the energy at which we study the localization transition. For a random potential with a bounded support, as the one considered in Eq.~\eqref{eq:gamma_epsilon}, a minimal amount of disorder is required to start to induce localized states at the spectral edge, as rigorously shown in Ref.~\cite{warzel2012} (see also Ref.~\cite{tonetti2026geometrylocalizationprobinglocalization}).}
    \label{fig:phase_diagram}
\end{figure}

Figure~\ref{fig:phase_diagram} sketches the phase diagram in the $W$-$E$ plane, showing the mobility edge that separates the delocalized (metallic) region from the localized (insulating) region~\cite{Biroli_Tarzia_semerjan_2010,tonetti2025testing,tonetti2026geometrylocalizationprobinglocalization}. For simplicity and concreteness, from now on we focus on the band centre $E=0$.

As before, Eqs.~\eqref{eq:Re_cavity_bl} and \eqref{eq:Im_cavity_bl} must be interpreted as self-consistent integral equations for the joint probability distribution. In the following, in order to study the linear stability of the cavity equation with respect to a small imaginary part in absence of the regulator, we omit the factor $1$ in the right hand side of Eq.~\eqref{eq:Im_cavity_bl}:
\begin{multline}
    P(g, \hat{g}) = \int d\varepsilon\,\gamma(\varepsilon) \prod_{i=1}^k \left[ dg_i\,d\hat{g}_i\,P(g_i,\hat{g}_i) \right] \\
    \times \delta\!\left( g - \frac{1}{E - \varepsilon - t^2 \sum_{i=1}^k g_i} \right)
    \delta\!\left( \hat{g} -  t^2 g^2  \! \sum_{i=1}^k \hat{g}_i \right).
\end{multline}
Writing the second Dirac delta in its integral representation,
\begin{equation}
    \delta\!\left( \hat{g} -  t^2 g^2  \!  \sum_{i=1}^k \hat{g}_i \right) = \int \frac{dq}{2\pi}\, e^{iq\left(\hat{g} -  t^2 g^2 \sum_i \hat{g}_i \right)},
\end{equation}
we recognize the partial Fourier transform of the joint probability distribution with respect to $\hat{g}$. Introducing $\hat{P}(g,q) = \int d\hat{g}\, e^{-iq\hat{g}} P(g,\hat{g})$, we obtain
\begin{align}
    \hat{P}(g,q) &= \int d\varepsilon\,\gamma(\varepsilon) \prod_{i=1}^k \left[ dg_i\, \frac{dq_i}{2\pi} \hat{P}(g_i,q_i) e^{i\hat{g}_i(q_i - q t^2 g^2)} \right] \delta\!\left( g - \frac{1}{E - \varepsilon - t^2 \sum_i g_i} \right) \nonumber \\
    &= \int d\varepsilon\,\gamma(\varepsilon) \prod_{i=1}^k \left[ dg_i\, \hat{P}\bigl(g_i, q t^2 g^2\bigr) \right] \delta\!\left( g - \frac{1}{E - \varepsilon - t^2 \sum_i g_i} \right).
\end{align}
Assuming that for large $\hat{g}$ (i.e. small $q$) the distribution of imaginary parts at fixed real part develops power‑law tails of the form $P(g,\hat g) \sim f(g)\,|\hat g|^{-(1+\beta)}$ \cite{abou1973selfconsistent,mirlin1991localization,tikhonov2019critical}, we have,
\begin{equation} \label{eq:ansatz}
    \hat{P}(g,q) \simeq P_0(g) - f(g)\,|q|^\beta,
\end{equation}
where $P_0(g)$ is the distribution of the real parts satisfying Eq.~\eqref{eq:Re_cavity_bl}:
\begin{equation}\label{eq:P0def}
    P_0(g) = \int d\varepsilon\,\gamma(\varepsilon) \prod_{i=1}^k \left[ dg_i\,P_0(g_i) \right] \delta\!\left( g - \frac{1}{E - \varepsilon - t^2 \sum_{i=1}^k g_i} \right).
\end{equation}

\emph{A subtle point: two different tails.} The tail exponent $1+\beta$ introduced in Eq.~\eqref{eq:ansatz} should not be confused with the exponent $3/2$ of Eq.~\eqref{eq:dist_ImG_loc}. If one looks at the \emph{marginal} distribution of $\operatorname{Im}G$, i.e., integrated over all values of $\operatorname{Re}G$, one always finds a tail $(\operatorname{Im}G)^{-3/2}$ throughout the localized phase, exactly as in the infinite-disorder example of Eq.~\eqref{eq:dist_ImG_loc_inf}. The exponent $1+\beta$, instead, only appears in the distribution of $\operatorname{Im}G$ at a \emph{fixed} value of $\operatorname{Re}G$, which is what the joint distribution $P(g,\hat g)$ in Eq.~\eqref{eq:ansatz} describes. The mechanism behind this difference is rather subtle, and it is worth explaining it in some detail.

Let us write the inverse of the cavity Green's function as $1/G = m - i\eta\hat m$. From Eqs.~\eqref{eq:Re_cavity_bl}--\eqref{eq:Im_cavity_bl}, $m = E - \varepsilon - t^2\sum_j g_j$ is the real renormalized energy of the site measured from $E$, while $\hat m = 1 + t^2 \sum_j \hat g_j$ collects the imaginary parts coming from the $k$ branches below it. Then
\begin{equation}\label{eq:G_m_mhat}
\operatorname{Re}G = \frac{m}{m^2+\eta^2\hat m^2}, \qquad \operatorname{Im}G = \frac{\eta\hat m}{m^2+\eta^2\hat m^2}.
\end{equation}
There are thus two distinct ways of producing a large value of $\operatorname{Im}G$:
\begin{itemize}
\item[(a)] \emph{Local resonances.} The renormalized energy of the site itself is close to $E$, $|m| \ll 1$. Then $\operatorname{Im}G \approx \eta\hat m/m^2$ is large even if $\hat m = O(1)$, and $\operatorname{Re}G \approx 1/m$ is large as well. This is the same mechanism as in the infinite-disorder limit, where $m = E-\varepsilon$ and $\hat m = 1$.
\item[(b)] \emph{Resonances inherited from far away.} The site itself is not resonant, $m = O(1)$, so that $\operatorname{Re}G \approx 1/m = O(1)$, but $\hat m$ is large because a resonance located deeper in the tree has been transmitted to the site through the branches below it, via the multiplicative factors $t^2g^2$ of Eq.~\eqref{eq:Im_cavity_bl}. Then $\operatorname{Im}G \approx \eta\,g^2\,\hat m$.
\end{itemize}
Mechanism (a) yields the universal $3/2$ tail. Let $\mathcal{Q}(m,\hat m)$ denote the joint distribution of $m$ and $\hat m$ (the analogue, for the cavity Green's functions, of the distribution $Q(m,\hat m)$ introduced in Sec.~\ref{sec:IPR}) below. For $y = \operatorname{Im}G \gg \eta$, Eq.~\eqref{eq:G_m_mhat} requires $|m|\simeq\sqrt{\eta\hat m/y} \ll 1$, and a simple change of variables gives
\begin{equation}\label{eq:tail_marginal}
P(y) \simeq \frac{\sqrt{\eta}}{y^{3/2}} \int_0^{1/(\eta y)} d\hat m\; \mathcal{Q}(0,\hat m)\,\sqrt{\hat m}, \qquad \eta \ll y \ll \eta^{-1}.
\end{equation}
The exponent $3/2$ only stems from the fact that the distribution of $m$ is smooth and finite around $m=0$, and is therefore independent of the details of the problem. For instance, for $t=0$ one has $\mathcal{Q}(m,\hat m) = \gamma(E-m)\,\delta(\hat m - 1)$, and Eq.~\eqref{eq:dist_ImG_loc_inf} is recovered. Mechanism (b), instead, transfers to $\operatorname{Im}G$ at fixed $\operatorname{Re}G$ the tail of the distribution of $\hat m$. This is the tail that propagates along the tree, and that is governed by the linear operator constructed below: $P(\hat g\,|\,g) \sim \hat g^{-(1+\beta)}$, as in Eq.~\eqref{eq:ansatz}. Consistently, $\mathcal{Q}(0,\hat m)\sim \hat m^{-(1+\beta)}$ at large $\hat m$, as confirmed numerically in Sec.~\ref{sec:IPR} (see Fig.~\ref{fig:IPR}).

Why does the $(1+\beta)$ tail not show up in the marginal distribution? Its contribution to $P(y)$ is of order $\eta^{\beta}\,y^{-(1+\beta)}$, and its ratio with the $3/2$ tail of Eq.~\eqref{eq:tail_marginal} is $(\eta/y)^{\beta - 1/2}$. This is vanishingly small for $y\gg\eta$, since $\beta>1/2$ throughout the localized phase. The heavy tail of $\hat m$ is thus completely hidden in the marginal distribution, which is dominated by local resonances. Its only trace is in the prefactor of Eq.~\eqref{eq:tail_marginal}: the integral $\int d\hat m\, \mathcal{Q}(0,\hat m)\sqrt{\hat m}$ converges at large $\hat m$ precisely because $\beta > 1/2$. At the transition, $\beta\to 1/2$, the two mechanisms become of the same order, and the integral diverges logarithmically with its upper cutoff $1/(\eta y)$. The tail exponent $\beta$, which controls the transition, can therefore only be detected by conditioning on the real part of the Green's function. This is also why in the numerical procedure of Sec.~\ref{sec:IPR} it is convenient to condition on $m=0$, i.e., to focus on the distribution $Q(0,\hat m)$.

Finally, the same mechanism (a) is responsible for the $3/2$ tail found in the delocalized phase, Eq.~\eqref{eq:dist_ImG_deloc}, where the role of $\eta\hat m$ is played by the self-generated imaginary part of the self-energy, of typical order $\Lambda_c^{-1}$.

Before deriving the explicit form of the operator, let us explain why the tail of the distribution of $\hat g$ is the natural object to look at. In the localized phase, $\operatorname{Im}G$ is typically of order $\eta$, and large values of $\hat g$ are produced by rare resonances, i.e., by sites whose renormalized energy happens to be very close to $E$. Equation~\eqref{eq:Im_cavity_bl} shows how a resonance is transmitted from a site to its parent: $\hat g$ is multiplied by the random factor $t^2 g^2$ and summed over the $k$ branches. Whether resonances proliferate or die out as one moves along the tree is therefore decided by a competition between the exponential number of branches and the typical smallness of the factors $t^2 g^2$. The exponent $\beta$ encodes how heavy the tail of resonances is, and the linear operator constructed below describes how this tail is transformed by one step of the recursion. Its largest eigenvalue $\lambda_\beta$ can thus be thought of as the growth factor of resonances per generation, weighted by the power $\beta$ of their amplitude.

To see how this tail is propagated by one cavity iteration, one inserts Eq.~\eqref{eq:ansatz} into the Fourier-transformed recursion. The leading term in the small-$q$ expansion gives back the fixed-point equation for $P_0(g)$. The next term, proportional to $|q|^\beta$, is obtained by selecting one of the $k$ neighboring branches to carry the tail amplitude $f(\tilde g)$, while the remaining $k-1$ branches are sampled from the
typical real distribution $P_0(g)$. 

A small-$q$ expansion then defines a linear integral operator acting on the tail amplitude $f(g)$. More generally, we write its eigenvalue equation as
\begin{equation}\label{eq:operator}
    \lambda_\beta f(g) =  \int d\tilde{g}\, K_\beta(g,\tilde{g})\, f(\tilde{g}) .
\end{equation}
The eigenvalue $\lambda_\beta$ measures how the amplitude of the large-$\hat g$ tail is transformed by one cavity iteration. The condition $\lambda_\beta=1$ therefore corresponds to a stable, self-consistent tail amplitude where the power-law tail is reproduced with the same amplitude after the iteration.

The kernel is explicitly given by
\begin{equation}
    \label{eq:kernel}
    K_\beta(g,\tilde{g}) = k\,|t g|^{2\beta} \int d\varepsilon\,\gamma(\varepsilon) \prod_{i=1}^{k-1} \left[ dg_i\,P_0(g_i) \right]
    \delta\!\left( g - \frac{1}{E - \varepsilon - t^2\bigl(\tilde{g} + \sum_{i=1}^{k-1} g_i\bigr)} \right).
\end{equation}
A more compact form is obtained by introducing the distribution of the sum of $k-1$ real parts,
\begin{equation} \label{eq:Rtilde}
    R(\tilde{g}) = \int \prod_{i=1}^{k-1} \left[ dg_i\,P_0(g_i) \right] \delta\!\left( \tilde{g} - \sum_{i=1}^{k-1} g_i \right).
\end{equation}
In terms of $R(\tilde{g})$, Eq.~\eqref{eq:operator} becomes
\begin{equation}\label{eq:eigenfunction_f}
    f(g) = k\,|t g|^{2(\beta-1)} \int d\varepsilon\,\gamma(\varepsilon) \, R\!\left( \frac{E-\varepsilon}{t^2} - \frac{1}{t^2 g} - \tilde{g} \right) f(\tilde{g})\, d\tilde{g}.
\end{equation}
An equivalent representation of the eigenfunction equation~\eqref{eq:eigenfunction_f} can be obtained by working with the self-energy rather than the cavity resolvent. This approach, originally used in Ref.~\cite{abou1973selfconsistent}, turns out to be more convenient for analytical calculations, especially in the limit of large connectivity (see Sec.~\ref{sec:largek} below).

The recursion for the real part of the self-energy (Eqs.~\eqref{eq:cavselfE_def} and \eqref{eq:selfE_def}) involves the quantity $\varepsilon + t^2 g$ at each branch. Its distribution is given by
\begin{equation}
    S(x) = \int d\varepsilon\,\gamma(\varepsilon) \int d\tilde{g}\, R(\tilde{g})\, \delta\!\bigl(x - (\varepsilon + t^2 \tilde{g})\bigr).
\end{equation}
In terms of $S$, and setting $E=0$ for simplicity, the eigenfunction $f(g)$ of the integral operator (Eq.~\eqref{eq:eigenfunction_f}) can be rewritten as
\begin{equation}
    f(g) = k\,|t g|^{2\beta} \int dx\, S(x) \int dg'\, f(g')\, \delta\!\left(g + \frac{1}{x + t^2 g'}\right).
\end{equation}
Integrating over $x$ and performing the change of variables $g \to 1/g$ and $g' \to 1/y$ leads to
\begin{equation}
    |g|^{2(\beta-1)} f\!\left(\frac{1}{g}\right) = k t^{2\beta} \int \frac{dy}{y^2}\; S\!\left(-g - \frac{t^2}{y}\right) f\!\left(\frac{1}{y}\right).
\end{equation}
Defining $\varphi(x) = |x|^{2(\beta-1)} f(1/x)$, we obtain a compact form of the integral operator:
\begin{equation}\label{eq:eigenfunction_varphi}
    \varphi(x) = k t^{2\beta} \int dy\; S\!\left(-x - \frac{t^2}{y}\right) \frac{\varphi(y)}{|y|^{2\beta}}.
\end{equation}
This expression is an equivalent representation of the integral operator in Eq.~\eqref{eq:eigenfunction_f}, and therefore the eigenvalues and eigenfunctions are the same. 

Notice that for $\beta=0$, Eq.~\eqref{eq:eigenfunction_f} reduces to the equation for $P_0(g)$ multiplied by $k$ (see equations \eqref{eq:P0def}-\eqref{eq:kernel}); hence $P_0(g)$ is the right eigenvector of the kernel $K_0(g,\tilde{g})$ with eigenvalue $k$. Since $P_0(g)$ is stable for any initial condition, $k$ must be the largest eigenvalue of $K_0(g,\tilde{g})$. 

\begin{figure}
    \centering
    \includegraphics[width=0.5\linewidth]{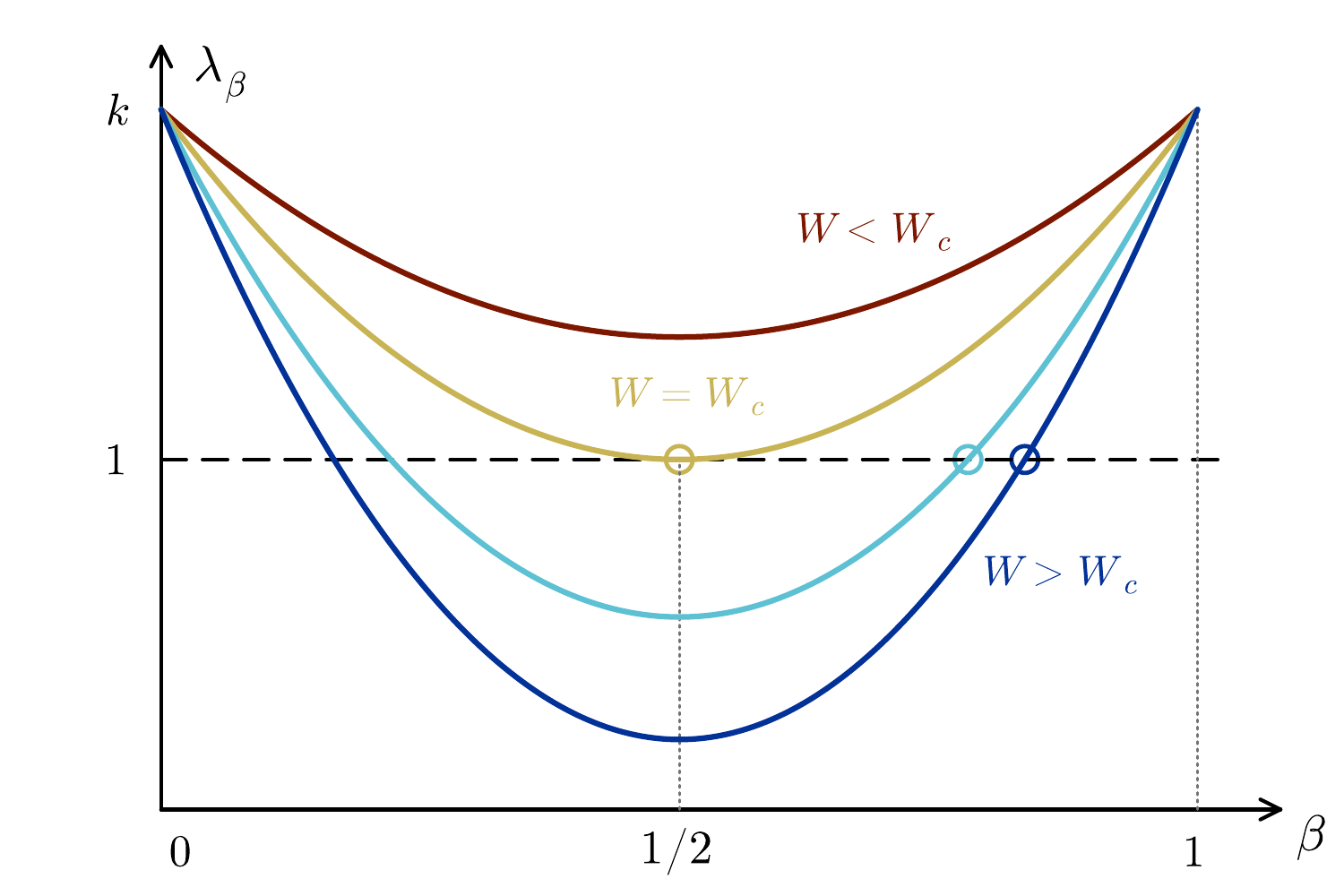}
    \caption{Illustration of the largest eigenvalue of the integral operator (Eq.~\eqref{eq:operator}) as a function of $\beta \in [0,1]$. The dark red curve corresponds to $W<W_c$ (unstable localized phase). The yellow curve represents the critical point $W=W_c$, and the other curves correspond to $W>W_c$ (stable localized phase). The intersections $\lambda_\beta=1$ determine the value of $\beta$ for each disorder strength.}
    \label{fig:lambda_beta}
\end{figure}

It can be shown \cite{abou1973selfconsistent, mirlin1991localization, tikhonov2019critical, parisi2019anderson, rizzo2024localized} that given a right eigenvector $\varphi(x)$ of the operator $K_\beta$, $\psi(x)=\frac{1}{|x|^{2(1-\beta)}}\varphi(\tfrac{t^2}{x})$ is a right eigenvector of the Kernel $K_{1-\beta}$ with the same eigenvalue. Therefore, the largest eigenvalue is symmetric around $\beta = 1/2$, $\lambda_\beta = \lambda_{1-\beta}$. Moreover, $\lambda_\beta$ is a convex function of $\beta$, with $\lambda_0 = \lambda_1 = k > 1$; together with the symmetry, this implies that its minimum is at $\beta=1/2$ (see Fig.~\ref{fig:lambda_beta}).

A stationary power-law tail requires that it be reproduced with the same amplitude after one iteration, i.e., $\lambda_\beta = 1$. In the localized phase, $\lambda_{1/2}<1$, and this condition admits two real solutions, $\beta$ and $1-\beta$, which fix the tail exponent for any given disorder $W>W_c$. Since the strong-disorder limit gives $\beta=1$, the physical solution is the one with $\beta>1/2$. Upon decreasing the disorder, the two solutions approach each other and merge at $\beta=1/2$ when $\lambda_{1/2}=1$: this condition defines the critical disorder $W_c$. For $W<W_c$, $\lambda_\beta>1$ for all real $\beta$, no stationary tail exists, and a small imaginary part grows under iteration: the localized solution is unstable.

The special role of $\beta=1/2$ has a simple interpretation, which will be made precise below (see Eq.~\eqref{eq:Mellin} and Sec.~\ref{sec:connection_DP}): $\lambda_{1/2}^r$ counts, up to subexponential factors, the number of paths of length $r$ along which the product of the factors $|t g|^2$ remains of order one, i.e., along which a resonance can propagate without being suppressed. The Anderson transition is the point at which the number of these resonant paths stops growing exponentially with their length.

The largest eigenvalue can be obtained numerically with high precision by discretizing the kernel in Eq.~\eqref{eq:kernel} \cite{tikhonov2019critical, parisi2019anderson}. The critical behavior is encoded in the dependence of $\lambda_\beta(W)$ close to the critical point \cite{zirnbauer1986anderson,zirnbauer1986localization,verbaarschot1988graded,
tikhonov2019critical,tikhonov2019statistics,mirlin1991localization,mirlin1994statistical}. Expanding $\lambda_\beta$ near $W_c$ and $\beta=1/2$ gives
\begin{equation} \label{eq:lambda_exp}
    \lambda_{\beta} (W)
    \simeq
    1 - c_1 (W - W_c)
    + c_2 \left( \beta - \frac{1}{2} \right)^2 ,
\end{equation}
with $W_c$ fixed by the condition $\lambda_{1/2}(W_c)=1$. For $k=2$, the numerical values are \cite{tikhonov2019critical}
\begin{equation}
    W_c\simeq 18.17,    \qquad    c_1\simeq 0.0308,    \qquad    c_2\simeq 3.18 .
\end{equation}

It is possible to show that the same operator~\eqref{eq:operator} also controls the spatial decay of off-diagonal Green's functions in the localized phase, and therefore of the localization length (see also Sec.~\ref{sec:DPRM} for a tightly related analysis). Consider two sites $0$ and $r$ at distance $r$. Since the Bethe lattice is a tree, there is a unique path connecting them. In the localized phase, the imaginary parts of the cavity resolvents can be neglected when extracting the leading decay, and the off-diagonal resolvent factorizes as
\begin{equation}
    \left| G_{0r} \right|^{2m} \simeq |g_{rr}|^{2m}\prod_{s=0}^{r-1}|t\,g_s^{(s+1)}|^{2m},
\end{equation}
with $s$ labeling the sites along the path connecting $0$ and $r$. The endpoint factor $|g_{rr}|^{2m}$ does not change the exponential dependence on $r$, so asymptotically the decay is controlled by the statistics of the product
\begin{equation}
    \chi_r =
    \prod_{s=0}^{r-1}|t g_s|^2,
\end{equation}
where we have relabeled $g_s^{(s+1)}\to g_s$ for simplicity. However, the variables entering this product are not independent. Consecutive real cavity fields along the path are related by the real cavity recursion, so the joint distribution of the sequence does not factorize into independent copies of $P_0(g)$. Therefore, the disorder average of $\chi_r^m$ must be computed by propagating the real cavity field along the path while
multiplying by the factor $|tg|^{2m}$ at each step. This is precisely the role of the kernel $K_\beta$ (Eq.~\eqref{eq:kernel})---with $m=\beta$---, except that here we follow a single path rather than the $k$ forward branches (see also Eq.~\eqref{eq:gproduct} below for an explicit derivation). Thus,
\begin{equation}
    \left\langle \chi_r^\beta \right\rangle = \int dg_0\cdots dg_r\, P_0(g_0)\prod_{s=0}^{r-1}
    \frac{K_\beta(g_s,g_{s+1})}{k}.
\end{equation}
Here the division by $k$ removes the branching factor already included in $K_\beta$. The final variable $g_r$ is only the output of the last real-cavity update and is integrated over.

For a fixed value of $\beta$, the right-hand side is the $r$-fold application of the single-path tilted operator $K_\beta/k$. Therefore, in the large-distance limit, its exponential dependence on $r$ is controlled by the largest eigenvalue of this operator. Denoting by $\lambda_\beta$ the largest eigenvalue of the branched operator $K_\beta$, one has
\begin{equation}
    \left\langle \chi_r^\beta \right\rangle
    \simeq
    A_\beta\left(\frac{\lambda_\beta}{k}\right)^r,
\end{equation}
where $A_\beta$ contains endpoint and other factors that are subexponential in $r$.

To analyze the behavior of this quantity, we follow the elegant approach introduced in Ref.~\cite{parisi2019anderson}. We write $\lambda_\beta = e^{\mu(\beta)}$, and introduce the rescaled variable $\theta = (\ln \chi_r)/r$, so that $\chi_r = e^{r\theta}$. Resonant paths, along which $\chi_r$ is not exponentially small in $r$, correspond to $\theta = 0$. Since there are $k^r$ paths of length $r$ emanating from a given site, $k^r \langle \chi_r^\beta \rangle \simeq \lambda_\beta^r = e^{r\mu(\beta)}$ is the generating function of the number $\mathcal{N}_r(\theta)$ of paths with a given value of $\theta$: $k^r \langle \chi_r^\beta \rangle = \int d\theta\, \mathcal{N}_r(\theta)\, e^{r\beta\theta}$. Inverting this (Mellin) transform, at large $r$ one obtains
\begin{equation} \label{eq:Mellin}
    \mathcal{N}_r(\theta)
    \sim
    \int \frac{d\beta}{2\pi i}
    \exp\left\{
    r\left[\mu(\beta)-\beta\theta\right]
    \right\},
\end{equation}
up to prefactors that do not affect the exponential dependence on $r$. This integral is dominated by a saddle point $\beta_\ast$, which satisfies $\mu'(\beta_\ast)=\theta$. For resonant paths, $\theta=0$, the saddle point is fixed by $\mu'(\beta_\ast)=0$; owing to the symmetry $\lambda_\beta=\lambda_{1-\beta}$, this gives $\beta_\ast = 1/2$, and the number of resonant paths of length $r$ behaves as $\mathcal{N}_r(0) \sim \lambda_{1/2}^r$. Resonances therefore proliferate if $\lambda_{1/2}>1$ (delocalized phase) and become exponentially rare if $\lambda_{1/2}<1$ (localized phase), which confirms that the transition is located at $\lambda_{1/2}=1$.

This observation is crucial for the correlation function: although Eq.~\eqref{eq:corr_def} involves $\langle |G_{0r}|^2\rangle$, its long-distance behavior in the localized phase is dominated by resonant paths, and is therefore governed by $\lambda_{1/2}$.
This is tightly related to the rigorous proofs of localization on tree graphs based on analysis of the fractional moments of the Green function~\cite{aizenman1993localization,aizenman1994localization,aizenman2011extended}.

Finally, the asymptotic large-distance decay of the off-diagonal elements of the Green's function in the localized phase is given by (see also Ref.~\cite{tikhonov2019statistics} and Sec.~\ref{sec:IPR} for a detailed discussion of the proper procedure to take the $\eta \to 0^+$ limit in the localized phase):
\begin{equation}
    \langle |G_{0r}|^2\rangle \approx \frac{1}{\eta} \left(\frac{\lambda_{1/2}}{k}\right)^r r^{-3/2}.
\end{equation}
The factor $r^{-3/2}$ is the subleading correction associated with the corrections to the exponential behavior due to the continuous part of the spectrum~\cite{zirnbauer1986localization,zirnbauer1986anderson,verbaarschot1988graded,mirlin1991localization,mirlin1991universality,fyodorov1991localization,fyodorov1992novel,mirlin1994statistical}. It is convenient to define the localization length $\xi$ by
\begin{equation}
    e^{-1/\xi}=\lambda_{1/2},
    \qquad
    \xi=-\frac{1}{\log\lambda_{1/2}} .
\end{equation}
Using Eq.~\eqref{eq:corr_def}, the correlation function describing the decay of the correlation of the eigenstates' amplitudes behaves as:
\begin{equation}
    C(r) \approx  k^{-r}e^{-r/\xi}r^{-3/2}.
\end{equation}
The factor $k^{-r}$ is a geometric contribution reflecting the exponential growth of the number of sites at distance $r$ on the Bethe lattice, while the factor $e^{-r/\xi}$ represents the additional exponential suppression due to localization. By setting $\beta=1/2$ and expanding to leading order close to the localization threshold, Eq.~\eqref{eq:lambda_exp} immediately yields
\begin{equation} \label{eq:xiloc}
    \xi \simeq \frac{1}{c_1(W-W_c)},
\end{equation}
corresponding to the critical exponent $\nu_{\rm loc}=1$. Similarly, one can determine the critical behavior of the tail exponent in the localized phase,
\begin{equation} \label{eq:beta}
    \beta
    \simeq
    \frac{1}{2}
    +
    \sqrt{\frac{c_1}{c_2}} \sqrt{W - W_c} .
\end{equation}
The positive branch is selected because the strong-disorder limit corresponds to $\beta\to1$. 

Finally, it has recently been shown~\cite{rizzo2024localized} that the inverse participation ratios also exhibit a singular behavior, characterized by a finite jump at the transition (as predicted by supersymmetry analyses~\cite{efetov1985anderson, efetov1987anderson, efetov1987density, zirnbauer1986anderson, zirnbauer1986localization, verbaarschot1988graded, mirlin1991localization, mirlin1991universality, fyodorov1991localization, fyodorov1992novel, mirlin1994statistical}), followed by a square-root behavior at larger disorder (see Sec.~\ref{sec:IPR} and Eq.~\eqref{eq:IPRsingularity} below).

In the delocalized phase the linearization of the self-consistent cavity equation~\eqref{eq:Re_cavity_bl} and \eqref{eq:Im_cavity_bl} ceases to be justified. Nonetheless, close to the transition ($W \lesssim W_c$), the imaginary parts of the Green’s functions remain small on most sites. In this regime, the operator~\eqref{eq:operator} captures the characteristic scale, known as the correlation volume $\Lambda_c$, beyond which the linear approximation breaks down~\cite{mirlin1994distribution,mirlin1991localization,mirlin1994statistical,biroli2022critical}. 

In the delocalized phase the localization length $\xi$ defined on the insulating side is no longer meaningful. It can be shown~\cite{efetov1985anderson,efetov1987anderson,efetov1987density,
zirnbauer1986anderson,zirnbauer1986localization,verbaarschot1988graded,
mirlin1991localization,mirlin1991universality,fyodorov1991localization,
fyodorov1992novel,mirlin1994statistical} that the correlation function is instead controlled by the correlation volume $\Lambda_c$ introduced above, and behaves as
\begin{equation}
    \langle |G_{0r}|^2\rangle \sim \Lambda_c \, k^{-r} r^{-3/2}.
\end{equation}
Hence, according to Eq.~\eqref{eq:corr_def}, after multiplying by $\eta$ and taking the limit $\eta \to 0^+$, $C(r)$ vanishes. Nevertheless, its prefactor diverges exponentially fast as the critical point is approached from the delocalized side.

The asymptotic behavior of $\Lambda_c$ is obtained from the imaginary part of the solution of $\lambda_\beta(W) = 1$ for $W \lesssim W_c$ when $\beta$ is analytically continued to the complex plane \cite{mirlin1991localization,mirlin1994statistical,tikhonov2019critical} yielding
\begin{equation} \label{eq:Lambdac}
    \Lambda_c
    \propto
    \exp\left(
        \pi \sqrt{\frac{c_2}{c_1}}
        \frac{1}{\sqrt{W_c-W}}
    \right).
\end{equation}
Equivalently, defining a correlation length by $k^{\xi_{\rm corr}}=\Lambda_c$, one obtains
\begin{equation}
    \xi_{\rm corr}
    \simeq
    \frac{\pi}{\ln k} \sqrt{\frac{c_2}{c_1}}
        \frac{1}{\sqrt{W_c-W}} ,
\end{equation}
corresponding to the critical exponent $\nu_{\rm del}=1/2$. The same scale controls transport on the delocalized side. In particular, the diffusion coefficient, $D$, vanishes with the inverse critical singularity of the correlation volume, $D\propto  \Lambda_c^{-1}$. This reflects the fact that $\Lambda_c$ is the volume scale above which
extended, diffusive behavior is established. As this volume diverges at the transition, transport becomes critically suppressed and $D$ goes to zero.

\begin{figure}
    \centering
    \includegraphics[width=0.65\linewidth]{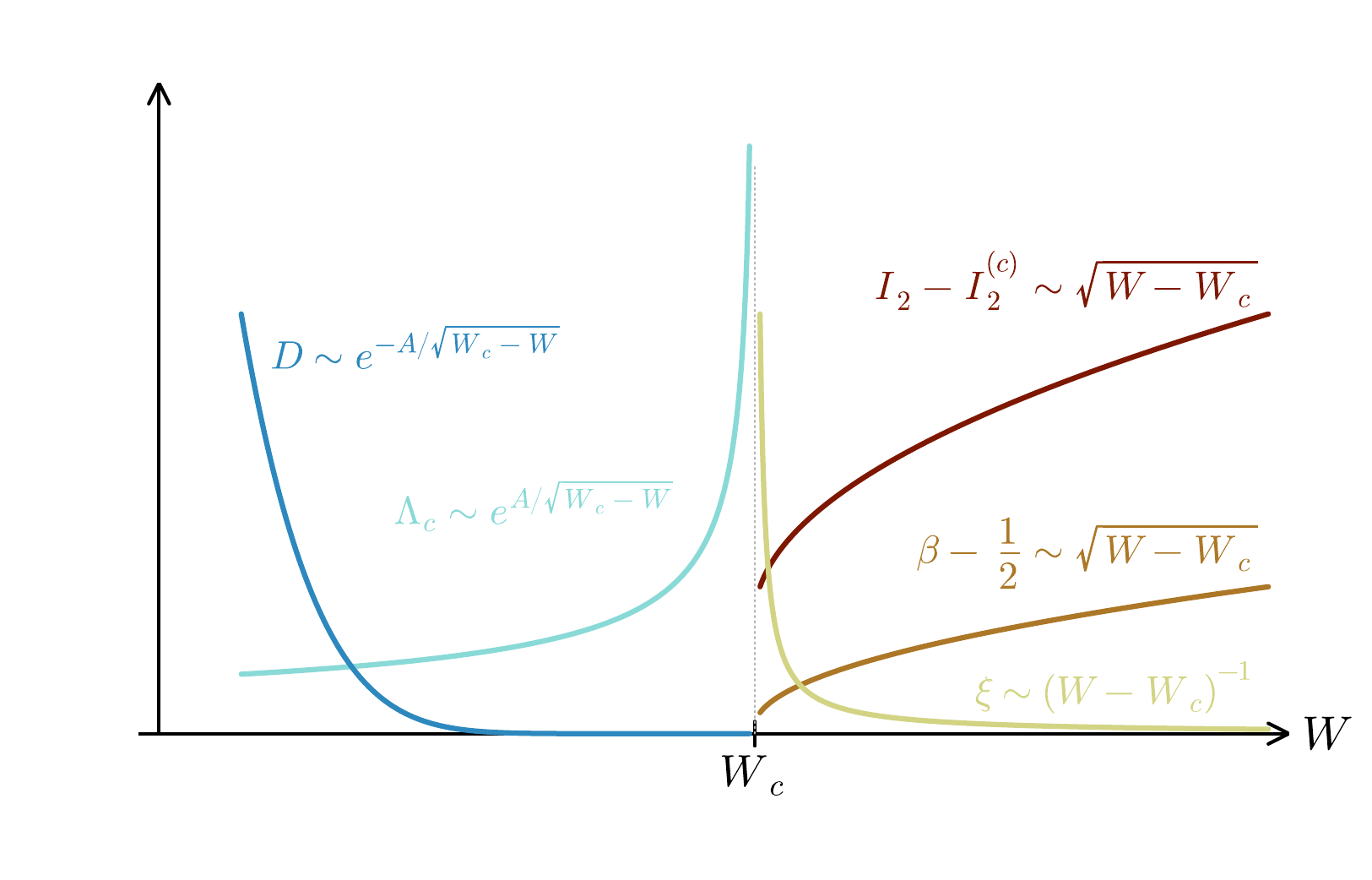}
    \caption{Sketch of the critical behavior near the Anderson transition on the Bethe
    lattice~\cite{abou1973selfconsistent, evers2008anderson, rizzo2024localized, tikhonov2019statistics, tikhonov2019critical, efetov1985anderson,efetov1987density,efetov1987anderson,zirnbauer1986localization,zirnbauer1986anderson,verbaarschot1988graded,mirlin1991localization,mirlin1991universality,fyodorov1991localization,fyodorov1992novel,mirlin1994statistical,mirlin1994distribution}. On the localized side, $W\gtrsim W_c$, the length $\xi$ diverges as $\xi \sim (W-W_c)^{-1}$ (Eq.~\eqref{eq:xiloc}), the tail exponent approaches its critical value as $\beta-1/2\sim\sqrt{W-W_c}$, Eq.~\eqref{eq:beta}, and the IPR approaches its finite critical value as $I_2(W)-I_2^{(c)}\sim\sqrt{W-W_c}$ (see Sec.~\ref{sec:IPR} and Eq.~\eqref{eq:IPRsingularity}). On the delocalized side, $W\lesssim W_c$, the correlation volume diverges as $\Lambda_c\propto \exp[A/\sqrt{(W_c-W)}]$ (Eq.~\eqref{eq:Lambdac}), while the diffusion coefficient vanishes as $D \propto \Lambda_c^{-1}$.
    }
    \label{fig:critical_behav}
\end{figure}

Figure~\ref{fig:critical_behav} summarizes these critical behaviors on both sides of the transition, showing that they are markedly different from those found on Euclidean lattices in any finite dimension and predicted by scaling arguments~\cite{abrahams1979scaling}. This difference stems from the fact that the upper critical dimension of Anderson localization is infinite~\cite{tarquini2016anderson, mirlin1994distribution, Castellani1986, baroni2024corrections}, so that the Bethe lattice limit represents a singular limit of the problem.

The main results of this section are collected in Table~\ref{tab:critical}. They should be contrasted with the conventional critical behavior found in finite dimensions, where the correlation length diverges as a power law on both sides of the transition with the same exponent $\nu$, the conductivity vanishes as $(W_c-W)^{\nu(d-2)}$, and the IPR vanishes continuously at $W_c$.

\begin{table}[htbp]
\centering
\small
\begin{tabular}{|p{4.6cm}|p{6cm}|p{1.3cm}|}
\hline
\textbf{Quantity} & \textbf{Critical behavior} & \textbf{Eq.} \\
\hline
Critical point & $\lambda_{1/2}(W_c)=1$ & \eqref{eq:lambda_exp} \\
\hline
Localization length ($W>W_c$) & $\xi \simeq [c_1(W-W_c)]^{-1}$, $\nu_{\rm loc}=1$ & \eqref{eq:xiloc} \\
\hline
Tail exponent of $P(\operatorname{Im}G)$ at fixed $\operatorname{Re}G$ ($W>W_c$) & $\beta - \frac12 \simeq \sqrt{c_1/c_2}\,\sqrt{W-W_c}$ & \eqref{eq:beta} \\
\hline
IPR ($W>W_c$) & finite jump at $W_c$, $I_2 - I_2^{(c)} \propto \sqrt{W-W_c}$ & \eqref{eq:IPRsingularity} \\
\hline
Correlation volume ($W<W_c$) & $\ln \Lambda_c \simeq \pi\sqrt{c_2/c_1}\,(W_c-W)^{-1/2}$, $\nu_{\rm del}=1/2$ & \eqref{eq:Lambdac} \\
\hline
Diffusion coefficient ($W<W_c$) & $D \propto \Lambda_c^{-1}$, exponentially small & \\
\hline
\end{tabular}
\caption{Summary of the critical behavior of the Anderson transition on the Bethe lattice. The coefficients $c_1$ and $c_2$ are defined in Eq.~\eqref{eq:lambda_exp}; for $k=2$ and $E=0$, $W_c \simeq 18.17$, $c_1 \simeq 0.0308$, and $c_2 \simeq 3.18$~\cite{tikhonov2019critical}.}
\label{tab:critical}
\end{table}

\subsection{Analytic results in the large connectivity limit}\label{sec:largek}

In the limit of large connectivity, $k \to \infty$, the largest eigenvalue of the integral operator with kernel $K_\beta$ (Eq.~\eqref{eq:kernel}) can be computed analytically~\cite{abou1973selfconsistent,bapst2014largekAL,herre2023ergodicity,prado2026anderson,tarquini2016level}. This simplification arises because, at large $k$, the localization transition occurs at very strong disorder.

Recall that the critical disorder $W_c$ is determined by the condition that this largest eigenvalue equals unity for the relevant value of the tail exponent, namely $\lambda_{\beta=1/2}=1$, as discussed in the previous section. We start from the eigenvalue equation in the form of Eq.~\eqref{eq:eigenfunction_varphi}. A useful approximation is that $S$ and $\varphi$ vary slowly on scales of order $t$:
\begin{itemize}
    \item $\varphi(x)\approx\varphi(0)$ for $|x| < t$,
    \item $S(y - t^2/x)\approx S(y)$ for $|x| > t$.
\end{itemize}

Using these approximations, we split the integral in Eq.~\eqref{eq:eigenfunction_varphi} into two contributions:
\begin{multline}
    \varphi(x) = k t^{2\beta} \varphi(0) \int_{-t}^{t} S\!\left(-x - \frac{t^2}{y}\right) |y|^{-2\beta} dy \\
    + k t^{2\beta} S(-x) \left[ \int_{-\infty}^{-t} \frac{\varphi(y)}{|y|^{2\beta}} dy + \int_{t}^{\infty} \frac{\varphi(y)}{|y|^{2\beta}} dy \right].
\end{multline}
Changing variables $y' = t^2/y$ in the first integral gives
\begin{multline}
    t^{2\beta}\int_{-t}^{t} S\!\left(-x - \frac{t^2}{y}\right) |y|^{-2\beta} dy \\
    = t^{2(1-\beta)} \left[ \int_{-\infty}^{-t} S(-x - y) |y|^{{2(\beta-1)}} dy + \int_{t}^{\infty} S(-x - y) |y|^{2(\beta-1)} dy \right].
\end{multline}
This suggests the ansatz
\begin{equation}
    \varphi(x) \sim 
    \begin{cases}
        \varphi(0), & x < t,\\
        S(-x), & x > t,
    \end{cases}
\end{equation}
which can be written as
\begin{equation}
    \varphi(x) = \frac{\varphi(0)}{S(0)}\bigl(S(-x) + h(x)\bigr),
\end{equation}
with $h(x)$ chosen such that $h(0)=0$.

Evaluating the equation at $x=0$ yields
\begin{multline}
    \varphi(0) = k t^{2(1-\beta)}\varphi(0)\left[\int_{-\infty}^{-t} S(-y) |y|^{2(\beta-1)} dy + \int_{t}^{\infty} S(-y) |y|^{2(\beta-1)} dy \right] \\
    + k t^{2\beta} S(0) \left[ \int_{-\infty}^{-t} \frac{\varphi(y)}{|y|^{2\beta}} dy + \int_{t}^{\infty} \frac{\varphi(y)}{|y|^{2\beta}} dy \right].
\end{multline}
Inserting the ansatz for $\varphi(y)$ and simplifying, we obtain
\begin{multline}\label{eq:der_int_x0}
    1 = k \int_t^{\infty} dy \bigl[S(y)+S(-y)\bigr] \left( \frac{t^{2(1-\beta)}}{|y|^{2(1-\beta)}} + \frac{t^{2\beta}}{|y|^{2\beta}} \right) 
    + k t^{2\beta} \int_t^{\infty} dy \frac{h(y)+h(-y)}{|y|^{2\beta}}.
\end{multline}
For $x\neq 0$, a similar manipulation leads to
\begin{multline}
    \bigl(S(-x)+h(x)\bigr)\frac{\varphi(0)}{S(0)}
    = k t^{2(1-\beta)}\varphi(0) \int_t^{\infty} dy \frac{S(-x+y)+S(-x-y)}{|y|^{2(1-\beta)}} \\
    + k t^{2\beta} S(-x) \left[ \int_t^{\infty} dy \frac{S(y)+S(-y)}{|y|^{2\beta}} + \int_t^{\infty} dy \frac{h(y)+h(-y)}{|y|^{2\beta}} \right] \frac{\varphi(0)}{S(0)}.
\end{multline}
Using Eq.~\eqref{eq:der_int_x0} to eliminate the last integral, we solve for $h(x)$:
\begin{multline}\label{eq:der_int_hx}
    h(x) = k t^{2(1-\beta)} S(0) \int_t^{\infty} dy \frac{S(-x+y)+S(-x-y)}{|y|^{2(1-\beta)}}\\
    - k t^{2(1-\beta)} S(-x) \int_t^{\infty} dy \frac{S(y)+S(-y)}{|y|^{2(1-\beta)}}.
\end{multline}
Inserting this expression back into Eq.~\eqref{eq:der_int_x0} gives a closed equation:
\begin{multline}
    1 = k \int_t^{\infty} \bigl[S(y)+S(-y)\bigr] \left( \frac{t^{2(1-\beta)}}{|y|^{2(1-\beta)}} + \frac{t^{2\beta}}{|y|^{2\beta}} \right) dy \\
    + k^2 t^{2} \int_t^{\infty} dy \int_t^{\infty} dx \Bigg[ S(0) \frac{S(-y+x)+S(-y-x)+S(y+x)+S(y-x)}{|x|^{2(1-\beta)}|y|^{2\beta}} \\
    - \frac{(S(y)+S(-y))(S(x)+S(-x))}{|x|^{2(1-\beta)}|y|^{2\beta}} \Bigg].
\end{multline}

We now argue that the transition occurs at very strong disorder. In that regime, $S(x)$ is well approximated by the uniform distribution of the on‑site energies, because the correction due to the contributions of the real part of the cavity green's function become negligible compared to the bare disorder. Thus
\begin{equation}
    S(x) \approx \gamma(x) = \frac{1}{W} \Theta\!\left(\frac{W}{2} - |x|\right).
\end{equation}
Consequently, the combination $S(0)kt \sim kt/W$ is small, and we may neglect terms of order $(kt/W)^2$ and higher. Under this approximation, Eq.~\eqref{eq:der_int_x0} reduces to
\begin{equation}\label{eq:der_int_self1}
    1 \simeq k \int_t^{\infty} dy \bigl[S(y)+S(-y)\bigr] \left( \frac{t^{2(1-\beta)}}{|y|^{2(1-\beta)}} + \frac{t^{2\beta}}{|y|^{2\beta}} \right).
\end{equation}
The right‑hand side of Eq.~\eqref{eq:der_int_self1} is precisely the largest eigenvalue $\lambda_\beta(W)$ of the integral operator. Integrating by parts (or evaluating the integral explicitly with the uniform distribution) gives, for $\beta \neq 1/2$,
\begin{equation}\label{eq:der_lambdabeta}
    \lambda_\beta(W) = \frac{k}{W(\beta - \tfrac12)} \left[ t^{2(1-\beta)}\left(\frac{W}{2}\right)^{2\beta-1} - t^{2\beta}\left(\frac{W}{2}\right)^{1-2\beta} \right].
\end{equation}
For the marginal case $\beta = 1/2$, we have
\begin{equation}
\begin{aligned}
    \lambda_\beta(W) &\simeq k \int_t^{\infty} dy \bigl[S(y)+S(-y)\bigr] \frac{2t}{|y|} \\
    &= k \left[ \bigl[S(y)+S(-y)\bigr] 2t \ln\frac{|y|}{t} \right]_{t}^{\infty}
       - k \int_t^{\infty} dy \bigl[S'(y)-S'(-y)\bigr] 2t \ln\frac{|y|}{t} \\
    &= \frac{4kt}{W} \ln\left(\frac{W}{2t}\right),
\end{aligned}
\end{equation}
where we used $S(\infty)=0$ and $S'(x) = \frac{1}{W}\delta(x+W/2) - \frac{1}{W}\delta(x-W/2)$.

The condition $\beta = 1/2$ is precisely the critical condition $\lambda_{1/2}(W_c)=1$. Hence, from the expression above,
\begin{equation}
    1 = \frac{4kt}{W_c}\ln\left(\frac{W_c}{2t}\right),
\end{equation}
or equivalently
\begin{equation}
    W_c = 4kt \ln\left(\frac{W_c}{2t}\right).
\end{equation}

This self‑consistent equation provides a very good estimate even for moderate $k$; for example, $k=2$ gives $W_c \approx 17.23$, while the exact numerical value is $18.17$ \cite{tikhonov2019critical}. The approximation improves as $k$ increases. 

Finally, Eq.~\eqref{eq:der_lambdabeta} can be expanded around $W\approx W_c$ and $\beta\approx 1/2$. Consider $W=W_c(1+\delta)$ and $\beta=1/2+\epsilon$,
\begin{equation}
\begin{aligned}
    \lambda_\beta(W)&=\frac{tk}{W\epsilon}\Big[\left(\frac{W}{2t}\right)^{2\epsilon}-\left(\frac{W}{2t}\right)^{-2\epsilon}\Big]\\
    &\approx 1 - \frac{W-W_c}{W_c}\left(1-\frac{4kt}{W_c}\right)+\frac{W_c^2}{24k^2t^2}\left(\beta-\frac12\right)^2.
\end{aligned}
\end{equation}
Comparing the equation above with Eq.~\eqref{eq:lambda_exp} immediately yields the coefficients $c_1$ and $c_2$ in the large-$k$ expansion of the operator, and thus the prefactors governing the asymptotic critical behavior of the localization length~\eqref{eq:xiloc}, the tail exponent~\eqref{eq:beta}, and the correlation volume~\eqref{eq:Lambdac}.

%=== POPULATION DYN ===
\section{Population Dynamics algorithm(s)}\label{sec:popdyn}
As discussed above, the cavity equations derived in Sec.~\ref{sec:cavity} for the Anderson model on the Bethe lattice must be interpreted as  self-consistent integral equations for the probability distribution of the cavity Green's functions. These equations are given by Eqs.~\eqref{eq:Pprob_cavity} and \eqref{eq:Qprob_G}. The local density of states, the inverse participation ratio, as well as all other relevant observables, are obtained from the moments of these distributions.

Eqs.~\eqref{eq:Pprob_cavity} and \eqref{eq:Qprob_G} can be efficiently solved numerically, with arbitrary precision, using the population dynamics (PD) algorithm, also known as sampled density evolution.
This algorithm was first introduced in the context of spin glasses and glassy systems on the Bethe lattice~\cite{mezard2001bethe,mezard2009information}. The idea consists in approximating $P(G)$ by the empirical distribution of a large pool of $\Omega$ complex numbers $\{G_\alpha\}_{\alpha=1}^\Omega$:
\begin{equation}
    P(G) \approx \frac{1}{\Omega}\sum_{\alpha=1}^\Omega \delta(G-G_\alpha).
\end{equation}
The algorithm proceeds as follows:
\begin{enumerate}
    \item Initialize the pool with $\Omega$ independent random complex numbers (e.g., drawn from a simple distribution, avoiding zero values).
    \item For a large number of iterations $N_{\text{eq}}$ (or until convergence), repeatedly:
        \begin{itemize}
            \item Draw a random on-site energy $\varepsilon$ from the uniform distribution $[-W/2, W/2]$;
            \item Select $k$ distinct elements $G_1,\dots,G_k$ uniformly at random from the pool;
            \item Compute a new value $G_{\text{new}}$ using the recursion
            \begin{equation} \label{eq:cavity_popdyn}
            G_{\text{new}} = \frac{1}{z - \varepsilon - t^2\sum_{j=1}^{k} G_j}.
            \end{equation}
            \item Replace a randomly chosen element of the pool by $G_{\text{new}}$.
        \end{itemize}
    \item Convergence is monitored by tracking the convergence of a few moments (e.g., $\langle \operatorname{Im} G\rangle$, $\langle \ln \operatorname{Im} G\rangle$, $\langle |G|^2\rangle$) over the pool. When these averages stabilize within statistical fluctuations (of order $1/\sqrt{\Omega}$), the empirical distribution is considered stationary.
\end{enumerate}

Once the stationary distribution $P(G)$ is obtained, the distribution $Q(G)$ of the full diagonal resolvent is estimated by an analogous procedure: draw $k+1$ independent values from the pool, draw an on-site energy, and compute
\begin{equation}
    G_{\text{full}} = \frac{1}{z - \varepsilon - t^2\sum_{j=1}^{k+1} G_j}.
\end{equation}
The empirical histogram of these $G_{\text{full}}$ approximates $Q(G)$. Observables such as the typical local density of states,
\begin{equation}
    \rho_{\text{typ}} = \exp\bigl\langle \ln (\tfrac{1}{\pi}\operatorname{Im} G_{ii}) \bigr\rangle,
\end{equation}
or the inverse participation ratio,
\begin{equation}\label{eq:I2}
    I_2 = \lim_{\eta \to 0^+} \frac{\eta\,\langle |G_{ii}|^2\rangle}{\langle \operatorname{Im} G_{ii}\rangle},
\end{equation}
are then computed as empirical averages over the stationary pool.

Although this algorithm is conceptually simple, obtaining precise and reliable results close to the transition point is extremely challenging. As discussed above, in the delocalized phase the probability distribution becomes very broad and highly asymmetric when approaching the critical point, developing a power-law regime that extends to very large values of $\operatorname{Im} G$. In the localized phase, instead, the distribution acquires power-law tails with a cutoff at $1/\eta$, and several key observables, such as the IPR, are dominated by this cutoff. Accurately resolving such broad distributions is therefore difficult and leads to strong finite-pool-size effects.

To illustrate this, consider running the algorithm with a large pool size $\Omega$ and accumulating statistics (after reaching stationarity) over $N_{\rm iter}$ iterations. The statistical accuracy of the resulting histograms scales as $O((\Omega N_{\rm iter})^{-1})$. With current computational resources, typical values are $\Omega \simeq 5 \cdot 10^8$ and $N_{\rm iter} \simeq 10^3$, implying that events with probabilities smaller than $\sim 2 \cdot 10^{-12}$ are essentially never sampled. As a consequence, when the distributions become sufficiently broad, their tails—crucial for determining several observables—are not properly captured. The impact of these finite-$\Omega$ effects has been analyzed in detail in~\cite{tikhonov2019critical}.

A first manifestation of the strong finite-$\Omega$ effects is the very slow and systematic drift of the estimated localization threshold as the population size is increased. As discussed in Sec.~\ref{sec:critical}, in the localized phase the imaginary part of the Green’s function vanishes with the regulator~$\eta$. Anderson localization can therefore be analyzed via the linear stability of Eqs.~\eqref{eq:cavity_bl} with respect to a small imaginary component, leading to the simplified recursion relations~\eqref{eq:Re_cavity_bl}–\eqref{eq:Im_cavity_bl}. Equation~\eqref{eq:Re_cavity_bl} determines the distribution of the real parts independently of the (small) imaginary components, while Eq.~\eqref{eq:Im_cavity_bl} is linear in the imaginary parts in the absence of the regulator. As a result, when $\eta=0$, the typical value of the imaginary part either grows exponentially (in the delocalized phase) or decays exponentially (in the localized phase) under iteration. The corresponding rate is controlled by the largest eigenvalue~$\lambda_{1/2}$ of the linear integral operator~\eqref{eq:operator},
\begin{equation}
\operatorname{Im} G_{\rm typ} \propto \lambda_{1/2}^{N_{\rm iter}} \, .
\end{equation}
This quantity can be computed numerically using the population dynamics algorithm. The procedure is as follows. One initializes a pool of $\Omega$ pairs $\{ \operatorname{Re} G, \operatorname{Im} G \}$. First, Eq.~\eqref{eq:Re_cavity_bl} is iterated for the real parts alone until their distribution reaches stationarity. Then, both real and imaginary parts are updated according to Eqs.~\eqref{eq:Re_cavity_bl}–\eqref{eq:Im_cavity_bl}. 

The only modification with respect to the standard algorithm is that, in order to monitor the exponential growth or decay of the imaginary part, at each iteration step $\tau+1$ a completely new pool of $\Omega$ elements is generated from the pool at iteration $\tau$, replacing it entirely. After a transient regime, the typical value of $\operatorname{Im} G$ exhibits a clear exponential growth or decay, depending on whether the system is in the delocalized or localized phase. The growth exponent is extracted by fitting this behavior, and the procedure is repeated several times to improve statistical accuracy.

The main limitation of this approach arises from the fact that the joint distribution of $(\operatorname{Re} G, \operatorname{Im} G)$ develops broad power-law tails at large arguments, as discussed in Sec.~\ref{sec:critical}. As a consequence, finite-size effects due to the finite pool size $\Omega$ are particularly strong. These effects induce a systematic shift of the apparent localization threshold towards larger values of $W$ as $\Omega$ increases~\cite{tonetti2025testing}, following
\begin{equation} \label{eq:fit_Ec}
W_c(\Omega) = W_c - \frac{A}{(\ln \Omega)^B} \, .
\end{equation}
The extremely slow convergence with $\Omega$ makes an accurate determination of $W_c$ using population dynamics particularly challenging~\cite{tikhonov2019critical}.

These strong finite-$\Omega$ effects have led, over the past decade, to a number of numerical studies~\cite{biroli2012difference,de2013ergodicity,Kravtsov2018nonergodic,pino2020scaling,bera2018return,de2020subdiffusion} suggesting the existence of an intermediate delocalized but non-ergodic phase, characterized by multifractal eigenfunctions over a broad range of disorder preceding the localization transition. While such a phase would be highly intriguing—also in view of its connection to many-body localization~\cite{altshuler1997quasiparticle}—it appears to be in tension with analytical predictions based on the supersymmetric approach for the Anderson model on the Bethe lattice~\cite{tikhonov2019statistics, efetov1985anderson,efetov1987density,efetov1987anderson,zirnbauer1986localization,zirnbauer1986anderson,verbaarschot1988graded,mirlin1991localization,mirlin1991universality,fyodorov1991localization,fyodorov1992novel,mirlin1994statistical,mirlin1994distribution}. Moreover, its existence in the thermodynamic limit has been ruled out by subsequent studies~\cite{tikhonov2016anderson,biroli2018delocalization,garcia2017scaling,tikhonov2019critical,biroli2022critical}.

In the following, we discuss two numerical improvements of the population dynamics algorithm that are specifically designed to handle broad probability distributions and accurately sample their tails, thereby significantly enhancing numerical precision.

\subsection{Large deviation approach in the delocalized phase}
Here we describe an advanced large-deviation algorithm that enables highly accurate sampling of the distribution $P(G)$ of cavity Green's functions in the delocalized phase ($W<W_c$) of the Anderson model on the Bethe lattice~\cite{biroli2022critical}. Since we restrict ourselves to the extended phase---where the probability distribution converges to a stable and well-defined limit as $\eta \to 0$---we can set $\eta=0$ from the outset, provided that the population is initialized with a finite imaginary part.

Because we will explicitly track the imaginary part of $G$, we rewrite Eq.~\eqref{eq:cavity_popdyn} in the following form (setting $t=1$ and $E=0$):
\begin{equation}
G_{\rm new} \equiv g + {\rm i}\hat{g}
= \frac{\left(-\epsilon - \sum_{j=1}^k g_j\right) + {\rm i}\sum_{j=1}^k \hat{g}_j}
{\left(-\epsilon - \sum_{j=1}^k g_j\right)^2 + \left(\sum_{j=1}^k \hat{g}_j\right)^2}
\equiv f_{A,B}(\epsilon),
\label{eq:anderson:local:field}
\end{equation}
where we have defined $A = \sum_{j=1}^k g_j$ and $B = \sum_{j=1}^k \hat{g}_j$, and introduced the shorthand
\begin{equation}
f_{A,B}(\epsilon) = \frac{(-\epsilon - A) + {\rm i}B}{(-\epsilon - A)^2 + B^2}.
\label{eq:transfer}
\end{equation}

The key idea of the method is the following. For a given set of randomly selected cavity Green’s functions $\{G_j\}$, the only remaining randomness comes from the on-site energy $\epsilon$, which is uniformly distributed in $[-W/2, W/2]$. One can therefore compute the conditional probability density of the imaginary part $\hat{g}$ given $A$ and $B$. Using standard properties of the Dirac delta function, one obtains
\begin{equation}
P_{A,B}(\hat{g}) = \int_{-W/2}^{W/2}
\delta\!\left(\hat{g} - \hat{f}_{A,B}(\tilde{\epsilon})\right)\frac{d\tilde{\epsilon}}{W}
= \frac{1}{W} \sum_{l:\,\hat{f}_{A,B}(\tilde{\epsilon}_l)=\hat{g}}
\frac{1}{\left|\hat{f}_{A,B}'(\tilde{\epsilon}_l)\right|},
\label{eq:PAB}
\end{equation}
where $\hat{f}_{A,B} = \Im f_{A,B}$, and the $\tilde{\epsilon}_l$ are the real solutions of $\hat{g} = \hat{f}_{A,B}(\tilde{\epsilon})$ within the interval $[-W/2, W/2]$. These solutions are easily obtained by solving a quadratic equation, yielding $\tilde{\epsilon}_l = -A \pm \sqrt{B/\hat{g} - B^2}$.

Suppose now that we wish to evaluate $P(\hat{g})$ at a fixed value of $\hat{g}$. The requirement that the roots $\tilde{\epsilon}_l$ be real immediately implies $\hat{g} \leq 1/B$, so that $P_{A,B}(\hat{g})=0$ whenever $\hat{g} > 1/B$. This observation suggests a strategy for importance sampling: when estimating $P(\hat{g})$, one should preferentially sample configurations such that $B = \sum_{j=1}^k \hat{g}_j \leq 1/\hat{g}$. A simple implementation consists in restricting the sampling to values satisfying $\hat{g}_j \leq 1/\hat{g}$, since larger values would automatically lead to $P_{A,B}(\hat{g})=0$. 

Because $B$ is a sum, some sampled configurations will still violate the condition $B \leq 1/\hat{g}$ and must be rejected, but this occurs less frequently than in an unconstrained sampling. We therefore restrict the sampling of all $k$-tuples to the region $\hat{g}_j \leq 1/\hat{g}$ and include a corresponding bias factor $[\int_0^{1/\hat{g}} P(x)\,dx]^k$, estimated from the empirical distribution. In practice, this is implemented by sorting the population according to $\hat{g}_j$ and sampling uniformly within the allowed range. If $j_{\max}$ denotes the largest index such that $\hat{g}_{j_{\max}} \leq 1/\hat{g}$, the bias factor is simply $(j_{\max}/\Omega)^k$. For each value of $\hat{g}$, the estimate of $P(\hat{g})$ is obtained by averaging over $N_{\rm est}$ independent realizations.

\begin{figure}[ht]
    \centering
    \includegraphics[width=0.5\textwidth]{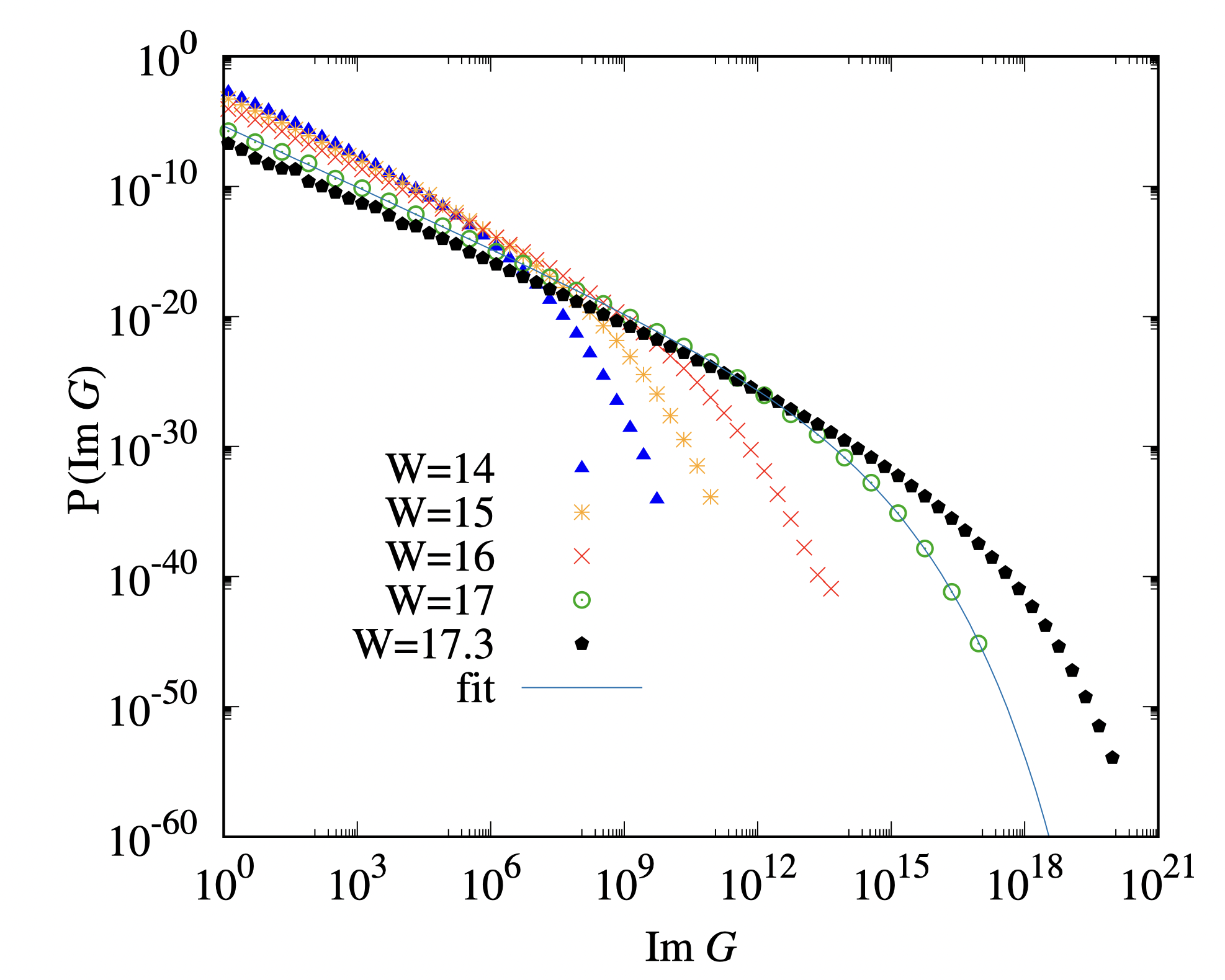}
    \caption{Distribution $P({\rm Im} G)$ of the imaginary part of the cavity Green's function for several values $W\in[13,17.3]$. The lines correspond to fits using Eq.~\eqref{eq:shape_distr} to extract the correlation volume (see text).} 
    \label{fig:distr_g_neu}
\end{figure}

This large-deviation approach was applied in Ref.~\cite{biroli2022critical} to compute the distribution of the cavity Green’s function for the Anderson model on the Bethe lattice with connectivity $k+1=3$ at $E=0$. The resulting distributions $P(\hat{g})$ are shown in Fig.~\ref{fig:distr_g_neu}. Remarkably, this method allows one to access probability densities as small as $10^{-50}$ with high precision, far beyond the reach of standard population dynamics.

The tails of the distributions are found to be in excellent agreement with the analytical predictions of the supersymmetric approach~\cite{mirlin1994distribution}. In particular, over many decades, they are well described by the asymptotic form
\begin{equation}\label{eq:shape_distr}
    P(\hat{g}) \simeq c\, \hat{g}^{-3/2} \exp\!\left[-\left(\frac{\hat{g}}{\Lambda_c}\right)^{\alpha}\right],
\end{equation}
where $\Lambda_c$ is the correlation volume introduced in Sec.~\ref{sec:critical}. This functional form captures the crossover from an apparent power-law behavior to a stretched-exponential cutoff at large $\hat{g}$.

By fitting the logarithm of the distribution tails to Eq.~\eqref{eq:shape_distr} for different values of $W$, we extract the cutoff scale $\Lambda_c$ as a function of disorder. The exponent $\alpha$ is found in the range $\alpha \in [0.163(2),\,0.208(2)]$. The resulting values of $\Lambda_c$ are in good agreement with the asymptotic predictions of the supersymmetric approach (see Eq.~\eqref{eq:Lambdac}), providing a stringent test of its validity.

\subsection{Computing the IPR in the localized phase} \label{sec:IPR}

In this section we show how the linearized cavity recursion equations~\eqref{eq:Re_cavity_bl}–\eqref{eq:Im_cavity_bl} can be efficiently exploited to compute relevant observables in the localized phase. For clarity, we focus on the inverse participation ratio. The extension to two-point correlation functions is straightforward and will not be discussed in detail. The approach was first introduced in Ref.~\cite{rizzo2024localized}.

The IPR, Eq.~\eqref{eq:I2}, is related to the second moment of $|G_{ii}|$, which is broadly distributed in the localized phase. The power-law tails of its probability distribution would, in principle, lead to divergent expressions for $\langle |G_{ii}|^2 \rangle$. In practice, this divergence is regularized by a cutoff at large values of the imaginary part, of order $\eta^{-1}$, which marks the limit of validity of the linearized equations. Nevertheless, evaluating $\langle |G_{ii}|^2 \rangle$ requires probing the regime of large imaginary parts of the Green’s function, rather than the region of typical finite values. At first sight, this might suggest that the linearized equations are not suitable for this purpose.

Fortunately, this is not the case. To see this, we define (restricting for simplicity to $E=0$, i.e., the center of the band, and setting $t=1$):
\begin{equation} \label{eq:mii1}
M_{ii} \equiv  m_{ii} - {\rm i} \eta \, \hat{m}_{ii} \equiv \frac{1}{G_{ii}} = - \epsilon_{i} - {\rm i} \eta - \!\! \sum_{m \in \partial i} G_{m}^{(i)}  \, .
\end{equation}
From this, one immediately obtains
\begin{equation}
\langle |G_{ii}|^2 \rangle = \int \mathrm{d}m \, \mathrm{d}\hat{m} \; Q(m,\hat{m}) \, \frac{1}{m^2 + \eta^2 \hat{m}^2} \, ,
\end{equation}
where $Q(m,\hat{m})$ denotes the joint probability density of $m$ and $\hat{m}$. Similarly, using ${\rm Im} G_{ii} = \eta \hat{m}_{ii}/(m_{ii}^2 + \eta^2 \hat{m}_{ii}^2)$, one finds
\begin{equation}
\langle {\rm Im} G_{ii} \rangle = \int \mathrm{d}m \, \mathrm{d}\hat{m} \; Q(m,\hat{m}) \, \frac{\eta \hat{m}}{m^2 + \eta^2 \hat{m}^2} \, .
\end{equation}
Since $\hat{m}$ is strictly positive, we perform the change of variables $m = \eta \hat{m} x$, which yields
\begin{eqnarray}
\langle |G_{ii}|^2 \rangle &=& \int \mathrm{d}x \, \mathrm{d}\hat{m} \; Q(\eta \hat{m} x, \hat{m}) \, \frac{(\eta \hat{m})^{-1}}{1 + x^2} \, , \label{eq:gii} \\
\langle {\rm Im} G_{ii} \rangle &=& \int \mathrm{d}x \, \mathrm{d}\hat{m} \; Q(\eta \hat{m} x, \hat{m}) \, \frac{1}{1 + x^2} \, . \label{eq:img}
\end{eqnarray}
In the limit $\eta \to 0^+$, we can approximate $Q(\eta \hat{m} x, \hat{m}) \simeq Q(0,\hat{m})$ and perform the integration over $x$ explicitly. Substituting Eqs.~\eqref{eq:gii} and~\eqref{eq:img} into Eq.~\eqref{eq:I2}, one finally obtains~\cite{mirlin1994statistical,rizzo2024localized}:
\begin{equation} \label{eq:IpQ}
I_2 = \frac{ \int \mathrm{d}\hat{m} \; Q(0,\hat{m}) \, \hat{m}^{-1} }
{\int \mathrm{d}\hat{m} \; Q(0,\hat{m}) } \, .
\end{equation}

In summary, although the moments of the local Green’s function are governed by rare events where $|G_{ii}| = O(1/\eta)$ with probability $O(\eta)$, they can be expressed in terms of the \emph{typical} values of $M_{ii}$, whose real part is $O(1)$ and whose imaginary part is $O(\eta)$. This observation allows one to compute observables in the localized phase such as the IPR using the linearized equations, and enables the use of highly efficient numerical methods that strongly reduce finite-population effects.

A suitable modification of the population dynamics algorithm (see Refs.~\cite{rizzo2024localized,tonetti2026geometrylocalizationprobinglocalization} for details) makes it possible to efficiently extrapolate $Q(m,\hat{m})$ to $m=0$, and thus evaluate both the numerator and denominator of Eq.~\eqref{eq:IpQ} with high precision. From Eq.~\eqref{eq:mii1}, we have $m_{ii} = -\epsilon_i - \sum_{m \in \partial i} g_m^{(i)}$. Therefore, the condition $m_{ii}=0$ is equivalent to $\epsilon_i = -\sum_{m \in \partial i} g_m^{(i)}$, which occurs with probability density $1/W$ if $|\sum_{m \in \partial i} g_m^{(i)}| < W/2$, and zero otherwise.

Based on this observation, the numerical procedure is as follows. For a given $W$ in the localized phase, one first runs standard population dynamics to obtain the stationary distribution $P(g,\hat{g})$ solving Eqs.~\eqref{eq:Re_cavity_bl}–\eqref{eq:Im_cavity_bl}. One then extracts $k+1$ elements from the pool and computes $m$ and $\hat{m}$ via Eq.~\eqref{eq:mii1}, defining $S = -\sum_{i=1}^{k+1} g_i$. If (and only if) $|S| < W/2$, one adds $\hat{m}^{-1}/W$ to the numerator and $1/W$ to the denominator in Eq.~\eqref{eq:IpQ}. This procedure is repeated many times, and the numerator and denominator are normalized by the total number of trials. The pool is periodically refreshed by a few standard population dynamics updates, and the process is iterated until the desired accuracy on $I_2$ is reached.

This method yields accurate results even very close to the critical point, and can be readily extended to compute general moments of eigenfunction amplitudes as well as two-point correlation functions.

\begin{figure}
\includegraphics[width=0.352\textwidth]{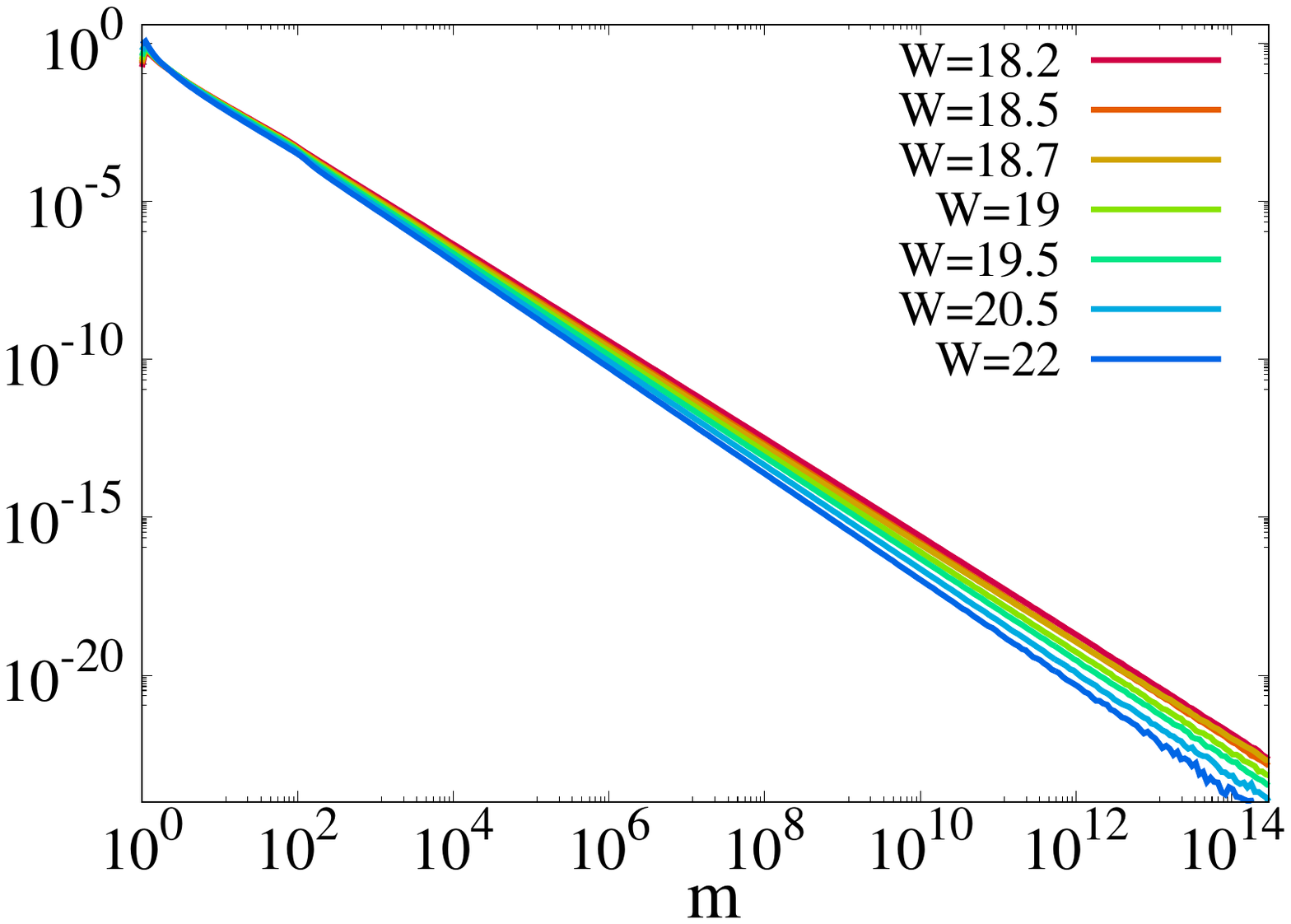} 
\put(-162,58){\rotatebox[origin=c]{90}{\small $Q(0,\hat{m})/\avg{\rho}$}} \put(-71,1){\small \^}
\includegraphics[width=0.352\textwidth]{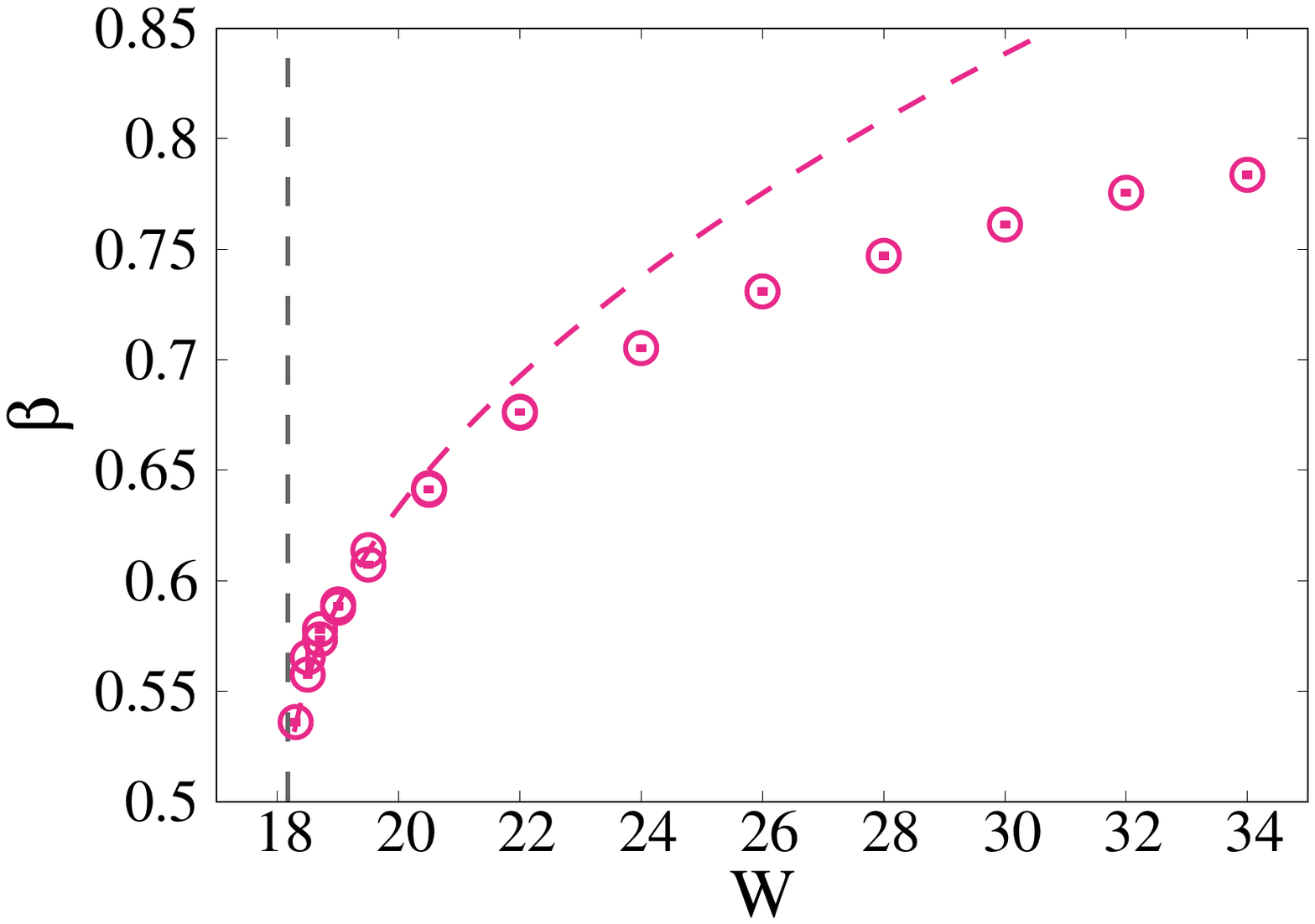} 
\includegraphics[width=0.352\textwidth]{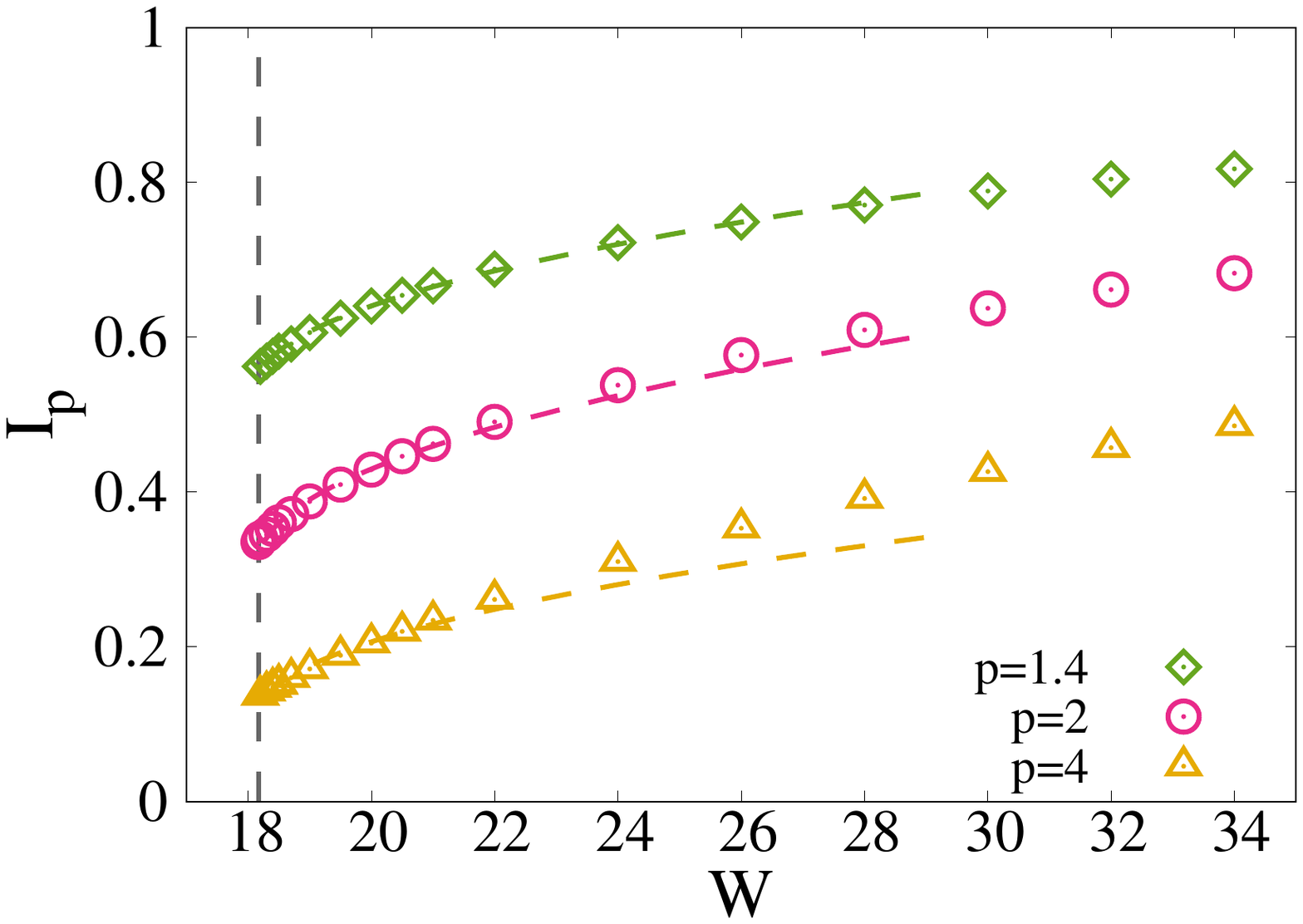} 
\caption{Left: Probability distribution $Q(0,\hat{m})/\avg{\rho}$ obtained by conditioning on $m=0$. Middle: Exponent $\beta$ of the power-law tails as a function of $W$. The dashed curve shows the prediction of Eq.~\eqref{eq:beta} near $W_c$ from Ref.~\cite{tikhonov2019critical}. Right: Generalized inverse participation ratios $I_p$ (for $\eta=0$ and $N \to \infty$) as a function of $W$ in the localized phase for $p=1.4$, $p=2$, and $p=4$. The dashed lines are fits of the form $I_p \simeq I_p^{\rm (c)} + a_p \sqrt{W - W_c}$.}
\label{fig:IPR}
\end{figure}

In Fig.~\ref{fig:IPR} we report representative numerical results obtained using this procedure (see Ref.~\cite{rizzo2024localized} for details). The left panel shows the distribution $Q(0,\hat{m})$, obtained with very high precision, including a clean characterization of its power-law tails and exponent $1+\beta$ (see Eq.~\eqref{eq:ansatz} in Sec.~\ref{sec:critical}). The values of $\beta$ extracted from these fits (middle panel) are in excellent agreement with the square-root critical behavior predicted by Eq.~\eqref{eq:beta}, with coefficients $c_1$ and $c_2$ obtained from the numerical diagonalization of the integral operator~\eqref{eq:operator} near the critical point~\cite{tikhonov2019critical}.

The right panel shows the generalized IPR $I_p$ as a function of $W > W_c$ for $p=1.4$, $p=2$, and $p=4$. The data clearly exhibit a jump to a finite value at $W_c$, confirming the discontinuity of the IPR at the localization transition predicted by the supersymmetric approach~\cite{efetov1985anderson,efetov1987density,efetov1987anderson,zirnbauer1986localization,zirnbauer1986anderson,verbaarschot1988graded,mirlin1991localization,mirlin1991universality,fyodorov1991localization,fyodorov1992novel,mirlin1994statistical,tikhonov2019statistics}. Moreover, the data suggest that just above the transition the IPR follows a square-root behavior,
\begin{equation} \label{eq:IPRsingularity}
I_p \simeq I_p^{\rm (c)} + a_p \sqrt{W - W_c} \, .
\end{equation}
A fit for $p=2$ yields $I_2^{\rm (c)} \simeq 0.304$ and $a_2 \simeq 0.094$. This square-root singularity had not been previously reported and was later confirmed within the NL$\sigma$M framework on the Bethe lattice~\cite{rizzo2024localized}, which provides an effective description of Anderson localization and is more tractable than the original tight-binding model.

%=== DIRECTED POLYMERS ===
\section{Connection with Derrida's directed polymers in random media}\label{sec:connection_DP}
Before concluding this review, we would like to discuss the close relationship between Anderson localization on the Bethe lattice and a paradigmatic problem of classical disordered systems, namely the freezing glass transition of directed polymers in random media (DPRM), first introduced and studied by Derrida~\cite{derrida1988polymers}. The mapping between Anderson localization and DPRM has been discussed in many recent papers~\cite{lemarie2019glassy,biroli2012difference,sonner2017multifractality,tikhonov2016fractality,monthus2008anderson,monthus2011anderson,chakrabarti2022traveling,biroli2020anomalous,swain2026two}, and its interest is not only that it provides a clear and physically appealing interpretation of Anderson localization in terms of the freezing of the paths along which delocalization can occur, but also that it offers a set of analytical tools that can be exploited to obtain precise information on localization phenomena, in particular concerning the role of rare disorder realizations~\cite{biroli_large-deviation_2024,miranda2025largedeviationsmanybodylocalization}.

\subsection{A brief review of DPRM on the Bethe lattice} \label{sec:DPRM}
Before discussing the mapping in detail, we briefly recall the basic properties of DPRM, focusing specifically on the infinite Bethe lattice case~\cite{derrida1988polymers}. The aim of this short section is not to provide an exhaustive review, but just to introduce the basic tools and ideas that we will use below, when mapping this problem onto Anderson localization.

Let us consider an infinite Bethe lattice with fixed branching ratio $k$. On each edge $(i_m \leftrightarrow i_{m+1})$ of the lattice, there is a random potential $\omega_{i_m \leftrightarrow i_{m+1}}^m$, independently and identically distributed according to a certain probability density $\gamma(\omega)$. A directed polymer on the tree is a self-avoiding walk of $n$ steps starting from the root at $0$. The total number of paths (\ie, the total number of configurations of the polymer of length $n$) is equal to the number of leaves, \ie, $(k+1) k^{n-1}$. By definition, the total energy of a walk is the sum of the random potentials on the edges crossed by the path ${\cal P}$:
\begin{equation}
    E_{\cal P} = \omega_{0 \leftrightarrow i_1}^0 + \omega_{i_1 \leftrightarrow i_2}^1 + \ldots + \omega_{i_{n-1} \leftrightarrow i_n}^{n-1} \, , 
\end{equation}
where the indices $i_m = 1, \ldots, (k+1) k^{m-1}$ denotes the nodes of the $m$-th generation. To describe the statistical properties of the directed polymer we thus need to compute the partition function:
\begin{equation} \label{eq:DPRM}
Z = \sum_{{\cal P}} e^{- \beta E_{\cal P}} =  \sum_{{\cal P}} e^{- \beta \sum_{(i_m \leftrightarrow i_{m+1} ) \in {\cal P}}  \omega^m_{i_m \leftrightarrow i_{m+1}}}\, .
\end{equation}
Assuming that the random potentials have zero mean and a finite variance $\sigma_\omega^2$, in the limit of large trees the distribution of the $E_{\cal P}$ converges to a Gaussian with zero mean and variance $n \sigma_\omega^2$. In this respect the partition function~\eqref{eq:DPRM} is very similar to the one of the celebrated Random Energy Model (REM)~\cite{derrida1980random}, which is the simplest and most paradigmatic model of mean-field structural glasses, featuring a low-temperature glassy phase described by a one-step replica symmetry breaking~\cite{mezard1987spin}. 

Thanks to the hierarchical structure of the lattice, the cavity partition function restricted to a branch of the tree originating from a node of the $m$-th generation in absence of the node of the previous generation satisfies the following exact relation in terms of the partition functions restricted to the branches originating from the $k$ neighboring nodes of the next generation:
\begin{equation}
    Z_{i_m}^{(i_{m-1})} = e^{- \beta \omega^{m-1}_{i_{m-1} \leftrightarrow i_{m}}} \!\!\!\! \sum_{i_{m+1} \in \partial i_m \setminus i_{m-1}} \! \! Z_{i_{m+1}}^{(i_m)} \, .
\end{equation}

An exact solution of these exact recursion relations in the thermodynamic limit can  be found following the mapping proposed in Ref.~\cite{derrida1988polymers} onto the study of a certain nonlinear equation of reaction-diffusion type, the so-called Kolmogorov-Petrovsky-Piscounov (KPP) equation~\cite{kpp1937}.

Below, we discuss instead a simplified approximate solution which nonetheless contains all the physical properties of the exact solution. The partition function~\eqref{eq:DPRM} is a sum over $\sim k^n$ terms. There is therefore a competition between an entropic factor (large number of terms) and an energetic factor (the Boltzmann weight). At high temperature the polymer visits an exponential number of configurations from which $Z$ receives a significant contribution. At low temperature, instead, in the glassy phase, the polymer can only visit a few $O(1)$ configurations having very large weights (\ie~very low energies), and the sum over the paths is dominated by these few outliers. 

It is clear that such a glass transition is driven by the condensation of the Gibbs' measure on the extreme values of the (correlated) random variable $e^{-\beta E_{\cal P}}$. Neglecting, at first, the correlations of the $E_{\cal P}$'s on different paths, one obtains a REM-like problem. 
The total number of configurations of the polymer at  energy (per unit length) $\epsilon_{\cal P} = E_{\cal P}/n$ in the large $n$ limit is:
\begin{equation}
    \Omega(\epsilon_{\cal P}) = (k+1) k^{n-1} \sqrt{\frac{n}{2 \pi \sigma_\omega^2}} e^{-n \epsilon_{\cal P}^2/(2 \sigma_\omega^2)} \, ,
\end{equation}

corresponding to a configurational entropy (per unit length) equal to
\begin{equation}
    \Sigma (\epsilon_{\cal P}) = \frac{1}{n} \ln \Omega(\epsilon_{\cal P}) \simeq \ln k - \frac{\epsilon_{\cal P}^2}{2 \sigma_\omega^2} \, 
\end{equation}
(where we have set $k_B=1$). This expression is only valid for intensive energies such that $\Sigma > 0$, \ie~for $\epsilon_{\cal P}$ within the interval $|\epsilon_{\cal P}| \le \epsilon_\star = \sqrt{2 \ln k} \, \sigma_\omega$. For $|\epsilon_{\cal P}| > \sqrt{2 \ln k} \, \sigma_\omega$, instead, the entropy $\Sigma$ is negative, implying that the number of walks at those values of the energy is exponentially small in $n$.

The partition function~\eqref{eq:DPRM} can be then rewritten in the large $n$ limit as an integral over all paths giving a contribution characterized by a value of the energy between $\epsilon_{\cal P}$ and $\epsilon_{\cal P} + {\rm d} \epsilon_{\cal P}$ times the number of such paths:
\begin{equation} \label{eq:Zint}
Z = \int \!\! {\rm d} \epsilon_{\cal P} \, e^{-n \beta \epsilon_{\cal P} + n \Sigma(\epsilon_{\cal P})} \, .
\end{equation}
In the limit $n \to \infty$ the value of the energy $\overline{\epsilon}_{\cal P}$ that dominates the integral is given by the saddle-point condition:
\begin{equation}
    \left . \frac{{\rm d} \Sigma}{{\rm d} \epsilon_{\cal P}} \right \vert_{\epsilon_{\cal P}= \overline{\epsilon}_{\cal P}} \!\!\!\!\!\!\!\! = \beta \, \Longrightarrow \, \overline{\epsilon}_{\cal P} = - \beta \sigma_\omega^2  \, .
\end{equation}

Yet, this solution is only correct provided that $\overline{\epsilon}_{\cal P} \ge - \epsilon_\star$, where the entropy vanishes. This condition yields the critical temperature $\beta_\star = \sqrt{2 \ln k} / \sigma_\omega$ below which the energy of the polymer freezes at the minimal value $-\epsilon_\star$. The corresponding (quenched) free-energy (per unit length) is thus:
\begin{equation} \label{eq:FDPRM}
f (\beta) = - \lim_{n \to \infty} \frac{\langle \ln Z \rangle}{\beta n}  = 
\left \{
\begin{array}{ll}
- \beta \sigma_\omega^2/ 2 - \ln k/ \beta  \,\, & \textrm{for~} \beta< \beta_\star \, , \\
-\epsilon_\star & \textrm{for~} \beta \ge \beta_\star \, .
\end{array}
\right .
\end{equation}
The transition taking place at $\beta_\star$ corresponds therefore to a transition to a frozen phase. The free-energy varies with temperature until it reaches its maximum, and sticks there at lower temperatures. Also the shape of the whole free-energy distribution does not change with temperature in the frozen phase~\cite{derrida1988polymers}. 

The configurational entropy, which encodes the exponential number of paths typically visited by the polymer, is related to the free-energy through a Legendre transform:
\begin{equation} \label{eq:ScDPRM}
\Sigma (\beta) = \beta^2 \frac{{\rm d} f}{{\rm d} \beta} = \left \{
\begin{array}{ll}
\ln k - \beta^2 \sigma_\omega^2/ 2   \,\, & \textrm{for~} \beta< \beta_\star \, , \\
0 & \textrm{for~} \beta \ge \beta_\star \, ,
\end{array}
\right .
\end{equation}
corresponding to the fact that in the high-temperature phase, $\beta<\beta_\star$, the partition function receives contribution from an exponential number of paths, while in the low-temperature phase, $\beta>\beta_\star$, only a few, specific, disorder-dependent paths, corresponding to the extreme values of $E_{\cal P}$ out of the $k^n$ realizations, dominate the sum. Note that for $\beta=0$ the configurational entropy attains its maximum value $\Sigma = \ln k$, signaling that all $O(k^n)$ configurations contribute equally to $Z$.

As explained above, the problem of DPRM is slightly more involved than that of the REM. Although the $\omega_{i_m}$ are i.i.d.\ random variables, the path energies $E_{\cal P}$ are correlated, since different walks typically share a finite number of nodes. Nevertheless, DPRM undergoes a low-temperature freezing (glass) transition of the same nature as that of the REM described above. By mapping the DPRM onto the KPP equation~\cite{kpp1937}, Derrida obtained the exact result for the quenched free energy per unit length:
\begin{equation} \label{eq:fqDPRM}
f (\beta) = - \lim_{n \to \infty} \frac{\langle \ln Z \rangle}{\beta n}  = 
\left \{
\begin{array}{ll}
- \frac{1}{\beta}\ln \left ( k \langle e^{-\beta \omega} \rangle \right)  & \text{for } \beta< \beta_\star \, , \\
- \frac{1}{\beta_\star}\ln \left ( k  \langle e^{-\beta_\star \omega} \rangle \right ) & \text{for } \beta \ge \beta_\star \, ,
\end{array}
\right .
\end{equation}
where $\beta_\star$ is determined by the condition $\left.\partial f(\beta) / \partial \beta \right|_{\beta_\star} = 0$.

Note that this expression for the free energy coincides with the one obtained from a direct one-step replica symmetry breaking (1-RSB) ansatz, as we now show. To this end, consider $p$ identical copies (or replicas) of the system (i.e., all with the very same realization of the disorder), and assume that they split into $p/m$ groups of $m$ replicas such that, within each group, the polymer occupies the same configuration $\mu$. At low temperature, when the polymer freezes on a few paths, these configurations are particularly favorable, \ie,  they correspond to paths with very low energy. Typically, two such low-energy paths are uncorrelated and occupy distinct edges, so that the corresponding energies $E_\mu$ associated with different groups of replicas can be treated as independent (the validity of this assumption can be checked self-consistently by computing the overlap distribution in the frozen phase, see below and Ref.~\cite{derrida1988polymers}). One thus has
\begin{equation}
\langle Z^p \rangle =  \left \langle \sum_{\mu = 1}^{k^n} e^{- \beta m E_\mu} \right \rangle^{p/m} = \left( k \langle e^{- \beta m \omega } \rangle \right)^{n p/m} \, .
\end{equation}
Taking the limit $p \to 0$, one immediately obtains
\begin{equation}
\langle Z^p \rangle  \simeq 1 + p \langle \ln Z \rangle = 1 + \frac{p n}{m} \ln \left ( k \langle e^{- \beta m \omega } \rangle \right) \, .
\end{equation}
The replicated free energy then reads
\begin{equation}
\phi(m) = - \frac{\langle \ln Z \rangle}{\beta n} = - \frac{1}{\beta m} \ln \left ( k \langle e^{-\beta m \omega } \rangle \right ) \, .
\end{equation}
Optimizing with respect to the Parisi variational parameter $m$ yields
\begin{equation}
m_\star = \left \{
\begin{array}{ll}
1 & \text{for } \beta \le \beta_\star, \\
\beta_\star/\beta & \text{for } \beta > \beta_\star,
\end{array}
\right.
\end{equation}
The critical temperature at which freezing occurs is therefore obtained for $m_\star = 1$. Plugging the expression of $m_\star$ into the replicated free energy, one recovers the expression~\eqref{eq:fqDPRM}.

It is easy to check that for the particular case in which $\gamma(\omega)$ is a Gaussian distribution with zero mean and variance $\sigma_\omega^2$ one recovers the results~\eqref{eq:FDPRM} and~\eqref{eq:ScDPRM}---with $\beta_\star = \sqrt{2 \ln k}/\sigma_\omega$---corresponding to the approximation in which one neglects the correlations between the energies of the different paths. For a generic probability density the results are quantitatively slightly different, although the physical picture is still exactly the same.

The low-temperature phase of DPRM is very similar to the one of mean-field glasses (such as the REM), where the phase space breaks up into many disconnected pieces separated by free-energy barriers~\cite{mezard1987spin}. The order parameter function associated with such many-valley structure is the distribution of the overlap $q({\cal P},{\cal P}^\prime)$ between two polymer's configurations ${\cal P}$ and ${\cal P}^\prime$, defined as the fraction of edges that the two walks have in common for a given disorder realization. As shown in Ref.~\cite{derrida1988polymers}, for $n \to \infty$ $p(q)$ behaves identically as in the REM~\cite{derrida1980random}: In the high-temperature phase the overlap is identically equal to zero, while in the low-temperature phase the overlap is either zero or one (with a relative height of the two $\delta$-peaks corresponding to $m_\star$ and $1-m_\star$), corresponding, in the spin-glass jargon, to a one-step replica symmetry breaking ansatz~\cite{mezard1987spin}.

The freezing transition of the paths is associated with the condensation of the Boltzmann weights on the configurations that correspond to the extreme values  of the probability distributions of the energies $E_{\cal P}$. Such anomalously large values of the Boltzmann weights have strong fluctuations from one sample to another, thereby producing a broad distribution $P(Z)$ of the partition function, which exhibits power-law tails (see Ref.~\cite{derrida1988polymers}).

\subsection{Relationship with Anderson localization on the Bethe lattice}

We now show that Anderson localization on the Bethe lattice can be formally mapped onto Derrida's DPRM problem described above (see Refs.~\cite{biroli2012difference,lemarie2019glassy,biroli2020anomalous,biroli_large-deviation_2024} for more details). Below, for simplicity, we limit the discussion of the mapping to the localized phase, although it can be generalized to the delocalized phase as well. Upon linearizing the exact recursion relations the cavity Green's functions with respect to the imaginary part, Eqs.~\eqref{eq:Re_cavity_bl}-\eqref{eq:Im_cavity_bl}, and setting the imaginary regulator to zero, one obtains (for $t=1$ and $E=0$): 
\begin{eqnarray} 
{g}_{i}^{(j)} &=& \frac{1}{- \epsilon_i - \sum_{m \in \partial i \setminus j} \, {g}_{m}^{(i)} }\, , \label{eq:real} \\
\hat{g}_{i}^{(j)} &= &\big({g}_{i}^{(j)}\big)^2  \sum_{m \in \partial i \setminus j} \hat{g}_{m}^{(i)} \, . \label{eq:imag}
\end{eqnarray}
As discussed above, these equations have two important features: i) The equation for the real part does not depend on the imaginary part; ii) The equation for the imaginary part is linear. This latter properties implies that the typical value of $\hat{g}$ under iteration can either grow or decrease exponentially according to the value of the largest eigenvalue of the linear operator associated to Eq.~\eqref{eq:imag}.

From Eq.~\eqref{eq:imag} we also obtain that the imaginary part of the Green's function in the linearized regime is formally given by exactly the same expression as the partition function of a directed polymer on a tree (and at inverse temperature $\beta=1)$:
\begin{equation} \label{eq:DP}
\hat{g}_{i}^{(j)} = \sum_{\mu = 1}^{k^n} e^{- E_\mu } \, ,
\end{equation}
where the random energies $E_\mu$ correspond to the $\mu$-th path that goes from the node $i$ to one of the $k^n$ nodes at distance $n$ from $i$ and are given by :
\begin{equation}
E_\mu = - \sum_{\ell=1}^{n-1} \ln \left( {g}_{m^{(\mu)}_\ell}^{(m^{(\mu)}_{\ell-1})} \right)^2  - \ln \hat{g}_{m^{(\mu)}_n}^{(m^{(\mu)}_{n-1})} \, .
\end{equation}
There are however two differences with respect to the directed polymer case described in the previous section:
\begin{enumerate}
    \item The local energies $\epsilon_m^{(\mu)} \equiv -\ln ( {g}_{m^{(\mu_\ell)}}^{(m^{(\mu)}_{\ell-1})})^2$ defined on each node of the tree have nonzero average;
    \item Most importantly, the $\epsilon_m^{(\mu)}$'s are strongly correlated, since the $g$ on a given node of the graph depend on the $g$'s on its neighboring nodes via Eq.~\eqref{eq:real}. Due to these correlations the central limit theorem does not apply to the $E_\mu$'s which in this case are not (necessarily) Gaussian distributed.
\end{enumerate}
Still, the problem can be solved using the same approach as the one put forward by Derrida in the directed polymer case with uncorrelated random local energies. 

To make the analogy with DPRM more explicit, let us define the partition function:
\begin{equation}
Z = \sum_{\mu = 1}^{k^n} \prod_{\ell = 1}^n \left | {g}_{m^{(\mu)}_\ell}^{(m^{(\mu)}_{\ell-1})} \right |^2 \, .
\end{equation}
According to Eq.~\eqref{eq:imag}, the typical value of $Z$ gives the asymptotic scaling at large $n$ (which here plays the role of the number of iterations of the recursive equations) of $\hat{g}$. We thus apply the replica approach to compute $\langle \ln Z \rangle$, inspired by the problem of DPRM (see above): We consider $p$ identical copies of the system and assume that these replicas clusterize in $p/m$ groups of $m$ replicas which will occupy the same path $\mu$:
\begin{equation} \label{eq:Zn}
\langle Z^p \rangle = \left \langle  \sum_{\mu = 1}^{k^n} \prod_{\ell = 1}^n \left | {g}_{m^{(\mu)}_\ell}^{(m^{(\mu)}_{\ell-1})} \right|^{2 m} \right \rangle^{p/m}  = \left[ k^n \left \langle  \prod_{\ell = 1}^n \left| {g}_{m^{(\mu)}_\ell}^{(m^{(\mu)}_{\ell-1})} \right|^{2 m} \right \rangle \right]^{p/m} \, .
\end{equation}
Let us now consider a given path of length $n$. From Eq.~\eqref{eq:real}, and using the definitions of the probability distributions~\eqref{eq:P0def} and~\eqref{eq:Rtilde}, one immediately obtains:
\begin{equation} \label{eq:gproduct}
\begin{aligned}
& \left \langle \left | g_{1}^{(0)} \right |^{2 m}  \left | g_{2}^{(1)} \right |^{2 m} \cdots \left | g_{n}^{(n-1)} \right |^{2 m} \right \rangle = \\
& \qquad = \int \de P_0 (g_n) \, |g_n|^{2 m} \, \de \gamma(\epsilon_{n-1}) \,  \de R(\tilde{g}_{n-1}) \, \delta\left( g_{n-1} + \frac{1}{\epsilon_{n-1} + \tilde{g}_{n-1} + g_n} \right) |g_{n-1}|^{2 m} \\
& \qquad \;\; \times \de \gamma(\epsilon_{n-2}) \, \de R(\tilde{g}_{n-2}) \, \delta\left( g_{n-2} + \frac{1}{\epsilon_{n-2} + \tilde{g}_{n-2} + g_{n-1}} \right) |g_{n-2}|^{2 m} \\
& \qquad \;\; \times \cdots \times \de \gamma(\epsilon_1) \, \de R(\tilde{g}_1) \, \delta\left( g_{1} + \frac{1}{\epsilon_1 + \tilde{g}_1 + g_{2}} \right) |g_{1}|^{2 m} \\
& = \int \de g_n P_0 (g_n) \, \frac{K_m (g_n, g_{n-1})}{k} \, \de g_{n-1} \, \frac{K_m (g_{n-1}, g_{n-2})}{k}  \times \cdots \times \de g_2  \, \frac{K_m (g_2,g_1)}{k}  \\
& \propto (K_m/k)^n \cdot P_0 \, ,
\end{aligned}
\end{equation}
where the kernel $K_m$ is precisely the kernel of the integral operator defined in Eqs.~\eqref{eq:operator} and~\eqref{eq:kernel}. In the large $n$ limit the asymptotic behavior is dominated by the largest eigenvalue of the integral operator. Therefore one has:
\begin{equation}
\ln \left \langle \left | g_{1}^{(0)} \right |^{2 m}  \left | g_{2}^{(1)} \right |^{2 m} \cdots \left | g_{n}^{(n-1)} \right |^{2 m} \right \rangle = n \ln \left( \frac{\lambda_{\rm max}(m)}{k} \right) + O(\ln n) \, .
\end{equation}
Plugging this result into Eq.~\eqref{eq:Zn}, and using the fact that in the limit $p \to 0$ one has $\langle \ln Z \rangle \simeq (\langle Z^p \rangle - 1)/p$, we obtain
\begin{equation}
\phi(m) = \frac{\langle \ln Z \rangle}{n} = \frac{1}{m} \ln \lambda_{\rm max}(m)\, .
\end{equation}

The function $\phi(m)$ must then be optimized with respect to the Parisi replica parameter $m$. Recall that Anderson localization occurs when the typical value of ${\rm Im}\,G$ (that is, the typical value of $Z$) changes from exponential growth to exponential decay, \ie, when $\langle \ln Z \rangle$ crosses zero. These two conditions read
\begin{equation}
\left \{
\begin{array}{l}
\dfrac{\de \phi(m)}{\de m}\bigg|_{m_\star} = 0 \, , \\[0.3cm]
\phi(m_\star) = 0 \, ,
\end{array}
\right.
\end{equation}
which, as discussed above, occurs at $m_\star = 1/2$, within the frozen phase of $Z$.  (Note that this procedure is equivalent---although expressed in a different language and derived along a different line of reasoning---to the saddle-point method used to evaluate the Mellin transform of the probability distribution of the off-diagonal elements of the Green's function, Eq.~\eqref{eq:Mellin}, and is tightly related to the rigorous proofs of localization on tree graphs based on analysis of the fractional moments of the Green function~\cite{aizenman1993localization,aizenman1994localization,aizenman2011extended}.)

This formal analogy with the freezing transition of DPRM provides an appealing and transparent physical interpretation of Anderson localization on the Bethe lattice. In the delocalized phase, a particle created on a given node of the graph can escape to infinity through an exponential number of paths. The Anderson-localized phase instead corresponds to the frozen regime, in which the particle can propagate only through a few $O(1)$ paths where rare resonances are encountered. Yet, even along these rare paths, the particle cannot escape to infinity, since the typical value of $Z$ (and thus of ${\rm Im}\,G$) decreases exponentially with distance. This picture corresponds to the strong multifractal behavior of localized eigenstates in the Anderson model on the Bethe lattice~\cite{de2013ergodicity}.

%=== CONCLUSIONS ===
\section{Conclusions and perspectives}\label{sec:conclusions}

\subsection{Summary}

More than six decades after Philip W. Anderson's seminal discovery~\cite{anderson1958absence}, the problem of quantum localization driven by disorder continues to reveal new frontiers, challenges, and subtleties. This pedagogical review has been dedicated to exploring this phenomenon within a unique and highly singular framework: the infinite Bethe lattice. Possessing an exponential volume growth that reflects an effectively infinite spatial dimensionality ($d \to \infty$), yet remaining locally sparse and tree-like, the Bethe lattice serves as a crucial bridge between exact analytical tractability and the complex physical behavior of localization phenomena.

We have introduced the Anderson model and discussed standard diagnostics of localization, including key observables related to transport, spectral statistics, and the scaling of eigenfunction moments. We have systematically unpacked the resolvent (or Green's function) formalism, demonstrating how the fundamental signatures of the localization transition are entirely encoded in the statistical properties of the resolvent elements. The full probability distribution of the local density of states emerges as the natural order parameter of the transition, allowing one to distinguish between an absolutely continuous spectrum (metallic, extended phase) and a pure point spectrum (insulating, localized phase). In the localized phase, the typical LDOS collapses to zero proportionally to the external regularizing broadening $\eta$, while rare spatial resonances manifest as singular peaks of height $\mathcal{O}(\eta^{-1})$. Conversely, in the delocalized phase, an analogous statistical structure persists close to the mobility edge, but the role of $\eta$ is replaced by an intrinsically self-generated energy scale $\Lambda_c^{-1}$, which vanishes exponentially upon approaching the transition.

Crucially, the tree structure of the Bethe lattice makes it possible to derive exact self-consistent cavity equations for the diagonal Green's functions. These equations provide a closed description of the localization transition, allowing for analytical results for both the transition point and the associated critical behavior~\cite{abou1973selfconsistent, evers2008anderson, biroli2010anderson, rizzo2024localized, tikhonov2019statistics, tikhonov2019critical, efetov1985anderson, efetov1987density, efetov1987anderson, zirnbauer1986localization, zirnbauer1986anderson, verbaarschot1988graded, mirlin1991localization, mirlin1991universality, fyodorov1991localization, fyodorov1992novel, mirlin1994statistical, mirlin1994distribution}.

These self-consistent equations can be solved numerically via population dynamics with very high accuracy~\cite{biroli2022critical, rizzo2024localized}. Furthermore, by analyzing the cavity recursion relations close to the critical point (with a special focus on the large-connectivity limit, which is exactly solvable), we have reviewed the highly unconventional critical behavior that distinguishes localization on sparse graphs from its counterpart in finite dimensions. Because the upper critical dimension of the Anderson transition is infinite~\cite{Castellani1986, tarquini2016anderson, mirlin1994distribution, baroni2024corrections}, the Bethe lattice does not exhibit the standard scaling behavior observed on Euclidean lattices and predicted by scaling arguments~\cite{abrahams1979scaling}. Instead, it represents a highly singular limit characterized by an unconventional ``hybrid'' or mixed transition. The inverse participation ratio (IPR) exhibits a discontinuous jump from $\mathcal{O}(1/N)$ to $\mathcal{O}(1)$ at the transition, while the correlation volume $\Lambda_c$ diverges via an essential singularity.

Importantly, the significance of the Bethe lattice extends far beyond its role as a toy model for non-interacting disordered systems; it has emerged as a cornerstone paradigm for understanding many-body localization. The Fock space (or configuration space) of an interacting quantum many-body system possesses an intrinsically high-dimensional and hierarchically organized geometry which, despite the presence of many short loops, shares key features with a Bethe lattice~\cite{altshuler1997quasiparticle}. In this context, interactions act as hopping matrix elements between different unperturbed many-body states. Consequently, the single-particle Anderson transition on the Bethe lattice provides a foundational mean-field proxy for the MBL transition in Fock space, shedding light on how a closed macroscopic quantum system can fail to thermalize and act as its own bath. While loops and correlations in true many-body Fock spaces introduce additional subtleties, the rigorous and numerical framework presented in these notes underscores the Bethe lattice as an indispensable reference paradigm for shaping our current and future understanding of quantum statistical mechanics~\cite{tikhonov2021anderson,tikhonov2021eigenstate,tarzia2020many,de2013ergodicity,biroli2017delocalized,biroli2020anomalous,logan2019many,garcia2022critical,herre2023ergodicity}.

Finally, we discussed the tight and profound relation between this problem and directed polymers in random media on the Bethe lattice~\cite{derrida1988polymers}. This analogy provides a transparent interpretation of localization in terms of a freezing transition~\cite{lemarie2019glassy,biroli2012difference,sonner2017multifractality,tikhonov2016fractality,monthus2008anderson,monthus2011anderson,chakrabarti2022traveling,biroli2020anomalous,swain2026two}: in the localized phase, propagation is dominated by a finite number of rare resonant paths, whereas in the delocalized phase an exponential number of paths contributes to the escape of the particle to infinity.

\emph{Main results at a glance.} We conclude this summary by collecting the main messages of these notes, which are also summarized quantitatively in Table~\ref{tab:main_results}:
\begin{itemize}
    \item The transition is not encoded in the average density of states, which is smooth on both sides, but in the full probability distribution of the local density of states, $\rho_i(\lambda) \propto \operatorname{Im} G_{ii}$. Its typical value vanishes in the localized phase and is finite in the delocalized one, while its average stays finite in both phases.
    \item In the localized phase, $\operatorname{Im} G_{ii}$ is dominated by rare resonances: its typical value is proportional to the infinitesimal broadening $\eta$. The marginal distribution of $\operatorname{Im} G_{ii}$ always has a tail with exponent $3/2$, due to local resonances, while its distribution at fixed $\operatorname{Re} G_{ii}$ has power-law tails with a non-universal exponent $1+\beta$, due to resonances transmitted along the tree, where $\beta \to 1/2$ at the transition.
    \item In the delocalized phase, the role of $\eta$ is played by a self-generated energy scale $\Lambda_c^{-1}$, where the correlation volume $\Lambda_c$ is the volume of the resonant clusters out of which extended eigenstates are built. $\Lambda_c$ diverges exponentially at the transition.
    \item The critical behavior is controlled by a single linear integral operator, whose largest eigenvalue $\lambda_\beta$ describes how resonances propagate along the branches of the tree. It yields a localization length exponent $\nu_{\rm loc} = 1$ on the localized side, an essential singularity on the delocalized side, and a discontinuous IPR at the transition.
    \item Localization can be interpreted as the freezing transition of a directed polymer on the tree: in the delocalized phase a particle escapes through an exponential number of paths, while in the localized phase only a few rare resonant paths survive, and even these cannot carry the particle to infinity.
\end{itemize}

\begin{table}[htbp]
\centering
\small
\begin{tabular}{|p{3.3cm}|p{4.4cm}|p{5cm}|}
\hline
 & \textbf{Delocalized side} ($W<W_c$) & \textbf{Localized side} ($W>W_c$) \\
\hline
Typical $\operatorname{Im} G_{ii}$ (as $\eta \to 0^+$) & finite, $\sim \Lambda_c^{-1}$ & $\propto \eta \to 0$ \\
\hline
Tail of $P(\operatorname{Im}G)$ & $(\operatorname{Im}G)^{-3/2}$, cut off at $\Lambda_c$ & marginal: $(\operatorname{Im}G)^{-3/2}$, cut off at $\eta^{-1}$; at fixed $\operatorname{Re}G$: $(\operatorname{Im}G)^{-(1+\beta)}$, with $\beta - \frac12 \sim \sqrt{W-W_c}$ \\
\hline
Characteristic scale & $\ln \Lambda_c \sim (W_c - W)^{-1/2}$ & $\xi \sim (W-W_c)^{-1}$ \\
\hline
Diffusion coefficient & $D \propto \Lambda_c^{-1}$ & $D=0$ \\
\hline
IPR $I_2$ & $\sim \Lambda_c/N \to 0$ & $O(1)$, with $I_2 - I_2^{(c)} \sim \sqrt{W-W_c}$ \\
\hline
Directed-polymer picture & exponential number of paths & frozen: few resonant paths \\
\hline
\end{tabular}
\caption{The main results for the Anderson transition on the Bethe lattice at a glance. They are derived in Sec.~\ref{sec:critical} (see Eqs.~\eqref{eq:xiloc}, \eqref{eq:beta}, \eqref{eq:Lambdac}, and~\eqref{eq:IPRsingularity}) and in Sec.~\ref{sec:connection_DP}.}
\label{tab:main_results}
\end{table}

\subsection{Perspectives}

Anderson localization in general, and Anderson localization on tree-like graphs in particular, remain very active fields of research and continue to reveal new facets and subtleties. Several important directions remain open for future research, which we outline below.

A central open challenge is to sharpen the quantitative analogy between Anderson localization on the Bethe lattice and many-body localization~\cite{tikhonov2021anderson,tikhonov2021eigenstate,tarzia2020many,de2013ergodicity,biroli2017delocalized,biroli2020anomalous,logan2019many,garcia2022critical,herre2023ergodicity}. This requires understanding the role of correlations and loops, which are absent on the Bethe lattice but ubiquitous in realistic many-body Hilbert spaces~\cite{roy2020fock, roy2020localization, scoquart2024role,logan2025multifractality}. Several efforts have been made in this direction. One approach consists in developing a systematic perturbative loop expansion, which yields corrections to the Bethe lattice solution in the form of topological loop diagrams contributing to single-point observables. Applied to the metallic side of the transition~\cite{baroni2024corrections}, this expansion reveals that loop corrections to Bethe lattice results are extremely large: they diverge exponentially fast upon approaching the critical point. This behavior stands in striking contrast with conventional phase transitions, where corrections to mean-field theory diverge only algebraically. The physical consequences are profound, as this implies that the exotic critical behavior found on the Bethe lattice is destroyed by loop corrections in any finite-dimensional system, providing quantitative support for the long-standing conjecture that the upper critical dimension for Anderson localization is infinite.

The natural---and considerably more ambitious---next step would be to compute the true non-mean-field critical exponents by resumming the diverging series of loop diagrams, in the spirit of field-theoretic renormalization group approaches~\cite{angelini2025bethe}. The main technical obstacle here is the presence of essential singularities at the critical point, which prevents a straightforward perturbative resummation. A promising route may be to organize the expansion in powers of the correlation volume $\Lambda_c$, rather than in powers of a conventional small parameter. Developing such a framework constitutes one of the most challenging and rewarding directions for future work.

A complementary and (possibly) technically more tractable direction is to find a systematic way to incorporate (at least approximately) the off-diagonal elements of the cavity Green's function in Eq.~\eqref{eq:Gj_block}, which encode the contributions of loops to the propagator. Accounting for these terms would allow one to go beyond the strict tree-level approximation while retaining much of the analytical tractability of the cavity method, and could provide important insights into how short loops modify the critical properties of the transition.

A new and particularly promising avenue is the study of Anderson localization on hierarchical graphs, which generalize the Bethe lattice by incorporating a finite density of local loops of tunable length~\cite{prado2026anderson}. These structures interpolate naturally between the loop-free Bethe lattice and more realistic finite-dimensional geometries, while remaining amenable to analytical treatment. They also provide a natural framework for introducing spatial correlations in the random on-site energies and/or hopping amplitudes, making them ideal model systems for studying the interplay between loops, correlations, and localization in a controlled setting. Preliminary results suggest that local loops enhance resonant processes, thereby lowering the critical disorder strength as their number and size increase. Simultaneously, loops promote local hybridization, leading to a spatial broadening of localized eigenstates. Taken together, these findings indicate that local loops are a crucial structural ingredient of realistic single-particle representations of many-body Hilbert spaces, and that hierarchical graphs may serve as a powerful bridge between the analytically solvable Bethe lattice limit and the full complexity of MBL.

Another set of fundamental open questions concerns what happens when one goes beyond the case of random graphs with fixed connectivity and constant hopping amplitudes. Recent works have addressed these issues and highlighted a number of unexpected results, considering Erd\H{o}s-Rényi graphs with fluctuating degree distributions~\cite{tarzia2022fully} and/or random hopping amplitudes~\cite{cugliandolo2024multifractal,Tapias2023,Silva2025spectral}. Broadly speaking, these results suggest that strong fluctuations in the graph topology---arising from rare nodes with much larger-than-typical degree and/or rare bonds with particularly small hopping amplitudes---can have a significant impact on the spectral properties of the system. In particular, they appear to induce a degree of multifractality on the delocalized side of the phase diagram. Understanding these phenomena more deeply, and developing a comprehensive framework that goes beyond these preliminary results, is certainly an important direction for future research.

\section*{Acknowledgements}
This review originated from a series of lecture notes delivered by one of the authors at the Oropa Summer School on ``Fundamental Problems in Statistical Physics XVI'', which took place in July 2025. First and foremost, the authors would like to warmly thank the organizers of the school, A. Gambassi, V. Ros, and T. Schilling for providing the opportunity to deliver these lectures and for encouraging the publication of this review, which would certainly not exist without their support. 

The material presented in this work is the fruit of numerous enlightening and stimulating discussions with friends and collaborators over the past few years. These illuminating exchanges were instrumental to the development of this review. In particular, the authors would like to express their deepest gratitude to G.~Biroli, L.~Cugliandolo, A. K. Hartmann, G.~Lemarié, A.~D. Mirlin, G. A. Miranda, T.~Rizzo, M. Schir\`o, G. Semerjian, L. Tonetti, and D.~Venturelli.

\bibliographystyle{iopart-num}
\bibliography{Bibliography} 

@article{anderson1958absence,
  title = {Absence of Diffusion in Certain Random Lattices},
  author = {Anderson, P. W.},
  journal = {Phys. Rev.},
  volume = {109},
  issue = {5},
  pages = {1492--1505},
  numpages = {0},
  year = {1958},
  month = {Mar},
  publisher = {American Physical Society},
  doi = {10.1103/PhysRev.109.1492},
  url = {https://link.aps.org/doi/10.1103/PhysRev.109.1492}
}

@article{lee1985disordered,
  title = {Disordered electronic systems},
  author = {Lee, Patrick A. and Ramakrishnan, T. V.},
  journal = {Rev. Mod. Phys.},
  volume = {57},
  issue = {2},
  pages = {287--337},
  numpages = {0},
  year = {1985},
  month = {Apr},
  publisher = {American Physical Society},
  doi = {10.1103/RevModPhys.57.287},
  url = {https://link.aps.org/doi/10.1103/RevModPhys.57.287}
}

@article{evers2008anderson,
  title = {Anderson transitions},
  author = {Evers, Ferdinand and Mirlin, Alexander D.},
  journal = {Rev. Mod. Phys.},
  volume = {80},
  issue = {4},
  pages = {1355--1417},
  numpages = {0},
  year = {2008},
  month = {Oct},
  publisher = {American Physical Society},
  doi = {10.1103/RevModPhys.80.1355},
  url = {https://link.aps.org/doi/10.1103/RevModPhys.80.1355}
}

@article{lagendijk2009fifty,
  title={{Fifty years of Anderson localization}},
  author={Lagendijk, Ad and Tiggelen, Bart van and Wiersma, Diederik S},
  journal={Physics today},
  volume={62},
  number={8},
  pages={24--29},
  year={2009},
  publisher={AIP Publishing},
  url = {https://physicstoday.aip.org/features/fifty-years-of-anderson-localization},
}

@article{gornyi2005interacting,
  title = {Interacting Electrons in Disordered Wires: Anderson Localization and Low-$T$ Transport},
  author = {Gornyi, I. V. and Mirlin, A. D. and Polyakov, D. G.},
  journal = {Phys. Rev. Lett.},
  volume = {95},
  issue = {20},
  pages = {206603},
  numpages = {4},
  year = {2005},
  month = {Nov},
  publisher = {American Physical Society},
  doi = {10.1103/PhysRevLett.95.206603},
  url = {https://link.aps.org/doi/10.1103/PhysRevLett.95.206603}
}

@article{basko2006metal,
  title = {Metal–insulator transition in a weakly interacting many-electron system with localized single-particle states},
  volume = {321},
  ISSN = {0003-4916},
  url = {http://dx.doi.org/10.1016/j.aop.2005.11.014},
  number = {5},
  journal = {Ann. Phys. (N. Y.)},
  publisher = {Elsevier BV},
  author = {Basko,  D.M. and Aleiner,  I.L. and Altshuler,  B.L.},
  year = {2006},
  month = may,
  pages = {1126–1205}
}

@article{Nandkishore2015,
  title = {Many-Body Localization and Thermalization in Quantum Statistical Mechanics},
  volume = {6},
  issn = {1947-5462},
  url = {https://doi.org/10.1146/annurev-conmatphys-031214-014726},
  number = {1},
  journal = {Annu. Rev. Condens. Mat. Phys.},
  publisher = {Annual Reviews},
  author = {Nandkishore,  Rahul and Huse,  David A.},
  year = {2015},
  pages = {15–38}
}

@article{Huse2014,
  title = {Phenomenology of fully many-body-localized systems},
  volume = {90},
  issn = {1550-235X},
  url = {http://dx.doi.org/10.1103/PhysRevB.90.174202},
  number = {17},
  journal = {Phys. Rev. B},
  publisher = {American Physical Society (APS)},
  author = {Huse,  David A. and Nandkishore,  Rahul and Oganesyan,  Vadim},
  year = {2014}
}

@article{biroli_large-deviation_2024,
  title = {Large-deviation analysis of rare resonances for the many-body localization transition},
  volume = {110},
  issn = {2469-9950, 2469-9969},
  url = {https://link.aps.org/doi/10.1103/PhysRevB.110.014205},
  
  number = {1},
  urldate = {2025-03-07},
  journal = {Phys. Rev.  B},
  author = {Biroli, Giulio and Hartmann, Alexander K. and Tarzia, Marco},
  year = {2024}
}

@article{abrahams1979scaling,
  title={Scaling theory of localization: Absence of quantum diffusion in two dimensions},
  author={Abrahams, Elihu and Anderson, Philip W and Licciardello, Donald C and Ramakrishnan, Tiruppattur V},
  journal={Physical Review Letters},
  volume={42},
  number={10},
  pages={673},
  year={1979},
  publisher={APS}
}

@article{wormald1999models,
  title={Models of random regular graphs},
  author={Wormald, Nicholas C and others},
  journal={London mathematical society lecture note series},
  pages={239--298},
  year={1999},
  publisher={Cambridge University Press}
}

@article{tarquini2017critical,
  title={Critical properties of the Anderson localization transition and the high-dimensional limit},
  author={Tarquini, Elena and Biroli, Giulio and Tarzia, Marco},
  journal={Physical Review B},
  volume={95},
  number={9},
  pages={094204},
  year={2017},
  publisher={APS}
}

@article{abou1973selfconsistent,
  doi = {10.1088/0022-3719/6/10/009},
  url = {https://doi.org/10.1088/0022-3719/6/10/009},
  year = {1973},
  month = {may},
  volume = {6},
  number = {10},
  pages = {1734},
  author = {R Abou-Chacra and D J Thouless and P W Anderson},
  title = {A selfconsistent theory of localization},
  journal = {Journal of Physics C: Solid State Physics},
}

@incollection{braunstein2023cavity,
  title={The cavity method: from exact solutions to algorithms},
  author={Braunstein, Alfredo and Semerjian, Guilhem},
  booktitle={Spin Glass Theory and Far Beyond: Replica Symmetry Breaking After 40 Years},
  pages={375--387},
  year={2023},
  publisher={World Scientific},
  url={https://arxiv.org/abs/2209.11499}
}

@book{mezard2009information,
  title={Information, physics, and computation},
  author={Mezard, Marc and Montanari, Andrea},
  year={2009},
  publisher={Oxford University Press},
  doi = {https://doi.org/10.1093/acprof:oso/9780198570837.001.0001},
  note = {Chap. 14, Sec. 14.6.4}
}

@article{mezard2001bethe,
  title={The Bethe lattice spin glass revisited},
   volume={20},
   ISSN={1434-6028},
   url={http://dx.doi.org/10.1007/PL00011099},
   DOI={10.1007/pl00011099},
   number={2},
   journal={The European Physical Journal B},
   publisher={Springer Science and Business Media LLC},
   author={Mézard, M. and Parisi, G.},
   year={2001},
   month=mar, pages={217–233}
}

@article{bapst2014largekAL,
   title={The large connectivity limit of the Anderson model on tree graphs},
   volume={55},
   ISSN={1089-7658},
   url={http://dx.doi.org/10.1063/1.4894055},
   DOI={10.1063/1.4894055},
   number={9},
   journal={Journal of Mathematical Physics},
   publisher={AIP Publishing},
   author={Bapst, Victor},
   year={2014},
   month=sep
}

@article{abanin2019MBLcolloquium,
  title = {Colloquium: Many-body localization, thermalization, and entanglement},
  author = {Abanin, Dmitry A. and Altman, Ehud and Bloch, Immanuel and Serbyn, Maksym},
  journal = {Rev. Mod. Phys.},
  volume = {91},
  issue = {2},
  pages = {021001},
  numpages = {26},
  year = {2019},
  month = {May},
  publisher = {American Physical Society},
  doi = {10.1103/RevModPhys.91.021001},
  url = {https://link.aps.org/doi/10.1103/RevModPhys.91.021001}
}

@article{alet2018many,
   title={Many-body localization: An introduction and selected topics},
   volume={19},
   ISSN={1878-1535},
   url={http://dx.doi.org/10.1016/j.crhy.2018.03.003},
   DOI={10.1016/j.crhy.2018.03.003},
   number={6},
   journal={Comptes Rendus. Physique},
   publisher={MathDoc/Centre Mersenne},
   author={Alet, Fabien and Laflorencie, Nicolas},
   year={2018},
   month=apr, pages={498–525}
}

@article{sierant2025many,
   title={Many-body localization in the age of classical computing*},
   volume={88},
   ISSN={1361-6633},
   url={http://dx.doi.org/10.1088/1361-6633/ad9756},
   DOI={10.1088/1361-6633/ad9756},
   number={2},
   journal={Reports on Progress in Physics},
   publisher={IOP Publishing},
   author={Sierant, Piotr and Lewenstein, Maciej and Scardicchio, Antonello and Vidmar, Lev and Zakrzewski, Jakub},
   year={2025},
   month=jan, pages={026502}
}

@article{imbrie2016diagonalization,
  title = {Diagonalization and Many-Body Localization for a Disordered Quantum Spin Chain},
  author = {Imbrie, John Z.},
  journal = {Phys. Rev. Lett.},
  volume = {117},
  issue = {2},
  pages = {027201},
  numpages = {5},
  year = {2016},
  month = {Jul},
  publisher = {American Physical Society},
  doi = {10.1103/PhysRevLett.117.027201},
  url = {https://link.aps.org/doi/10.1103/PhysRevLett.117.027201}
}

@article{tikhonov2021anderson,
   title={From Anderson localization on random regular graphs to many-body localization},
   volume={435},
   ISSN={0003-4916},
   url={http://dx.doi.org/10.1016/j.aop.2021.168525},
   DOI={10.1016/j.aop.2021.168525},
   journal={Annals of Physics},
   publisher={Elsevier BV},
   author={Tikhonov, K.S. and Mirlin, A.D.},
   year={2021},
   month=dec, pages={168525}
}

@article{altshuler1997quasiparticle,
  title={Quasiparticle Lifetime in a Finite System: A Nonperturbative Approach},
   volume={78},
   ISSN={1079-7114},
   url={http://dx.doi.org/10.1103/PhysRevLett.78.2803},
   DOI={10.1103/physrevlett.78.2803},
   number={14},
   journal={Physical Review Letters},
   publisher={American Physical Society (APS)},
   author={Altshuler, Boris L. and Gefen, Yuval and Kamenev, Alex and Levitov, Leonid S.},
   year={1997},
   month=apr, pages={2803–2806} 
}

@article{biroli2017delocalized,
  title = {Delocalized glassy dynamics and many-body localization},
  author = {Biroli, G. and Tarzia, M.},
  journal = {Phys. Rev. B},
  volume = {96},
  issue = {20},
  pages = {201114(R)},
  numpages = {5},
  year = {2017},
  month = {Nov},
  publisher = {American Physical Society},
  doi = {10.1103/PhysRevB.96.201114},
  url = {https://link.aps.org/doi/10.1103/PhysRevB.96.201114}
}

@article{biroli2020anomalous,
  title = {Anomalous dynamics on the ergodic side of the many-body localization transition and the glassy phase of directed polymers in random media},
  author = {Biroli, G. and Tarzia, M.},
  journal = {Phys. Rev. B},
  volume = {102},
  issue = {6},
  pages = {064211},
  numpages = {17},
  year = {2020},
  month = {Aug},
  publisher = {American Physical Society},
  doi = {10.1103/PhysRevB.102.064211},
  url = {https://link.aps.org/doi/10.1103/PhysRevB.102.064211}
}

@article{logan2019many,
  title={Many-body localization in Fock space: A local perspective},
   volume={99},
   ISSN={2469-9969},
   url={http://dx.doi.org/10.1103/PhysRevB.99.045131},
   DOI={10.1103/physrevb.99.045131},
   number={4},
   journal={Physical Review B},
   publisher={American Physical Society (APS)},
   author={Logan, David E. and Welsh, Staszek},
   year={2019},
   month=jan
}

@article{de2013ergodicity,
  title={Ergodicity breaking in a model showing many-body localization},
   volume={101},
   ISSN={1286-4854},
   url={http://dx.doi.org/10.1209/0295-5075/101/37003},
   DOI={10.1209/0295-5075/101/37003},
   number={3},
   journal={EPL (Europhysics Letters)},
   publisher={IOP Publishing},
   author={De Luca, A. and Scardicchio, A.},
   year={2013},
   month=feb, pages={37003}
}

@article{garcia2022critical,
  title = {{Critical properties of the Anderson transition on random graphs: Two-parameter scaling theory, Kosterlitz-Thouless type flow, and many-body localization}},
  author = {Garc\'{\i}a-Mata, I. and Martin, J. and Giraud, O. and Georgeot, B. and Dubertrand, R. and Lemari\'e, G.},
  journal = {Phys. Rev. B},
  volume = {106},
  issue = {21},
  pages = {214202},
  numpages = {33},
  year = {2022},
  month = {Dec},
  publisher = {American Physical Society},
  doi = {10.1103/PhysRevB.106.214202},
  url = {https://link.aps.org/doi/10.1103/PhysRevB.106.214202}
}

@article{roy2020localization,
  title = {Localization on Certain Graphs with Strongly Correlated Disorder},
  author = {Roy, Sthitadhi and Logan, David E.},
  journal = {Phys. Rev. Lett.},
  volume = {125},
  issue = {25},
  pages = {250402},
  numpages = {6},
  year = {2020},
  month = {Dec},
  publisher = {American Physical Society},
  doi = {10.1103/PhysRevLett.125.250402},
  url = {https://link.aps.org/doi/10.1103/PhysRevLett.125.250402}
}

@article{roy2020fock,
  title = {Fock-space correlations and the origins of many-body localization},
  author = {Roy, Sthitadhi and Logan, David E.},
  journal = {Phys. Rev. B},
  volume = {101},
  issue = {13},
  pages = {134202},
  numpages = {23},
  year = {2020},
  month = {Apr},
  publisher = {American Physical Society},
  doi = {10.1103/PhysRevB.101.134202},
  url = {https://link.aps.org/doi/10.1103/PhysRevB.101.134202}
}

@article{scoquart2024role,
  title = {Role of Fock-space correlations in many-body localization},
  author = {Scoquart, Thibault and Gornyi, Igor V. and Mirlin, Alexander D.},
  journal = {Phys. Rev. B},
  volume = {109},
  issue = {21},
  pages = {214203},
  numpages = {26},
  year = {2024},
  month = {Jun},
  publisher = {American Physical Society},
  doi = {10.1103/PhysRevB.109.214203},
  url = {https://link.aps.org/doi/10.1103/PhysRevB.109.214203}
}

@article{herre2023ergodicity,
   title={Ergodicity-to-localization transition on random regular graphs with large connectivity and in many-body quantum dots},
   volume={108},
   ISSN={2469-9969},
   url={http://dx.doi.org/10.1103/PhysRevB.108.014203},
   DOI={10.1103/physrevb.108.014203},
   number={1},
   journal={Physical Review B},
   publisher={American Physical Society (APS)},
   author={Herre, Jan-Niklas and Karcher, Jonas F. and Tikhonov, Konstantin S. and Mirlin, Alexander D.},
   year={2023},
   month=jul
}

@misc{biroli2012difference,
      title={Difference between level statistics, ergodicity and localization transitions on the Bethe lattice}, 
      author={G. Biroli and A. C. Ribeiro-Teixeira and M. Tarzia},
      year={2012},
      eprint={1211.7334},
      archivePrefix={arXiv},
      primaryClass={cond-mat.dis-nn},
      url={https://arxiv.org/abs/1211.7334}, 
}

@misc{biroli2018delocalization,
  title={Delocalization and ergodicity of the Anderson model on Bethe lattices}, 
  author={Giulio Biroli and Marco Tarzia},
  year={2018},
  eprint={1810.07545},
  archivePrefix={arXiv},
  primaryClass={cond-mat.dis-nn},
  url={https://arxiv.org/abs/1810.07545}
}

@article{tikhonov2021eigenstate,
  title = {Eigenstate correlations around the many-body localization transition},
  author = {Tikhonov, K. S. and Mirlin, A. D.},
  journal = {Phys. Rev. B},
  volume = {103},
  issue = {6},
  pages = {064204},
  numpages = {13},
  year = {2021},
  month = {Feb},
  publisher = {American Physical Society},
  doi = {10.1103/PhysRevB.103.064204},
  url = {https://link.aps.org/doi/10.1103/PhysRevB.103.064204}
}

@article{biroli2010anderson,
   title={Anderson Model on Bethe Lattices: Density of States, Localization Properties and Isolated Eigenvalue},
   volume={184},
   ISSN={0375-9687},
   url={http://dx.doi.org/10.1143/PTPS.184.187},
   DOI={10.1143/ptps.184.187},
   journal={Progress of Theoretical Physics Supplement},
   publisher={Oxford University Press (OUP)},
   author={Biroli, Giulio and Semerjian, Guilhem and Tarzia, Marco},
   year={2010},
   pages={187–199}
}

@article{rizzo2024localized,
  title = {Localized phase of the Anderson model on the Bethe lattice},
  author = {Rizzo, Tommaso and Tarzia, Marco},
  journal = {Phys. Rev. B},
  volume = {110},
  issue = {18},
  pages = {184210},
  numpages = {14},
  year = {2024},
  month = {Nov},
  publisher = {American Physical Society},
  doi = {10.1103/PhysRevB.110.184210},
  url = {https://link.aps.org/doi/10.1103/PhysRevB.110.184210}
}

@article{tarzia2020many,
   title={Many-body localization transition in Hilbert space},
   volume={102},
   ISSN={2469-9969},
   url={http://dx.doi.org/10.1103/PhysRevB.102.014208},
   DOI={10.1103/physrevb.102.014208},
   number={1},
   journal={Physical Review B},
   publisher={American Physical Society (APS)},
   author={Tarzia, M.},
   year={2020},
   month=jul 
}

@article{tikhonov2019statistics,
  title = {Statistics of eigenstates near the localization transition on random regular graphs},
  author = {Tikhonov, K. S. and Mirlin, A. D.},
  journal = {Phys. Rev. B},
  volume = {99},
  issue = {2},
  pages = {024202},
  numpages = {25},
  year = {2019},
  month = {Jan},
  publisher = {American Physical Society},
  doi = {10.1103/PhysRevB.99.024202},
  url = {https://link.aps.org/doi/10.1103/PhysRevB.99.024202}
}

@article{tikhonov2019critical,
  title = {Critical behavior at the localization transition on random regular graphs},
  author = {Tikhonov, K. S. and Mirlin, A. D.},
  journal = {Phys. Rev. B},
  volume = {99},
  issue = {21},
  pages = {214202},
  numpages = {10},
  year = {2019},
  month = {Jun},
  publisher = {American Physical Society},
  doi = {10.1103/PhysRevB.99.214202},
  url = {https://link.aps.org/doi/10.1103/PhysRevB.99.214202}
}

@article{wigner1958distribution,
  ISSN = {0003486X, 19398980},
  URL = {http://www.jstor.org/stable/1970008},
  author = {Eugene P. Wigner},
  journal = {Annals of Mathematics},
  number = {2},
  pages = {325--327},
  publisher = {[Annals of Mathematics, Trustees of Princeton University on Behalf of the Annals of Mathematics, Mathematics Department, Princeton University]},
  title = {On the Distribution of the Roots of Certain Symmetric Matrices},
  urldate = {2026-03-18},
  volume = {67},
  year = {1958}
}

@book{potters2020first,
  title={A first course in random matrix theory: for physicists, engineers and data scientists},
  author={Potters, Marc and Bouchaud, Jean-Philippe},
  year={2020},
  publisher={Cambridge University Press},
  doi = {https://doi.org/10.1017/9781108768900},
  note = {Chap. 4}
}

@article{mirlin1991localization,
  title={{Localization transition in the Anderson model on the Bethe lattice: spontaneous symmetry breaking and correlation functions}},
  author={Mirlin, Alexander D and Fyodorov, Yan V},
  journal={Nuclear Physics B},
  volume={366},
  number={3},
  pages={507--532},
  year={1991},
  publisher={Elsevier},
  doi = {https://doi.org/10.1016/0550-3213(91)90028-V},
  url = {https://www.sciencedirect.com/science/article/pii/055032139190028V},
}

@article{mirlin1994distribution,
  title = {Distribution of local densities of states, order parameter function, and critical behavior near the Anderson transition},
  author = {Mirlin, Alexander D. and Fyodorov, Yan V.},
  journal = {Phys. Rev. Lett.},
  volume = {72},
  issue = {4},
  pages = {526--529},
  numpages = {0},
  year = {1994},
  month = {Jan},
  publisher = {American Physical Society},
  doi = {10.1103/PhysRevLett.72.526},
  url = {https://link.aps.org/doi/10.1103/PhysRevLett.72.526}
}

@article{efetov1985anderson,
  title={{Anderson metal-insulator transition in a system of metal granules: existence of a minimum metallic conductivity and a maximum dielectric constant}},
  author={Efetov, KB},
  journal={Sov. Phys. JETP},
  volume={61},
  pages={606--617},
  year={1985},
  url = {https://jetp.ras.ru/cgi-bin/dn/e_061_03_0606.pdf}
}

@article{efetov1987density,
  title={{Density--Density Correlator in a Model of a Disordered Metal on a Bethe Lattice}},
  author={Efetov, KB},
  journal={Sov. Phys. JETP},
  volume={65},
  pages={360--370},
  year={1987},
  url= {https://jetp.ras.ru/cgi-bin/dn/e_065_02_0360.pdf}
}

@article{efetov1987anderson,
  title={{Anderson transition on a Bethe lattice (the symplectic and orthogonal ensembles)}},
  author={Efetov, KB},
  journal={Sov. Phys. JETP},
  volume={66},
  pages={634--642},
  year={1987},
  url= {https://jetp.ras.ru/cgi-bin/dn/e_066_03_0634.pdf}
}

@article{zirnbauer1986localization,
  title = {Localization transition on the Bethe lattice},
  author = {Zirnbauer, Martin R.},
  journal = {Phys. Rev. B},
  volume = {34},
  issue = {9},
  pages = {6394--6408},
  numpages = {0},
  year = {1986},
  month = {Nov},
  publisher = {American Physical Society},
  doi = {10.1103/PhysRevB.34.6394},
  url = {https://link.aps.org/doi/10.1103/PhysRevB.34.6394}
}

@article{zirnbauer1986anderson,
  title = {Anderson localization and non-linear sigma model with graded symmetry},
  journal = {Nuclear Physics B},
  volume = {265},
  number = {2},
  pages = {375-408},
  year = {1986},
  issn = {0550-3213},
  doi = {https://doi.org/10.1016/0550-3213(86)90316-0},
  url = {https://www.sciencedirect.com/science/article/pii/0550321386903160},
  author = {Martin R. Zirnbauer},
}

@article{verbaarschot1988graded,
  title = {Graded symmetry and Anderson localization on the Bethe lattice for time-reversal invariant systems},
  journal = {Nuclear Physics B},
  volume = {300},
  pages = {263-288},
  year = {1988},
  issn = {0550-3213},
  doi = {https://doi.org/10.1016/0550-3213(88)90598-6},
  url = {https://www.sciencedirect.com/science/article/pii/0550321388905986},
  author = {J.J.M. Verbaarschot},
}

@article{mirlin1991universality,
  title={Universality of level correlation function of sparse random matrices},
  author={Mirlin, AD and Fyodorov, Yan V},
  journal={Journal of Physics A: Mathematical and General},
  volume={24},
  number={10},
  pages={2273},
  year={1991},
  publisher={IOP Publishing},
  doi = {10.1088/0305-4470/24/10/016},
  url = {https://doi.org/10.1088/0305-4470/24/10/016},
}

@article{fyodorov1991localization,
  title = {Localization in ensemble of sparse random matrices},
  author = {Fyodorov, Yan V. and Mirlin, Alexander D.},
  journal = {Phys. Rev. Lett.},
  volume = {67},
  issue = {15},
  pages = {2049--2052},
  numpages = {0},
  year = {1991},
  month = {Oct},
  publisher = {American Physical Society},
  doi = {10.1103/PhysRevLett.67.2049},
  url = {https://link.aps.org/doi/10.1103/PhysRevLett.67.2049}
}

@article{mirlin1994statistical,
  title={{Statistical properties of one-point Green functions in disordered systems and critical behavior near the Anderson transition}},
  author={Mirlin, Alexander D and Fyodorov, Yan V},
  journal={Journal de Physique I},
  volume={4},
  number={5},
  pages={655--673},
  year={1994},
  publisher={EDP Sciences},
  doi = {https://doi.org/10.1051/jp1:1994168}
}

@article{baroni2024corrections,
  title = {Corrections to the Bethe lattice solution of Anderson localization},
  author = {Baroni, Matilde and Lorenzana, Giulia Garcia and Rizzo, Tommaso and Tarzia, Marco},
  journal = {Phys. Rev. B},
  volume = {109},
  issue = {17},
  pages = {174216},
  numpages = {23},
  year = {2024},
  month = {May},
  publisher = {American Physical Society},
  doi = {10.1103/PhysRevB.109.174216},
  url = {https://link.aps.org/doi/10.1103/PhysRevB.109.174216}
}

@article{biroli2022critical,
  title = {Critical behavior of the Anderson model on the Bethe lattice via a large-deviation approach},
  author = {Biroli, Giulio and Hartmann, Alexander K. and Tarzia, Marco},
  journal = {Phys. Rev. B},
  volume = {105},
  issue = {9},
  pages = {094202},
  numpages = {11},
  year = {2022},
  month = {Mar},
  publisher = {American Physical Society},
  doi = {10.1103/PhysRevB.105.094202},
  url = {https://link.aps.org/doi/10.1103/PhysRevB.105.094202}
}

@article{parisi2019anderson,
  title={Anderson transition on the Bethe lattice: an approach with real energies},
   volume={53},
   ISSN={1751-8121},
   url={http://dx.doi.org/10.1088/1751-8121/ab56e8},
   DOI={10.1088/1751-8121/ab56e8},
   number={1},
   journal={Journal of Physics A: Mathematical and Theoretical},
   publisher={IOP Publishing},
   author={Parisi, Giorgio and Pascazio, Saverio and Pietracaprina, Francesca and Ros, Valentina and Scardicchio, Antonello},
   year={2019},
   month=dec, pages={014003}
}

@article{tarquini2016level,
  title = {Level Statistics and Localization Transitions of L\'evy Matrices},
  author = {Tarquini, E. and Biroli, G. and Tarzia, M.},
  journal = {Phys. Rev. Lett.},
  volume = {116},
  issue = {1},
  pages = {010601},
  numpages = {5},
  year = {2016},
  month = {Jan},
  publisher = {American Physical Society},
  doi = {10.1103/PhysRevLett.116.010601},
  url = {https://link.aps.org/doi/10.1103/PhysRevLett.116.010601}
}

@misc{tonetti2025testing,
  title={Testing the Localization Landscape Theory on the Bethe Lattice}, 
  author={Lorenzo Tonetti and Leticia F. Cugliandolo and Marco Tarzia},
  year={2025},
  eprint={2512.04037},
  archivePrefix={arXiv},
  primaryClass={cond-mat.dis-nn},
  url={https://arxiv.org/abs/2512.04037}, 
}

@misc{tonetti2026geometrylocalizationprobinglocalization,
      title={Geometry and localization: Probing Localization Landscape Theory on the Bethe Lattice}, 
      author={Lorenzo Tonetti and Leticia F. Cugliandolo and Marco Tarzia},
      year={2026},
      eprint={2605.29745},
      archivePrefix={arXiv},
      primaryClass={cond-mat.dis-nn},
      url={https://arxiv.org/abs/2605.29745}, 
}

@article{swain2026two,
  title={Two-dimensional Anderson localization and KPZ subuniversality classes: Sensitivity to boundary conditions and insensitivity to symmetry classes},
  author={Swain, Nyayabanta and Adam, Shaffique and Lemari{\'e}, Gabriel},
  journal={Physical Review Research},
  volume={8},
  number={1},
  pages={013038},
  year={2026},
  publisher={APS}
}

@article{logan2025multifractality,
  title = {Multifractality in high-dimensional graphs induced by correlated radial disorder},
  author = {Logan, David E. and Roy, Sthitadhi},
  journal = {Phys. Rev. B},
  volume = {112},
  issue = {18},
  pages = {184202},
  numpages = {17},
  year = {2025},
  month = {Nov},
  publisher = {American Physical Society},
  doi = {10.1103/5vqd-v4b8},
  url = {https://link.aps.org/doi/10.1103/5vqd-v4b8}
}

@misc{miranda2025largedeviationsmanybodylocalization,
      title={Large deviations in the many-body localization transition: The case of the random-field XXZ chain}, 
      author={Greivin Alfaro Miranda and Fabien Alet and Giulio Biroli and Leticia F. Cugliandolo and Nicolas Laflorencie and Marco Tarzia},
      year={2025},
      eprint={2510.18545},
      archivePrefix={arXiv},
      primaryClass={cond-mat.dis-nn},
      url={https://arxiv.org/abs/2510.18545}, 
}

@article{bloch1928,
  author  = {Bloch, Felix},
  title   = {Über die Quantenmechanik der Elektronen in Kristallgittern},
  journal = {Zeitschrift für Physik},
  volume  = {52},
  pages   = {555--600},
  year    = {1928},
  doi     = {10.1007/BF01339455}
}

@article{Drude1900,
    author = {Drude, P.},
    title = {Zur Elektronentheorie der Metalle},
    journal = {Annalen der Physik},
    volume = {306},
    number = {3},
    pages = {566-613},
    doi = {https://doi.org/10.1002/andp.19003060312},
    url = {https://onlinelibrary.wiley.com/doi/abs/10.1002/andp.19003060312},
    eprint = {https://onlinelibrary.wiley.com/doi/pdf/10.1002/andp.19003060312},
    year = {1900}
}

@article{sommerfeld1927elektronentheorie,
  title={Zur elektronentheorie der metalle},
  author={Sommerfeld, Arnold},
  journal={Naturwissenschaften},
  volume={15},
  number={41},
  pages={825--832},
  year={1927},
  doi = {10.1007/BF01505083}
}

@misc{scardicchio2017perturbation,
      title={Perturbation theory approaches to Anderson and Many-Body Localization: some lecture notes}, 
      author={Antonello Scardicchio and Thimothée Thiery},
      year={2017},
      eprint={1710.01234},
      archivePrefix={arXiv},
      primaryClass={cond-mat.dis-nn},
      url={https://arxiv.org/abs/1710.01234}, 
}

@incollection{pascazio2023anderson,
  title={Anderson localization on the Bethe lattice},
  author={Pascazio, Saverio and Scardicchio, Antonello and Tarzia, Marco},
  booktitle={Spin Glass Theory and Far Beyond: Replica Symmetry Breaking After 40 Years},
  pages={335--352},
  year={2023},
  publisher={World Scientific}
}

@article{Mott1967,
	author = {N.F. Mott},
	doi = {10.1080/00018736700101265},
	eprint = { https://doi.org/10.1080/00018736700101265 },
	journal = {Advances in Physics},
	number = {61},
	pages = {49–144},
	publisher = {Taylor \& Francis},
	title = {Electrons in disordered structures},
	url = {https://doi.org/10.1080/00018736700101265},
	volume = {16},
	year = {1967}
}

@article{slevin1999corrections,
   title={Corrections to Scaling at the Anderson Transition},
   volume={82},
   ISSN={1079-7114},
   url={http://dx.doi.org/10.1103/PhysRevLett.82.382},
   DOI={10.1103/physrevlett.82.382},
   number={2},
   journal={Physical Review Letters},
   publisher={American Physical Society (APS)},
   author={Slevin, Keith and Ohtsuki, Tomi},
   year={1999},
   month=Jan, pages={382–385} 
}

@article{slevin2014critical,
   title={Critical exponent for the Anderson transition in the three-dimensional orthogonal universality class},
   volume={16},
   ISSN={1367-2630},
   url={http://dx.doi.org/10.1088/1367-2630/16/1/015012},
   DOI={10.1088/1367-2630/16/1/015012},
   number={1},
   journal={New Journal of Physics},
   publisher={IOP Publishing},
   author={Slevin, Keith and Ohtsuki, Tomi},
   year={2014},
   month=Jan, pages={015012} 
}

@article{hikami1992localization,
  author  = {Shinobu Hikami},
  title   = {Localization, Nonlinear $\sigma$ Model and String Theory},
  journal = {Progress of Theoretical Physics Supplement},
  volume  = {107},
  pages   = {213--227},
  year    = {1992},
  doi     = {10.1143/PTPS.107.213}
}

@article{wegner1979mobility,
  author  = {F. Wegner},
  title   = {The mobility edge problem: Continuous symmetry and a conjecture},
  journal = {Zeitschrift für Physik B Condensed Matter},
  volume  = {35},
  pages   = {207--210},
  year    = {1979},
  doi     = {10.1007/BF01311391}
}

@article{schafer1980disordered,
  author  = {L. Schäfer and F. Wegner},
  title   = {Disordered system with $n$ orbitals per site: Lagrange formulation, hyperbolic symmetry, and Goldstone modes},
  journal = {Zeitschrift für Physik B Condensed Matter},
  volume  = {38},
  pages   = {113--126},
  year    = {1980},
  doi     = {10.1007/BF01303700}
}

@article{John1984,
  title = {Electromagnetic Absorption in a Disordered Medium near a Photon Mobility Edge},
  author = {John, Sajeev},
  journal = {Phys. Rev. Lett.},
  volume = {53},
  issue = {22},
  pages = {2169--2172},
  numpages = {0},
  year = {1984},
  month = {Nov},
  publisher = {American Physical Society},
  doi = {10.1103/PhysRevLett.53.2169},
  url = {https://link.aps.org/doi/10.1103/PhysRevLett.53.2169}
}

@article{Anderson1985,
	author = {Philip W. Anderson},
	doi = {10.1080/13642818508240619},
	eprint = { https://doi.org/10.1080/13642818508240619 },
	journal = {Philosophical Magazine B},
	number = {3},
	pages = {505–509},
	publisher = {Taylor \& Francis},
	title = {The question of classical localization A theory of white paint?},
	url = {https://doi.org/10.1080/13642818508240619},
	volume = {52},
	year = {1985}
}

@article{billy2008direct,
  title={Direct observation of Anderson localization of matter waves in a controlled disorder},
  author={Billy, Juliette and Josse, Vincent and Zuo, Zhanchun and Bernard, Alain and Hambrecht, Ben and Lugan, Pierre and Cl{\'e}ment, David and Sanchez-Palencia, Laurent and Bouyer, Philippe and Aspect, Alain},
  journal={Nature},
  volume={453},
  number={7197},
  pages={891--894},
  year={2008},
  publisher={Nature Publishing Group},
  doi = {10.1038/nature07000}
}

@article{roati2008anderson,
  title={Anderson localization of a non-interacting Bose--Einstein condensate},
  author={Roati, Giacomo and D’Errico, Chiara and Fallani, Leonardo and Fattori, Marco and Fort, Chiara and Zaccanti, Matteo and Modugno, Giovanni and Modugno, Michele and Inguscio, Massimo},
  journal={Nature},
  volume={453},
  number={7197},
  pages={895--898},
  year={2008},
  publisher={Nature Publishing Group UK London},
  doi={https://doi.org/10.1038/nature07071}
}

@article{kondov2011three,
   title={Three-Dimensional Anderson Localization of Ultracold Matter},
   volume={334},
   ISSN={1095-9203},
   url={http://dx.doi.org/10.1126/science.1209019},
   DOI={10.1126/science.1209019},
   number={6052},
   journal={Science},
   publisher={American Association for the Advancement of Science (AAAS)},
   author={Kondov, S. S. and McGehee, W. R. and Zirbel, J. J. and DeMarco, B.},
   year={2011},
   month=Oct, pages={66–68} 
}

@article{jendrzejewski2012three,
   title={Three-dimensional localization of ultracold atoms in an optical disordered potential},
   volume={8},
   ISSN={1745-2481},
   url={http://dx.doi.org/10.1038/nphys2256},
   DOI={10.1038/nphys2256},
   number={5},
   journal={Nature Physics},
   publisher={Springer Science and Business Media LLC},
   author={Jendrzejewski, F. and Bernard, A. and Müller, K. and Cheinet, P. and Josse, V. and Piraud, M. and Pezzé, L. and Sanchez-Palencia, L. and Aspect, A. and Bouyer, P.},
   year={2012},
   month=Mar, pages={398–403} 
}

@article{semeghini2015measurement,
  title={Measurement of the mobility edge for 3D Anderson localization},
   volume={11},
   ISSN={1745-2481},
   url={http://dx.doi.org/10.1038/nphys3339},
   DOI={10.1038/nphys3339},
   number={7},
   journal={Nature Physics},
   publisher={Springer Science and Business Media LLC},
   author={Semeghini, G. and Landini, M. and Castilho, P. and Roy, S. and Spagnolli, G. and Trenkwalder, A. and Fattori, M. and Inguscio, M. and Modugno, G.},
   year={2015},
   pages={554–559}
}

@article{chabe2008experimental,
   title={Experimental Observation of the Anderson Metal-Insulator Transition with Atomic Matter Waves},
   volume={101},
   ISSN={1079-7114},
   url={http://dx.doi.org/10.1103/PhysRevLett.101.255702},
   DOI={10.1103/physrevlett.101.255702},
   number={25},
   journal={Physical Review Letters},
   publisher={American Physical Society (APS)},
   author={Chabé, Julien and Lemarié, Gabriel and Grémaud, Benoît and Delande, Dominique and Szriftgiser, Pascal and Garreau, Jean Claude},
   year={2008},
   month=Dec 
}

@article{hu2008localization,
  title={Localization of ultrasound in a three-dimensional elastic network},
   volume={4},
   ISSN={1745-2481},
   url={http://dx.doi.org/10.1038/nphys1101},
   DOI={10.1038/nphys1101},
   number={12},
   journal={Nature Physics},
   publisher={Springer Science and Business Media LLC},
   author={Hu, Hefei and Strybulevych, A. and Page, J. H. and Skipetrov, S. E. and van Tiggelen, B. A.},
   year={2008},
   month=Oct, pages={945–948}
}

@article{fyodorov1992novel,
  title = {A novel field theoretical approach to the {Anderson} localization: sparse random hopping model},
  volume = {2},
  ISSN = {1286-4862},
  url = {http://dx.doi.org/10.1051/jp1:1992229},
  DOI = {10.1051/jp1:1992229},
  number = {8},
  journal = {J. Phys. I France},
  publisher = {EDP Sciences},
  author = {Fyodorov,  Yan V. and Mirlin,  Alexander D. and Sommers,  Hans-J\"{u}rgen},
  year = {1992},
  month = aug,
  pages = {1571–1605}
}

@book{ashcroft1976solid,
  title={Solid state},
  author={Ashcroft, Neil W and Mermin, N David},
  volume={1},
  year={1976},
  publisher ={Harcourt College Publishers}
}

@article{Pietracaprina_2016,
   title={Forward approximation as a mean-field approximation for the Anderson and many-body localization transitions},
   volume={93},
   ISSN={2469-9969},
   url={http://dx.doi.org/10.1103/PhysRevB.93.054201},
   DOI={10.1103/physrevb.93.054201},
   number={5},
   journal={Physical Review B},
   publisher={American Physical Society (APS)},
   author={Pietracaprina, Francesca and Ros, Valentina and Scardicchio, Antonello},
   year={2016},
   month=feb 
}

@article{Mirlin_2000,
title = {Statistics of energy levels and eigenfunctions in disordered systems},
journal = {Phys. Rep.},
volume = {326},
pages = {259-382},
year = {2000},
issn = {0370-1573},
doi = {10.1016/S0370-1573(99)00091-5},
url = {https://www.sciencedirect.com/science/article/pii/S0370157399000915},
author = {Alexander D. Mirlin},
}

@article{rodriguez2011multifractal,
  title = {Multifractal finite-size scaling and universality at the {Anderson} transition},
  author = {Rodriguez, Alberto and Vasquez, Louella J. and Slevin, Keith and R\"omer, Rudolf A.},
  journal = {Phys. Rev. B},
  volume = {84},
  issue = {13},
  pages = {134209},
  numpages = {16},
  year = {2011},
  month = {Oct},
  publisher = {American Physical Society},
  doi = {10.1103/PhysRevB.84.134209},
  url = {https://link.aps.org/doi/10.1103/PhysRevB.84.134209}
}

@article{mott_twose_1961,
  author  = {Mott, N. F. and Twose, W. D.},
  title   = {The theory of impurity conduction},
  journal = {Advances in Physics},
  volume  = {10},
  number  = {38},
  pages   = {107--163},
  year    = {1961},
  publisher = {Taylor \& Francis},
  doi     = {10.1080/00018736100101271}
}

@article{borland1963nature,
  author  = {Borland, R. E.},
  title   = {The Nature of the Electronic States in Disordered One-Dimensional Systems},
  journal = {Proceedings of the Royal Society of London. Series A. Mathematical and Physical Sciences},
  volume  = {274},
  number  = {1356},
  pages   = {529--545},
  year    = {1963},
  publisher = {The Royal Society},
  doi     = {10.1098/rspa.1963.0119}
}

@article{ishii1973localization,
  author  = {K. Ishii},
  title   = {Localization of Eigenstates and Transport Phenomena in One-Dimensional Disordered Systems},
  journal = {Progress of Theoretical Physics Supplement},
  volume  = {53},
  pages   = {77--138},
  year    = {1973},
  doi = {https://doi.org/10.1143/PTPS.53.77}
}

@article{Aspect2009anderson,
  author  = {A. Aspect and M. Inguscio},
  title   = {Anderson localization of ultracold atoms},
  journal = {Physics Today},
  volume  = {62},
  number  = {8},
  pages   = {30--35},
  year    = {2009},
  doi     = {10.1063/1.3206098}
}

@article{Thouless1972ARB,
	author = {D J Thouless},
	doi = {10.1088/0022-3719/5/1/010},
	journal = {Journal of Physics C: Solid State Physics},
	month = {jan},
	number = {1},
	pages = {77},
	publisher = {},
	title = {A relation between the density of states and range of localization for one dimensional random systems},
	url = {https://doi.org/10.1088/0022-3719/5/1/010},
	volume = {5},
	year = {1972}
}

@article{comtet_lyapunov_2013,
  author       = {Alain Comtet and Jean-Marc Luck and Christophe Texier and Yves Tourigny},
  title        = {The {Lyapunov} exponent of products of random 2 × 2 matrices close to the identity},
  journal      = {Journal of Statistical Physics},
  volume       = {150},
  number       = {1},
  year         = {2013},
  pages        = {13--65},
  doi          = {10.1007/s10955-012-0615-5},
}

@article{Texier2013,
  author  = {Alain Comtet and Christophe Texier and Yves Tourigny},
  title   = {Lyapunov exponents, one-dimensional Anderson localisation and products of random matrices},
  journal = {Journal of Physics A: Mathematical and Theoretical},
  volume  = {46},
  number  = {25},
  pages   = {254003},
  year    = {2013},
  publisher = {IOP Publishing},
  doi     = {10.1088/1751-8113/46/25/254003}
}

@article{kramer1993localization,
  author  = {B. Kramer and A. MacKinnon},
  title   = {Localization: Theory and experiment},
  journal = {Reports on Progress in Physics},
  volume  = {56},
  number  = {12},
  pages   = {1469--1564},
  year    = {1993},
  doi     = {10.1088/0034-4885/56/12/001}
}

@article{abrahams1980critical,
  author  = {E. Abrahams and T. V. Ramakrishnan},
  title   = {Critical behavior of disordered metals},
  journal = {Philosophical Magazine},
  volume  = {42},
  pages   = {827--838},
  year    = {1980},
  doi     = {10.1080/14786438008218984}
}

@article{foster2009termination,
  title = {{Termination of typical wave-function multifractal spectra at the Anderson metal-insulator transition: Field theory description using the functional renormalization group}},
  author = {Foster, Matthew S. and Ryu, Shinsei and Ludwig, Andreas W. W.},
  journal = {Phys. Rev. B},
  volume = {80},
  issue = {7},
  pages = {075101},
  numpages = {21},
  year = {2009},
  month = {Aug},
  publisher = {American Physical Society},
  doi = {10.1103/PhysRevB.80.075101},
  url = {https://link.aps.org/doi/10.1103/PhysRevB.80.075101}
}

@article{GarcaGarca2007,
  title = {Dimensional dependence of the metal-insulator transition},
  volume = {75},
  ISSN = {1550-235X},
  url = {http://dx.doi.org/10.1103/PhysRevB.75.174203},
  DOI = {10.1103/physrevb.75.174203},
  number = {17},
  journal = {Physical Review B},
  publisher = {American Physical Society (APS)},
  author = {García-García,  Antonio M. and Cuevas,  Emilio},
  year = {2007},
  month = may 
}

@article{Castellani1986,
  title = {On the upper critical dimension in Anderson localisation},
  volume = {19},
  ISSN = {1361-6447},
  url = {http://dx.doi.org/10.1088/0305-4470/19/17/009},
  DOI = {10.1088/0305-4470/19/17/009},
  number = {17},
  journal = {Journal of Physics A: Mathematical and General},
  publisher = {IOP Publishing},
  author = {Castellani,  C and Castro,  C Di and Peliti,  L},
  year = {1986},
  month = dec,
  pages = {L1099–L1103}
}

@article{minami1996local,
  author       = {N. Minami},
  title        = {Local Fluctuation of the Spectrum of a Multidimensional Anderson Tight Binding Model},
  journal      = {Communications in Mathematical Physics},
  volume       = {177},
  pages        = {709--725},
  year         = {1996},
  doi          = {10.1007/BF02099282}
}

@article{elgart2009lifshitz,
  author    = {Alexander Elgart},
  title     = {Lifshitz tails and localization in the three-dimensional Anderson model},
  journal   = {Duke Mathematical Journal},
  volume    = {146},
  number    = {2},
  pages     = {331--360},
  year      = {2009},
  doi       = {10.1215/00127094-2008-068},
}

@article{klein1994absolutely,
  author       = {Abel Klein},
  title        = {Absolutely Continuous Spectrum in the Anderson Model on the Bethe Lattice},
  journal      = {Mathematical Research Letters},
  volume       = {1},
  number       = {4},
  pages        = {399--407},
  year         = {1994},
  doi          = {10.4310/MRL.1994.v1.n4.a1},
  url          = {https://www.intlpress.com/site/pub/files/_fulltext/journals/mrl/1994/0001/0004/MRL-1994-0001-0004-a001.pdf}
}

@article{aizenman2006absolutely,
  author       = {Michael Aizenman and Robert Sims and Simone Warzel},
  title        = {Absolutely Continuous Spectra of Quantum Tree Graphs with Weak Disorder},
  journal      = {Communications in Mathematical Physics},
  volume       = {264},
  number       = {2},
  pages        = {371--389},
  year         = {2006},
  doi          = {10.1007/s00220-006-0035-6},
  url          = {https://arxiv.org/abs/math-ph/0504039}
}

@article{froese2006transfer,
  author       = {R. Froese and D. Hasler and W. Spitzer},
  title        = {Transfer Matrices, Hyperbolic Geometry and Absolutely Continuous Spectrum for Some Discrete Schr{\"o}dinger Operators on Graphs},
  journal      = {Journal of Functional Analysis},
  volume       = {230},
  pages        = {184--221},
  year         = {2006}
}

@misc{aggarwal2025mobility,
      title={Mobility Edge for the Anderson Model on the Bethe Lattice}, 
      author={Amol Aggarwal and Patrick Lopatto},
      year={2025},
      eprint={2503.08949},
      archivePrefix={arXiv},
      primaryClass={math.PR},
      url={https://arxiv.org/abs/2503.08949}, 
}

@misc{liu2026mobility,
      title={Mobility Edge for the Anderson Model on Random Regular Graphs}, 
      author={Suhan Liu and Patrick Lopatto},
      year={2026},
      eprint={2603.14230},
      archivePrefix={arXiv},
      primaryClass={math.PR},
      url={https://arxiv.org/abs/2603.14230}, 
}

@article{efetov1983supersymmetry,
	author = {K.B. Efetov},
	doi = {10.1080/00018738300101531},
	eprint = { https://doi.org/10.1080/00018738300101531 },
	journal = {Advances in Physics},
	number = {1},
	pages = {53–127},
	publisher = {Taylor \& Francis},
	title = {Supersymmetry and theory of disordered metals},
	url = {https://doi.org/10.1080/00018738300101531},
	volume = {32},
	year = {1983}
}

@book{economou2006green,
  title={Green's Functions in Quantum Physics},
  author={Economou, E.N.},
  isbn={9783540288411},
  lccn={2006926231},
  series={Springer Series in Solid-State Sciences},
  url={https://books.google.fr/books?id=HdJDAAAAQBAJ},
  year={2006},
  publisher={Springer Berlin Heidelberg}
}

@article{CraigSimon1983,
  author  = {Craig, Walter and Simon, Barry},
  title   = {Log Hölder continuity of the integrated density of states for stochastic Jacobi matrices},
  journal = {Communications in Mathematical Physics},
  volume  = {90},
  number  = {2},
  pages   = {207--218},
  year    = {1983},
  doi     = {10.1007/BF01205312}
}

@article{Ising1925,
  author  = {Ising, Ernst},
  title   = {Beitrag zur Theorie des Ferromagnetismus},
  journal = {Zeitschrift f{\"u}r Physik},
  volume  = {31},
  number  = {1},
  pages   = {253--258},
  year    = {1925},
  doi     = {10.1007/BF01333358}
}

@article{erdos1959random1,
  author  = {Erd{\H{o}}s, P. and R\'{e}nyi, A.},
  title   = {On Random Graphs I},
  journal = {Publicationes Mathematicae Debrecen},
  year    = {1959},
  volume  = {6},
  pages   = {290--297},
  doi     = {https://doi.org/10.5486/pmd.1959.6.3-4.12}
}

@article{erdos1961connectedness,
	address = "Budapest, Hungary",
	author = "P. Erdős and A. Rényi",
	doi = "10.1007/bf02066689",
	journal = "Acta Mathematica Hungarica",
	number = "1-2",
	pages = "261–267",
	publisher = "Akadémiai Kiadó, co-published with Springer Science+Business Media B.V., Formerly Kluwer Academic Publishers B.V.",
	title = "On the strength of connectedness of a random graph",
	url = "https://www.akjournals.com/view/journals/10473/12/1-2/article-p261.xml",
	volume = "12",
	year = "1961"
}

@article{erdos1960evolutionRandomGraph,
  title={On the evolution of random graphs},
  author={Paul L. Erdos and Alfr{\'e}d R{\'e}nyi},
  journal={Transactions of the American Mathematical Society},
  year={1984},
  volume={286},
  pages={257-257},
  doi= {https://doi.org/10.1090/S0002-9947-1984-0756039-5}
}

@article{Bogacz2006,
   title={Homogeneous complex networks},
   volume={366},
   ISSN={0378-4371},
   url={http://dx.doi.org/10.1016/j.physa.2005.10.024},
   DOI={10.1016/j.physa.2005.10.024},
   journal={Physica A: Statistical Mechanics and its Applications},
   publisher={Elsevier BV},
   author={Bogacz, Leszek and Burda, Zdzisław and Wacław, Bartłomiej},
   year={2006},
   pages={587–607} 
}

@book{newman2018networks,
  title={Networks},
  author={Newman, Mark},
  year={2018},
  publisher={Oxford university press},
  doi = {https://doi.org/10.1093/oso/9780198805090.001.0001}
}

@book{dorogovtsev2022nature,
  title={The nature of complex networks},
  author={Dorogovtsev, Sergey N and Mendes, Jos{\'e} FF},
  year={2022},
  publisher={Oxford University Press},
  doi = {https://doi.org/10.1093/oso/9780199695119.001.0001}
}

@article{Marinari2004,
 author    = {Enzo Marinari and R{\'e}mi Monasson},
  title     = {Circuits in random graphs: from local trees to global loops},
  journal   = {Journal of Statistical Mechanics: Theory and Experiment},
  volume    = {2004},
  number    = {09},
  pages     = {P09004},
  year      = {2004},
  publisher = {IOP Publishing},
  doi       = {10.1088/1742-5468/2004/09/P09004}
}

@article{MarinariMonassonSemerjian2004,
  author    = {Enzo Marinari and R{\'e}mi Monasson and Guilhem Semerjian},
  title     = {An algorithm for counting circuits: application to real-world and random graphs},
  journal   = {Europhysics Letters},
  volume    = {73},
  number    = {1},
  pages     = {8--14},
  year      = {2006},
  publisher = {IOP Publishing},
  doi       = {10.1209/epl/i2005-10343-0}
}

@book{Bollobas2001,
   title={Random graphs},
   author={Bollob{\'a}s, B{\'e}la},
   year={2001},
   publisher={Cambridge University Press},
   doi={https://doi.org/10.1017/CBO9780511814068}
 }

@misc{krzakala2011belief,
  title={Belief Propagation for the (physicist) layman},
  author={Krzakala, Florent},
  journal={Lecture Notes},
  year={2011},
  url={http://www.lps.ens.fr/~krzakala/BP.pdf}
}

@article{Susca_Vivo_Kuhn_2021,
  author    = {Susca, V. A. R. and Vivo, P. and Kühn, R.},
  title     = {Cavity and replica methods for the spectral density of sparse symmetric random matrices},
  journal   = {SciPost Physics Lecture Notes},
  volume    = {33},
  pages     = {1--43},
  year      = {2021},
  doi       = {10.21468/SciPostPhysLectNotes.33},
}

@article{BordenaveLelarge2010,
  author       = {Charles Bordenave and Marc Lelarge},
  title        = {Resolvent of Large Random Graphs},
  journal      = {Random Structures \& Algorithms},
  volume       = {37},
  number       = {3},
  pages        = {332--352},
  year         = {2010},
  doi          = {10.1002/rsa.20313},
}

@article{arous2008spectrum,
  title={The spectrum of heavy tailed random matrices},
  author={Arous, G{\'e}rard Ben and Guionnet, Alice},
  journal={Commun. Math. Phys.},
  volume={278},
  number={3},
  pages={715--751},
  year={2008},
  publisher={Springer},
 url={https://doi.org/10.1007/s00220-007-0389-x}
}

@phdthesis{tarquini2016anderson,
  author       = {Elena Tarquini},
  title        = {Anderson Localization in High Dimensional Lattices},
  school       = {Université Paris-Saclay (COmUE)},
  year         = {2016},
  type         = {PhD thesis},
  language     = {English},
  eprint       = {tel-01727220},
  eprinttype   = {tel},
  note         = {NNT: 2016SACLS540},
  url          = {https://tel.archives-ouvertes.fr/tel-01727220}
}

@article{Biroli_Tarzia_semerjan_2010,
   title={Anderson Model on Bethe Lattices: Density of States, Localization Properties and Isolated Eigenvalue},
   volume={184},
   ISSN={0375-9687},
   url={http://dx.doi.org/10.1143/PTPS.184.187},
   DOI={10.1143/ptps.184.187},
   journal={Progress of Theoretical Physics Supplement},
   publisher={Oxford University Press (OUP)},
   author={Biroli, Giulio and Semerjian, Guilhem and Tarzia, Marco},
   year={2010},
   pages={187–199}
}

@book{Livan_2018,
   title={Introduction to Random Matrices},
   ISBN={9783319708850},
   ISSN={2197-1765},
   url={http://dx.doi.org/10.1007/978-3-319-70885-0},
   DOI={10.1007/978-3-319-70885-0},
   journal={SpringerBriefs in Mathematical Physics},
   publisher={Springer International Publishing},
   author={Livan, Giacomo and Novaes, Marcel and Vivo, Pierpaolo},
   year={2018} }

@article{molloy1995critical,
    author = {Molloy, Michael and Reed, Bruce},
    title = {A critical point for random graphs with a given degree sequence},
    journal = {Random Structures \& Algorithms},
    volume = {6},
    number = {2-3},
    pages = {161-180},
    doi = {https://doi.org/10.1002/rsa.3240060204},
    url = {https://onlinelibrary.wiley.com/doi/abs/10.1002/rsa.3240060204},
    eprint = {https://onlinelibrary.wiley.com/doi/pdf/10.1002/rsa.3240060204},
    year = {1995}
}

@article{Schur1917,
  author    = {Schur, Issai},
  title     = {Über Potenzreihen, die im Innern des Einheitskreises beschränkt sind},
  journal   = {Jahresbericht der Deutschen Mathematiker-Vereinigung},
  volume    = {25},
  pages     = {205--232},
  year      = {1917},
  doi       = {10.1515/crll.1917.147.205},
}

@book{morters2008growth,
  editor    = {Peter M{\"o}rters and Roger Moser and Mathew Penrose and Hartmut Schwetlick and Johannes Zimmer},
  title     = {Analysis and Stochastics of Growth Processes and Interface Models},
  publisher = {Oxford University Press},
  year      = {2008},
  address   = {Oxford, UK},
  isbn      = {978-0-19-923925-2},
  doi       = {10.1093/acprof:oso/9780199239252.001.0001}
}

@article{derrida1988polymers,
  title={Polymers on disordered trees, spin glasses, and traveling waves},
  author={Derrida, Bernard and Spohn, Herbert},
  journal={Journal of Statistical Physics},
  volume={51},
  number={5},
  pages={817--840},
  year={1988},
  publisher={Springer}
}

@article{monthus2011anderson,
  title={Anderson localization on the Cayley tree: multifractal statistics of the transmission at criticality and off criticality},
  author={Monthus, C{\'e}cile and Garel, Thomas},
  journal={Journal of Physics A: Mathematical and Theoretical},
  volume={44},
  number={14},
  pages={145001},
  year={2011},
  publisher={IOP Publishing}
}

@article{monthus2008anderson,
  title={Anderson transition on the Cayley tree as a traveling wave critical point for various probability distributions},
  author={Monthus, C{\'e}cile and Garel, Thomas},
  journal={Journal of Physics A: Mathematical and Theoretical},
  volume={42},
  number={7},
  pages={075002},
  year={2008},
  publisher={IOP Publishing}
}

@article{tikhonov2016fractality,
  title={Fractality of wave functions on a Cayley tree: Difference between tree and locally treelike graph without boundary},
  author={Tikhonov, Konstantin S and Mirlin, Alexander D},
  journal={Physical Review B},
  volume={94},
  number={18},
  pages={184203},
  year={2016},
  publisher={APS}
}

@article{sonner2017multifractality,
  title={Multifractality of wave functions on a Cayley tree: From root to leaves},
  author={Sonner, M and Tikhonov, KS and Mirlin, AD},
  journal={Physical Review B},
  volume={96},
  number={21},
  pages={214204},
  year={2017},
  publisher={APS}
}

@article{lemarie2019glassy,
  title={Glassy properties of anderson localization: Pinning, avalanches, and chaos},
  author={Lemari{\'e}, Gabriel},
  journal={Physical review letters},
  volume={122},
  number={3},
  pages={030401},
  year={2019},
  publisher={APS}
}

@article{chakrabarti2022traveling,
  title={Traveling/non-traveling phase transition and non-ergodic properties in the random transverse-field Ising model on the Cayley tree},
  author={Chakrabarti, Ankita and Martins, Cyril and Laflorencie, Nicolas and Georgeot, Bertrand and Brunet, {\'E}ric and Lemari{\'e}, Gabriel},
  journal={arXiv preprint arXiv:2212.13593},
  year={2022}
}

@book{mezard1987spin,
  title={Spin glass theory and beyond: An Introduction to the Replica Method and Its Applications},
  author={M{\'e}zard, Marc and Parisi, Giorgio and Virasoro, Miguel Angel},
  volume={9},
  year={1987},
  publisher={World Scientific Publishing Company}
}

@article{derrida1980random,
  title={Random-energy model: Limit of a family of disordered models},
  author={Derrida, Bernard},
  journal={Physical Review Letters},
  volume={45},
  number={2},
  pages={79},
  year={1980},
  publisher={APS}
}

@article{kpp1937,
  title={},
  author={Kolmogorov, A. and Petrovsky, I. and Piskounov, B.},
  journal={Moscow Univ. Bull. Math.},
  volume={1},
  pages={1},
  year={1937},
}

@article{Kravtsov2018nonergodic,
   title={Non-ergodic delocalized phase in Anderson model on Bethe lattice and regular graph},
   volume={389},
   ISSN={0003-4916},
   url={http://dx.doi.org/10.1016/j.aop.2017.12.009},
   DOI={10.1016/j.aop.2017.12.009},
   journal={Annals of Physics},
   publisher={Elsevier BV},
   author={Kravtsov, V.E. and Altshuler, B.L. and Ioffe, L.B.},
   year={2018},
   month=Feb, pages={148–191} }

@article{pino2020scaling,
  title = {Scaling up the {Anderson} transition in random-regular graphs},
  author = {Pino, M.},
  journal = {Phys. Rev. Res.},
  volume = {2},
  issue = {4},
  pages = {042031},
  numpages = {6},
  year = {2020},
  month = {Nov},
  publisher = {American Physical Society},
  doi = {10.1103/PhysRevResearch.2.042031},
  url = {https://link.aps.org/doi/10.1103/PhysRevResearch.2.042031}
}

@article{bera2018return,
  title = {Return probability for the {Anderson} model on the random regular graph},
  author = {Bera, Soumya and De Tomasi, Giuseppe and Khaymovich, Ivan M. and Scardicchio, Antonello},
  journal = {Phys. Rev. B},
  volume = {98},
  issue = {13},
  pages = {134205},
  numpages = {9},
  year = {2018},
  month = {Oct},
  publisher = {American Physical Society},
  doi = {10.1103/PhysRevB.98.134205},
  url = {https://link.aps.org/doi/10.1103/PhysRevB.98.134205}
}

@article{de2020subdiffusion,
  title = {Subdiffusion in the {Anderson} model on the random regular graph},
  author = {De Tomasi, Giuseppe and Bera, Soumya and Scardicchio, Antonello and Khaymovich, Ivan M.},
  journal = {Phys. Rev. B},
  volume = {101},
  issue = {10},
  pages = {100201},
  numpages = {7},
  year = {2020},
  month = {Mar},
  publisher = {American Physical Society},
  doi = {10.1103/PhysRevB.101.100201},
  url = {https://link.aps.org/doi/10.1103/PhysRevB.101.100201}
}

@article{tikhonov2016anderson,
  title = {Anderson localization and ergodicity on random regular graphs},
  author = {Tikhonov, K. S. and Mirlin, A. D. and Skvortsov, M. A.},
  journal = {Phys. Rev. B},
  volume = {94},
  issue = {22},
  pages = {220203},
  numpages = {6},
  year = {2016},
  month = {Dec},
  publisher = {American Physical Society},
  doi = {10.1103/PhysRevB.94.220203},
  url = {https://link.aps.org/doi/10.1103/PhysRevB.94.220203}
}

@article{garcia2017scaling,
  title = {Scaling Theory of the {Anderson} Transition in Random Graphs: Ergodicity and Universality},
  author = {Garc\'{\i}a-Mata, I. and Giraud, O. and Georgeot, B. and Martin, J. and Dubertrand, R. and Lemari\'e, G.},
  journal = {Phys. Rev. Lett.},
  volume = {118},
  issue = {16},
  pages = {166801},
  numpages = {5},
  year = {2017},
  month = {Apr},
  publisher = {American Physical Society},
  doi = {10.1103/PhysRevLett.118.166801},
  url = {https://link.aps.org/doi/10.1103/PhysRevLett.118.166801}
}

@article{angelini2025bethe,
  title={Bethe $ M $-layer construction for the percolation problem},
  author={Angelini, Maria Chiara and Palazzi, Saverio and Rizzo, Tommaso and Tarzia, Marco},
  journal={SciPost Physics},
  volume={18},
  number={1},
  pages={030},
  year={2025}
}

@article{cugliandolo2024multifractal,
  title={Multifractal phase in the weighted adjacency matrices of random Erd{\"o}s-R{\'e}nyi graphs},
  author={Cugliandolo, Leticia F and Schehr, Gr{\'e}gory and Tarzia, Marco and Venturelli, Davide},
  journal={Physical Review B},
  volume={110},
  number={17},
  pages={174202},
  year={2024},
  publisher={APS}
}

@article{Tapias2023,
  title = {Multifractality and statistical localization in highly heterogeneous random networks},
  volume = {144},
  ISSN = {1286-4854},
  url = {http://dx.doi.org/10.1209/0295-5075/ad1001},
  DOI = {10.1209/0295-5075/ad1001},
  number = {4},
  journal = {Europhys. Lett.},
  publisher = {IOP Publishing},
  author = {Tapias,  Diego and Sollich,  Peter},
  year = {2023},
  month = nov,
  pages = {41001}
}

@Article{Silva2025spectral,
	title={{Spectral properties, localization transition and multifractal eigenvectors of the Laplacian on heterogeneous networks}},
	author={Jeferson D. da Silva and Diego Tapias and Peter Sollich and Fernando L. Metz},
	journal={SciPost Phys.},
	volume={18},
	pages={047},
	year={2025},
	publisher={SciPost},
	doi={10.21468/SciPostPhys.18.2.047},
	url={https://scipost.org/10.21468/SciPostPhys.18.2.047},
}

@article{tarzia2022fully,
  title = {Fully localized and partially delocalized states in the tails of Erd\"os-R\'enyi graphs in the critical regime},
  author = {Tarzia, M.},
  journal = {Phys. Rev. B},
  volume = {105},
  issue = {17},
  pages = {174201},
  numpages = {23},
  year = {2022},
  month = {May},
  publisher = {American Physical Society},
  doi = {10.1103/PhysRevB.105.174201},
  url = {https://link.aps.org/doi/10.1103/PhysRevB.105.174201}
}

@article{prado2026anderson,
   title={Anderson localization on Husimi trees and its implications for many-body localization},
   volume={113},
   ISSN={2469-9969},
   url={http://dx.doi.org/10.1103/q3bz-97v6},
   DOI={10.1103/q3bz-97v6},
   number={14},
   journal={Physical Review B},
   publisher={American Physical Society (APS)},
   author={Prado Bandeira, Dafne and Tarzia, Marco},
   year={2026},
   month=Apr }

@article{vanoni2024renormalization,
  title={Renormalization group analysis of the Anderson model on random regular graphs},
  author={Vanoni, Carlo and Altshuler, Boris L and Kravtsov, Vladimir E and Scardicchio, Antonello},
  journal={Proceedings of the National Academy of Sciences},
  volume={121},
  number={29},
  pages={e2401955121},
  year={2024},
  publisher={National Academy of Sciences}
}

@article{sierant2023universality,
  title={Universality in Anderson localization on random graphs with varying connectivity},
  author={Sierant, Piotr and Lewenstein, Maciej and Scardicchio, Antonello},
  journal={SciPost Physics},
  volume={15},
  number={2},
  pages={045},
  year={2023}
}

@article{PhysRevResearch.2.012020,
  title = {Two critical localization lengths in the Anderson transition on random graphs},
  author = {Garc\'{\i}a-Mata, I. and Martin, J. and Dubertrand, R. and Giraud, O. and Georgeot, B. and Lemari\'e, G.},
  journal = {Phys. Rev. Res.},
  volume = {2},
  issue = {1},
  pages = {012020(R)},
  numpages = {7},
  year = {2020},
  month = {Jan},
  publisher = {American Physical Society},
  doi = {10.1103/PhysRevResearch.2.012020},
  url = {https://link.aps.org/doi/10.1103/PhysRevResearch.2.012020}
}

@incollection{warzel2012,
author = {S. Warzel},
title = {Surprises in the phase diagram of the Anderson model on the Bethe lattice},
booktitle = {XVIIth International Congress on Mathematical Physics},
pages = {239-253},
year = {2012},
publisher = {World Scientific},
address = {Singapore},
doi = {10.1142/9789814449243_0014},
URL = {https://www.worldscientific.com/doi/abs/10.1142/9789814449243_0014},
}

@article{aizenman1994localization,
  title={Localization at weak disorder: some elementary bounds},
  author={Aizenman, Michael},
  journal={Reviews in mathematical physics},
  volume={6},
  number={05a},
  pages={1163--1182},
  year={1994},
  publisher={World Scientific}
}

@article{aizenman1993localization,
  title={Localization at large disorder and at extreme energies: An elementary derivations},
  author={Aizenman, Michael and Molchanov, Stanislav},
  journal={Communications in Mathematical Physics},
  volume={157},
  number={2},
  pages={245--278},
  year={1993},
  publisher={Springer}
}

@article{aizenman2011extended,
  title={Extended states in a Lifshitz tail regime for random Schr{\"o}dinger operators on trees},
  author={Aizenman, Michael and Warzel, Simone},
  journal={Physical review letters},
  volume={106},
  number={13},
  pages={136804},
  year={2011},
  publisher={APS}
}

@article{frohlich1983absence,
  author  = {Fr{\"o}hlich, J{\"u}rg and Spencer, Thomas},
  title   = {Absence of diffusion in the {A}nderson tight binding model for large disorder or low energy},
  journal = {Communications in Mathematical Physics},
  volume  = {88},
  number  = {2},
  pages   = {151--184},
  year    = {1983},
  doi     = {10.1007/BF01209475}
}
\end{document}